\documentclass{aims}
\usepackage{amsmath}
  \usepackage{paralist}
  \usepackage{graphics} %% add this and next lines if pictures should be in esp format
  \usepackage{epsfig} %For pictures: screened artwork should be set up with an 85 or 100 line screen
\usepackage{graphicx}  \usepackage{epstopdf}%This is to transfer .eps figure to .pdf figure; please compile your paper using PDFLeTex or PDFTeXify.
 \usepackage[colorlinks=true]{hyperref}
 \usepackage{cleveref}
 \usepackage{multirow}
\hypersetup{urlcolor=blue, citecolor=red}

\def\currentvolume{X}
 \def\currentissue{X}
  \def\currentyear{200X}
   \def\currentmonth{XX}
    \def\ppages{X--XX}
     \def\DOI{10.3934/xx.xx.xx.xx}

\theoremstyle{definition}

\newcommand{\transp}{^{\rm T}}  

\title[Region of interest reconstruction in EIT]
{Estimation of conductivity changes in a region of interest with electrical impedance tomography}

\author[Liu Kolehmainen Siltanen Laukkanen and Sepp{\"a}nen]{}


 \email{dong.liu@uef.fi}
 \email{ville.kolehmainen@uef.fi}
 \email{samuli.siltanen@helsinki.fi}
  \email{anne-maria.laukkanen@uta.fi}
   \email{aku.seppanen@uef.fi}

\thanks{This work was supported by the Academy of Finland under
projects 119270, 134868, 140280, 250215 and 270174,
Finnish Doctoral Programme in Computational Sciences and 
Finnish Center of Excellence of Inverse Problems Research 2012-2017.}

\begin{document}
\maketitle

% Enter the first author's name and address:
\centerline{\scshape Dong liu and Ville Kolehmainen }
\medskip
{\footnotesize
% please put the address of the first author
 \centerline{Department
of Applied Physics, University of Eastern Finland, FIN-70211 Kuopio, Finland}
%    \centerline{Other lines}
%    \centerline{ Springfield, MO 65801-2604, USA}
} % Do not forget to end the {\footnotesize by the sign }

\medskip

\centerline{\scshape Samuli Siltanen}
\medskip
{\footnotesize
 % please put the address of the second  and third author
 \centerline{ Department of Mathematics and Statistics, University of Helsinki, FIN-00014 Helsinki, Finland}
%    \centerline{Other lines}
%    \centerline{Springfield, MO 65810, USA}
}

\medskip

\centerline{\scshape Anne-maria Laukkanen}
\medskip
{\footnotesize
 % please put the address of the second  and third author
 \centerline{ School of Education, University of Tampere, FIN-33014 Tampere, Finland}
%    \centerline{Other lines}
%    \centerline{Springfield, MO 65810, USA}
}

\medskip

\centerline{\scshape Aku Sepp{\"a}nen}
\medskip
{\footnotesize
 % please put the address of the second  and third author
 \centerline{ Department
of Applied Physics, University of Eastern Finland, FIN-70211 Kuopio, Finland}
%    \centerline{Other lines}
%    \centerline{Springfield, MO 65810, USA}
}

\bigskip

% The name of the associate editor will be entered by an editorial staff
% "Communicated by the associate editor name" is not needed for special issue.
 \centerline{(Communicated by the associate editor name)}

%The abstract of your paper
\begin{abstract}
This paper proposes a novel approach to reconstruct changes
in a target conductivity from electrical impedance tomography measurements. 
As in the conventional {\em difference imaging},
the reconstruction of the conductivity change is based on electrical potential measurements 
from the exterior boundary of the target before and after the change. 
In this paper, however, images of the conductivity 
before and after the change are reconstructed simultaneously based on the two data sets. 
The key feature of the approach is that the conductivity after the change is parameterized as a linear combination of the initial state and the change. 
This allows for modeling independently the spatial characteristics of the background conductivity
and the change of the conductivity - by separate regularization functionals. 
The approach also allows in a straightforward way the restriction of the conductivity change to a localized region of interest inside the domain. 
While conventional difference imaging reconstruction is based on a global linearization of the observation model, the proposed approach amounts to solving a non-linear inverse problem. The feasibility of the proposed reconstruction method is tested experimentally and with a simulation which demonstrates a potential new medical application of electrical impedance tomography: 
imaging of vocal folds in voice loading studies.
\end{abstract}

%The title of your section 1
\section{Introduction}

\label{sect.Introduction}
In electrical impedance tomography (EIT), 
electrical currents are injected into an object 
using a set of electrodes attached on the boundary of the object and the resulting electrode 
potentials are measured.
The  
conductivity of the object is reconstructed as a spatially distributed parameter 
based on the known currents and measured potentials.
EIT has been applied to e.g. geophysical exploration \cite{Daily1992},
biomedical imaging 
\cite{frerichs2002detection,Cherepenin2002,zou2003review,boverman2008robust,
tidswell2001three,bagshaw2003electrical},
industrial process monitoring and control \cite{ScottMcCann2005}
and non-destructive testing  \cite{Hou2007, Hou2009, Karhunen2010a, Karhunen2010c}.
%The proposed biomedical applications of EIT include monitoring of lung function
%\cite{frerichs2000electrical,frerichs2002detection}, 
%detection of breast cancer \cite{Cherepenin2002,zou2003review,boverman2008robust}
%and imaging of brain activity \cite{tidswell2001three,bagshaw2003electrical}. 
For reviews of EIT, see \cite{Cheney1999,kaipio2000,Brown2003,holder2005electrical,mueller2012}.

The EIT image reconstruction problem is an ill-posed inverse problem. 
The generalized Tikhonov regularization framework
\cite{vauhkonen1998b} is often used 
in practical applications.
%In the deterministic inversion framework, 
%the most commonly used method 
%is the generalized Tikhonov regularization
%\cite{vauhkonen1998b}. 
Other options 
within the deterministic inversion framework
include variational regularization with sparsity constraints \cite{JinMaass} 
and direct methods, such as the D-bar method  \cite{Siltanen2000,Isaacson2004,Isaacson2006,Knudsen2009}; see also \cite[Table 14.1]{mueller2012}.
For Bayesian approaches, see \cite{kaipio2000,kaipio2005}.
Reconstructing the distributed conductivity by
solving the non-linear inverse problem of EIT
is referred to as {\em absolute imaging}
\cite{Vauhkonen1999,zou2003review}.

Many applications of EIT aim at monitoring a time-varying object.
Often the main interest is to track {\em changes} in the conductivity between two measurements,
rather than estimate absolute conductivity values.
In {\em difference imaging}
\cite{Cheney1999,tidswell2001three,Brown2003,bagshaw2003electrical,frerichs2002detection,victorino2004imbalances,costa2008real},
the conductivity change is estimated based on difference data,
i.e. difference between the electrode potentials before and after the change.
A conventional approach to image reconstruction in difference imaging
is to linearize the mapping between the electrical conductivity and the electrode potentials
globally at some {\em a priori} defined conductivity,
and to solve a linear inverse problem such as a regularized linear least squares problem. 
Due to the linearization, the images are often only qualitative in nature 
and their spatial resolution may be low, 
especially if the target conductivity is highly inhomogeneous before and/or after the change.

We note that some exceptions to linearization-based approach exist:
the D-bar method can be used for both absolute and 
difference imaging
\cite{Isaacson2004,Isaacson2006}
without linearizing the model.
In the present paper, we formulate the reconstruction problem in the
generalized Tikhonov regularization framework.
Also in the proposed reconstruction method, the global linearization of the model is not needed.
We also note that the proposed method could be
formulated equivalently in the Bayesian inversion framework.

The conventional linearization-based reconstruction approach is widely used in difference imaging,
especially because the reconstructions are fast to compute, 
and because the reconstructions are to some extent 
tolerant of modeling errors because of the partial cancellation of the
errors in the subtraction of the measurement data from the data after the change.
%
%In medical applications of EIT, for example,
Typical sources of modeling errors in EIT are 
the unknown shape of the body,
inaccurately known electrode positions and
unknown contact impedances between the electrodes and the body surface.
If these uncertainties are not accounted for,
% in the reconstructions,
the solutions of the non-linear absolute imaging problem 
can be highly erroneous,
due to the ill-posedness of the inverse problem.
However, if all model uncertainties 
are modeled properly, the absolute imaging can give quantitative information on
the tissue properties.
Recently, several approaches for handling the above
uncertainties and other model inaccuracies have been developed.
For the recovery from unknown boundary shape and/or electrode locations,
alternative approaches have been proposed 
\cite{Kolehmainen05,kolehmainen2007inverse,kolehmainen2008electrical,
nissinen2011compensation,Nissinen11b,BoyleA12,DardeJeremi12,Darde13,Darde13b}.
All these reconstruction methods have also been evaluated with real measurement data
\cite{kolehmainen2008electrical,nissinen2011compensation,BoyleA12,Darde13b}.
For two alternative computational methods for recovering from 
unknown contact impedances in absolute imaging, see
\cite{vilhunen2002a,nissinen2009}.
In the latter paper, also modeling errors due to
discretization and domain truncation
were accounted for, by using the so-called
approximation error method.

If the application at hand does not require real-time imaging
and if quantitative information is desired,
a preferable choice is to handle the model uncertainties properly
and to compute solution to the non-linear problem instead of
using the linearization based approach.
% compute absolute images instead of difference images.
%If the model uncertainties are handled properly,
%then difference imaging approach will not be needed for reducing the effects of modeling errors.
%In such case, absolute imaging is a preferable choice, because it is not based on a linear approximation 
%and leads to quantitative reconstructions.
However, when the measurement data of the target before and after the change are available,
it might be possible to 
somehow combine the data sets within the absolute EIT imaging framework.
Our hypothesis is that 
this could improve the
quality of the absolute EIT reconstructions,
especially if 
the change of the conductivity is known to be restricted to a subvolume
(region of interest, ROI) inside the body.
Examples of such applications  are
monitoring of water ingress in soil \cite{Daily1992}
and cracking of concrete \cite{Hou2009,Karhunen2010c},
assessment of regional lung ventilation
\cite{victorino2004imbalances,lindgren2007,costa2009electrical,leonhardt2012,pulletz2012regional}
and monitoring of 
%cardiac stroke \cite{zlochiver2006} and
intraperitoneal bleeding \cite{wanjun2008}.
This is also the case in
a potential new application of EIT:
imaging of vocal folds in voice loading studies
\cite{kob2009,Seppanen2011,Hezard,Liu2012}.
Indeed,
the location of the glottis is known relatively well,
and
vocal folds are the most rapidly moving part in a human body;
hence,
during the movement of vocal folds,
the conductivity changes outside a relatively small volume around the glottis
are negligible.

In this paper,
we formulate the non-linear reconstruction problem
for a temporally varying target
so that the measurements before and after the target change
are concatenated into single measurement vector,
and the two conductivity distributions
are reconstructed simultaneously
based on the combined data.
The key feature in the approach is that conductivity after the change is represented 
as a linear combination of the initial state and the change of the conductivity. 
This
allows for modeling separately the spatial characteristics of the
background conductivity and the change of the conductivity
- by separate regularization functionals.
This also enables the
restriction of the conductivity change to a region of interest.

The rest of this paper is organized as follows. 
In Section \ref{eit}, a brief review of the EIT observation model is given,
and the properties of the absolute and difference imaging are outlined.
In Section \ref{lad}, the new approach for reconstructing changes in the region of interest is proposed.
The proposed approach is tested with experiments in Section
\ref{sec.experimentalstudies}
and with a simulation related to glottis imaging application
in Section \ref{sec.simulation}.
Conclusions are drawn in Section \ref{conclusion}.
%%%%%%%%%%%%%%%%%%%%%%%%%%%%%%%%%%%%%%%%%%%%%%%%%%%%%%%%%
%%%%%%%%%%%%%%%%%%%%%%%%%%%%%%%%%%%%%%%%%%%%%%%%%%%%%%%%%

\section{Absolute and difference imaging in EIT}
\label{eit}

\subsection{Forward Model}

Let $\Omega\subset\mathbb{R}^q, q=2,3$ denote the body to be imaged.
In EIT, $L$ contact electrodes are attached to the positions 
$e_{\ell}\subset \partial\Omega \,\ \ell=1,2,\ldots,L$
on the boundary $\partial\Omega$.
Electric currents are injected into the body using the electrodes, 
and the resulting electrode potentials are measured.
These measurements may consist of potential differences (voltages)
between a set of electrode pairs, 
or potentials of the electrodes with respect to a common ground.

The most accurate physically realizable model for such measurements is referred to as
the complete electrode model \cite{cheng1989electrode}  
\begin{align}
\nabla \cdot (\sigma(x) \nabla u(x) )&= 0 \ ,\  x \in \Omega 
\label{malli1} \\
u(x)+z_{\ell} \sigma(x) \frac{\partial u(x)}{\partial n}&=U_{\ell},\ x \in e_{\ell},\ \ell=1,...,L
\label{malli2} \\  
% \int_{e_{\ell}}\sigma(x) \frac{\partial u(x)}{\partial n} \hspace{2mm}dS&=&I_{\ell} \
% ,\ \ell=1,2,...,L
\int_{e_{\ell}}\sigma(x) \frac{\partial u(x)}{\partial n} \,\mathrm{d}S&=I_{\ell}, \
\ \ell=1,...,L
\label{malli3}  \\ 
\sigma(x) \frac{\partial u(x)}{\partial n}&=0, \ \  x\in
\partial\Omega \backslash \bigcup_{\ell=1}^{L} e_{\ell}  
\label{malli4} 
\end{align}
where $x\in\Omega$ is the spatial coordinate,  
$\sigma(x)$ is the conductivity, ${u(x)}$ is the electric potential distribution inside $\Omega$, 
$U_\ell$ and $I_\ell$ are the potential and current at electrode $\ell$, respectively,
$z_\ell$ is the contact impedance between the electrode $e_\ell$ and the body $\Omega$, 
and $n$ denotes the outward unit normal vector on the boundary $\partial\Omega$. 
The currents satisfy the charge conservation law
\begin{equation}
 \sum_{\ell=1}^L {I_\ell} =0,
 \label{malli5} 
\end{equation}
and a ground level for the potentials needs to be fixed,
for example by writing
\begin{equation}
 \sum_{\ell=1}^L {U_\ell} =0.
 \label{malli6} 
\end{equation}

The existence and uniqueness of the solution
of the model (\ref{malli1}-\ref{malli6}) was proven
and its weak form written in
\cite{somersalo1992existence}.
For the \textit{finite element} (FE) approximation of the model,
see \cite{Vauhkonen1999}.

Assuming an additive Gaussian noise model,
the FE approximation of (\ref{malli1}-\ref{malli6})
leads to the observation model
\begin{equation}
\label{eq.EITobs_eq}
   V = U(\sigma)+e
%\quad e\sim \mathcal{N}(e^*,\Gamma_e)
\end{equation}
where $V\in\mathbb{R}^M$ is the vector including all the measured electrode potentials.
Here, $M=m N_{\mathrm{inj}}$, where $N_{\mathrm{inj}}$ denotes the number of
current injections and $m$ is the number of measured potentials or voltages for each current injection.
Further,
$U$ is the mapping between the finite dimensional approximation of the conductivity $\sigma$
and the electrode potentials, and
$e\in \mathbb{R}^M$ is the Gaussian distributed measurement noise 
$e \sim \mathcal{N}(e^*,\Gamma_{e})$.
The mean $e^*\in \mathbb{R}^M$ and the covariance matrix $\Gamma_e\in \mathbb{R}^{M\times M}$
can usually be determined experimentally,
see \cite{heikkinen02}.

%%%%%%%%%%%%%%%%%%%%%%%%%%%%%%%%%%%%%%%%%%%%%%%%%%%%%%%%%
%%%%%%%%%%%%%%%%%%%%%%%%%%%%%%%%%%%%%%%%%%%%%%%%%%%%%%%%%

\subsection{Absolute imaging}
\label{subsec.absoluteEIT}

In absolute imaging,
the conductivity $\sigma$ is reconstructed
using a single set of potential measurements $V$
during which the target is assumed to be non-varying.

Within the {\em generalized Tikhonov regularization} framework,
the estimate for the conductivity is obtained as 
\begin{equation}
  \hat\sigma= {\rm arg} \min_{\sigma >0}\{\Vert L_{e}(V-U(\sigma))\Vert^2
+p_{ \sigma}(\sigma)\}
\label{eq.absolute.estimate}
\end{equation}
where
$L_e$ is a Cholesky factor of the noise precision matrix, i.e. 
%inverted covariance matrix $\Gamma_e$, i.e.
$L_e\transp L_e=\Gamma_e^{-1}$, and
$p_{ \sigma}(\sigma)$ is a regularization functional.
The functional $p_{ \sigma}(\sigma)$ is usually designed such that
it gives high penalty for unwanted/improbable features of $\sigma$.
Examples of widely used regularization functionals are smoothness regularization
$p_\sigma (\sigma) = \| L \sigma \|^2$
where $L\sigma$ is some (possibly spatially and directionally weighted) 
differential of the conductivity \cite{phillips62, kaipio2005,kaipio99c} and  
% spatially weighted $L_2$ norm {kaipio99c}, 
% spatially weighted inhomogeneous and anisotropic (structural) smoothness  \cite{kaipio99c},
% total variation (TV) priors \cite{dobson93,kaipio2000,BillsformerstudentTVcirca2010}, and
total variation (TV) regularization $p_\sigma (\sigma) = \alpha \| \nabla \sigma \|_1$ \cite{RudinOsherFatemi,dobson93,kaipio2000}.
% $L_1$ (impulse) priors \cite{kaipio2000}.
% and Besov priors \cite{Rantala06}. 
%The construction of the prior model depends on the
%prior information about the target. 
For example,
targets that are results of diffusion processes, are usually modeled with a
(homogeneous) smoothness regularization,
while piecewise regular targets which have 
sparse gradient image
might be modeled with a TV prior model.

We note that the estimate (\ref{eq.absolute.estimate}) can be intepreted
in the Bayesian inversion framework as the maximum a posteriori (MAP) estimate
from a posterior density model which is based on the observation model  
(\ref{eq.EITobs_eq}) and a Gibbs type prior model with prior potential (or functional)
$p_\sigma (\sigma)$, see \cite{kaipio2005}.

The minimization problem (\ref{eq.absolute.estimate}) can be solved iteratively,
for example with the Gauss-Newton method while
the positivity constraint on the conductivity can be taken into account by using interior point 
methods \cite{fiacco,nocedalbook}. 
For properties of different optimization methods in EIT,
see e.g. \cite{pvauhkonen04}.

%%%%%%%%%%%%%%%%%%%%%%%%%%%%%%%%%%%%%%%%%%%%%%%%%%%%%%%%%
%%%%%%%%%%%%%%%%%%%%%%%%%%%%%%%%%%%%%%%%%%%%%%%%%%%%%%%%%

\subsection{Difference imaging}
\label{subsec.diff.imaging}

% Conventionally, the difference images are computed as follows.
Consider
%Difference imaging is based on
two EIT measurement realizations $V_1$ and $V_2$ at time instants $t_1$ and $t_2$,
corresponding to conductivities
$\sigma_1$ and $\sigma_2$, respectively.
The observation models corresponding to the two EIT measurement realizations
can be written as in Equation
(\ref{eq.EITobs_eq})
\begin{eqnarray}
   V_1 = U(\sigma_1)+e_1 \label{eq.EITobs_eq_diff1}\\
   V_2 = U(\sigma_2)+e_2\label{eq.EITobs_eq_diff2}
\end{eqnarray}
where $e_{i}\sim \mathcal{N}(e^*,\Gamma_e),\ i=1,2$.
The aim in difference imaging is to reconstruct the change in conductivity $\delta \sigma = \sigma_2 - \sigma_1$ based on
the change $\delta V = V_2 - V_1$ in the data.

%In the conventional difference imaging reconstruction,
Conventionally, the image reconstruction in difference imaging is carried out as follows.
Models (\ref{eq.EITobs_eq_diff1}) and (\ref{eq.EITobs_eq_diff2})
are approximated by first order Taylor approximations as:
\begin{equation}
\label{eq.EITobs_eq_diff_lin}
V_i \approx U(\sigma_0) + J(\sigma_i-\sigma_0) + e_i,\ \ \ \ i=1,2
\end{equation}
where $\sigma_0$ is the linearization point,
and $J=\frac{\partial U}{\partial \sigma}(\sigma_0)$ is the Jacobian matrix
evaluated at $\sigma_0$.
Using the linearizations and subtracting $V_1$ from $V_2$ gives the {\em linear} observation model
%(\ref{eq.EITobs_eq_diff1}) from (\ref{eq.EITobs_eq_diff2})
%and writing the linearizations (\ref{eq.EITobs_eq_diff_lin}) yield
%
\begin{equation}
\label{eq.EITobs_eq_diff_lin2}
\delta V \approx J \delta\sigma + \delta e
\end{equation}
where
$\delta V = V_2-V_1$, 
$\delta \sigma = \sigma_2-\sigma_1$ and
$\delta e = e_2-e_1$.

Given the model (\ref{eq.EITobs_eq_diff_lin2}), 
% the objective
% in difference imaging 
%is to estimate
%the conductivity change $\delta \sigma$ 
%based on the change in the data $\delta V$.
%and the globally linearized observation model (\ref{eq.EITobs_eq_diff_lin2}).
the conductivity change $\delta \sigma$ can be reconstructed as
%This leads to a minimization problem
%
\begin{equation}
  \widehat{\delta\sigma}= {\rm arg} \min_{\delta\sigma}\{\Vert L_{\delta e}(\delta V-J\delta\sigma)\Vert^2
+p_{\delta\sigma}(\delta\sigma)\}
\label{eq.difference.estimate}
\end{equation}
where $p_{\delta\sigma}(\delta\sigma)$ is a regularization functional.
%corresponding to $p_{ \sigma}(\sigma)$ in absolute imaging.
The weighting matrix $L_{\delta e}$ is defined as
$L_{\delta e}\transp L_{\delta e}=\Gamma_{\delta e}^{-1}$,
where $\Gamma_{\delta e}$,
the covariance of the noise term $\delta e$ is 
$\Gamma_{\delta e} = \Gamma_{e_1} + \Gamma_{e_2}  = 2\Gamma_{e}$.

Note that the regularization 
functional $p_{\delta\sigma}(\delta\sigma)$ is often chosen to be of
the quadratic form $p_{\delta\sigma}(\delta\sigma) = \Vert L_{\delta\sigma} \delta\sigma\Vert^2$
where $L_{\delta\sigma} $ is a {\it regularization matrix}.
In such a case,
%$\hat{\delta\sigma}$ in
(\ref{eq.difference.estimate}) is of the form of a regularized linear least squares problem,
the solution of which can be computed with one step --
in contrast to iterative solution of (\ref{eq.absolute.estimate}) in absolute imaging.
The main benefit of the difference imaging, however,
is that when considering the difference data $\delta V$ at least part of the systematic errors in
the models/measurements are subtracted, and hence the estimates are often 
to some extent tolerant of systematic measurement errors and model uncertainties and inaccuracies.
A drawback of the approach is that
the difference images are usually only qualitative in nature 
and their spatial resolution can be weak,
because they rely on the global linearization of the non-linear observation model 
(\ref{eq.EITobs_eq}).
Moreover, the estimates depend on the selection of the linearization point $\sigma_0$.
Typically, $\sigma_0$ is selected as a homogeneous (spatially constant) estimate 
of the initial conductivity $\sigma_1$.
This choice can lead to significant errors in the reconstructions,
especially if the initial conductivity is highly inhomogeneous.

%%%%%%%%%%%%%%%%%%%%%%%%%%%%%%%%%%%%%%%%%%%%%%%%%%%%%%%%%
%%%%%%%%%%%%%%%%%%%%%%%%%%%%%%%%%%%%%%%%%%%%%%%%%%%%%%%%%

\section{Absolute imaging of the conductivity change in a region-of-interest}
\label{lad}

In this section, 
we formulate the reconstruction of the conductivity change in the absolute imaging framework,
in the case where two measurements $V_1$ and $V_2$, taken
before and after the change of target, are available.
Instead of considering the obvious approach of reconstructing $\sigma_1$ and $\sigma_2$ 
by solving minimization (\ref{eq.absolute.estimate}) {\em separately} for realizations $V_i, \ i=1,2$ 
and then subtracting $\delta \sigma = \sigma_2 - \sigma_1$, 
we propose an approach where $\delta \sigma$ is reconstructed together with $\sigma_1$ 
by using simultaneously measurements $V_1$ and
$V_2$. With this approach, we gain flexibility for modeling prior information in cases where
i) the spatial characteristics of initial state $\sigma_1$ and the change $\delta \sigma$ are different
(for example, smooth $\sigma_1$ but sparse $\delta \sigma$) and/or ii) the change in the conductivity is known
to be restricted to a localized subvolume (region of interest, ROI) of the body $\Omega$. 
%focus on a case where prior information on the position of the target change
%between two data sets is available,
%such that the change of the conductivity can be restricted to
%a region-of-interest (ROI).
Utilizing
such prior information is expected to improve the accuracy of the EIT reconstructions.

%\indent Let $\sigma_1$ and $\sigma_2$ denote the conductivity distribution of domain $\Omega$ at the reference time $t_1$ and actual time $t_2$, respectively.
%The observation model for EIT measurements can be written as
%\begin{equation}
%V_1=U(\sigma_1)+e_1,\quad   e_1\sim \mathcal{N}(e^*_1,{\Gamma_e}_1)
%\label{equ1}
%\end{equation}
%\begin{equation}
%V_2=U(\sigma_2)+e_2, \quad   e_2\sim \mathcal{N}(e^*_2,{\Gamma_e}_2)
%\label{equ2}
%\end{equation}

%As in Section \ref{subsec.diff.imaging}, consider 
%two EIT measurement realizations $V_1$ and $V_2$
%corresponding to two different conductivity distributions
%$\sigma_1$ and $\sigma_2$, respectively.
%In such a case,
The observation models for the two EIT data sets $V_1$ and $V_2$
are of the forms
(\ref{eq.EITobs_eq_diff1}-\ref{eq.EITobs_eq_diff2}).
Assume
that the conductivity change $\delta\sigma=\sigma_2-\sigma_1$ 
is known to be restricted to a region of interest $\Omega_{\mathrm{ROI}} \subseteq \Omega$,
and denote the conductivity change within $\Omega_{\mathrm{ROI}}$
by $\delta\sigma_{\mathrm{ROI}}$.
Then, $\delta\sigma=\mathcal{M}\delta\sigma_{\mathrm{ROI}}$ 
where $\mathcal{M}$ is the mapping $\mathcal{M}:\Omega_{\mathrm{ROI}}\rightarrow\Omega$ such that
\begin{equation}
 \mathcal{M}\delta\sigma_{\mathrm{ROI}}=
 \left\{
 \begin{array}{l}
  \delta\sigma_{\mathrm{ROI}},\quad x \in \Omega_{\mathrm{ROI}} \\
  0, \quad x\in \Omega \setminus \Omega_{\mathrm{ROI}}
 \end{array}
\right.
\end{equation}
and
$\sigma_2$,
the conductivity after the change, can be represented in the form
\begin{equation}
 \sigma_2=\sigma_1+\mathcal{M}\delta\sigma_{\mathrm{ROI}}.
\label{eq.sigma2}
 \end{equation}
Inserting expression (\ref{eq.sigma2}) to
Equation (\ref{eq.EITobs_eq_diff2}) and concatenating the measurement vectors $V_1$ and $V_2$
and the corresponding models in (\ref{eq.EITobs_eq_diff1}-\ref{eq.EITobs_eq_diff2}) into block vectors leads
to observation model
\begin{equation}
 \underbrace{\begin{bmatrix}
       V_1           \\[0.3em]
       V_2
     \end{bmatrix}}_{\bar{V}}
 = \underbrace{\begin{bmatrix}
       U(\sigma_1)         \\[0.3em]
       U(\sigma_1+\mathcal{M}\delta\sigma_{\mathrm{ROI}})
     \end{bmatrix} }_{\bar{U}(\bar{\sigma})}
     +  \underbrace{\begin{bmatrix}
       e_1        \\[0.3em]
       e_2
     \end{bmatrix}}_{\bar{e}}
\end{equation}
%\begin{center}
% $\Leftrightarrow V=U(\sigma)+e$
%\end{center}
or
\begin{equation}
\label{obs.eq.roi}
%V_{1,2} = \bar U (\sigma_{1,\delta}) + e_{1,2}
\bar V= \bar U (\bar{\sigma}) + \bar e,
\end{equation}
where 
$$
\bar{\sigma} =  \begin{bmatrix}
       \sigma_1           \\[0.3em]
       \delta \sigma_{\mathrm{ROI}}
     \end{bmatrix}.
$$
%$\bar V=[V_1\transp, V_2\transp]\transp$, 
%$\bar\sigma = [\sigma_1\transp, \delta\sigma_{\mathrm{ROI}}\transp]\transp$,
%$\bar U(\bar\sigma)=[U(\sigma_1)\transp, U(\sigma_1+\mathcal{M}\delta\sigma_{\mathrm{ROI}})\transp]\transp$
%and 
%$\bar e=[e_1\transp, e_2\transp]\transp$.
%
Here,
we have identified the 
conductivities
$\sigma_{1}, \delta\sigma_{\mathrm{ROI}}$ and mapping $\mathcal{M}$
with their finite dimensional approximations.

Based on the observation model (\ref{obs.eq.roi}),
the generalized Tikhonov regularized solution is written in the form
\begin{equation}
\begin{aligned}
  \hat{\bar{\sigma}}
  &= {\rm arg} \min_{\bar\sigma}\{\Vert L_{\bar e}(\bar V-\bar U(\bar{\sigma}))\Vert^2
+p_{\bar{\sigma}} (\bar{\sigma})\}.
  % +\Vert  L_{\sigma}(\sigma-\sigma^*) \Vert^2\}
\label{ladmain}
\end{aligned}
\end{equation}
Here, $L_{\bar e}\in \mathbb{R}^{2M\times 2M}$ is the Cholesky factor such that 
$L_{\bar e} \transp L_{\bar e} = \Gamma_{\bar e}^{-1}$, where
%
% is the covariance matrix of the stacked measurement noise vector $\bar e$
%
\begin{equation*}
\Gamma_{\bar e}=
\left [
\begin{array}{ cc }
 {\Gamma_{e_1}} & \mathbf{0}_{M\times M} \\
  \mathbf{0}_{M\times M} &{\Gamma_{e_2}}
\end{array}
\right ]
\end{equation*}
and $\mathbf{0}_{M\times M}\in \mathbb{R}^{M\times M}$ is an all-zero matrix.
Typically, the noise statistics can be modelled stationary, i.e., $\Gamma_{e_1} = \Gamma_{e_2} = \Gamma_e$.
Further,
$p_{\bar\sigma}(\bar\sigma)$ is a compound regularization functional
of the form 
$$p_{\bar\sigma}(\bar\sigma)=p_{\sigma_1}(\sigma_1)+p_{\delta\sigma_{\mathrm{ROI}}}(\delta\sigma_{\mathrm{ROI}})$$
which allows naturally the use of different spatial models for $\sigma_1$ and $\delta \sigma$.
Similarly as in Section
\ref{subsec.absoluteEIT},
the Tikhonov regularized solution (\ref{ladmain}) has to be computed iteratively.
In the iterations,
the Jacobian matrix $J_{\bar U}(\bar \sigma)=\frac{\partial \bar U}{\partial \bar\sigma}$ is needed;
the Jacobian is of the form
\begin{equation*}
J_{\bar U}(\bar \sigma)
= 
\left [
\begin{array}{ cc }
 J_U(\sigma_1)& \mathbf{0}_{M\times N_{\mathrm{ROI}}}\\
 J_U(\sigma_1+\mathcal{M}\delta\sigma_{\mathrm{ROI}})&
 J_U(\sigma_1+\mathcal{M}\delta\sigma_{\mathrm{ROI}})\mathcal{M}
\end{array}
\right ]
\end{equation*}
where $ J_U(\sigma)$ is the Jacobian matrix of the function $U(\sigma)$,
$\mathbf{0}_{M\times N_{\mathrm{ROI}}}\in \mathbb{R}^{M\times N_{\mathrm{ROI}}}$ 
is an all-zero matrix,
and $N_{\mathrm{ROI}}$ is the dimension of the vector $\delta\sigma_{\mathrm{ROI}}$.

%In the approach proposed above,
%the conductivity distribution before the change, $\sigma_1$, and the change of the target in {\it a priori} defined
%ROI, $\delta\sigma_{\mathrm{ROI}}$, are estimated simultanously on the basis of two data sets $V_1$ and $V_2$.
In contrast to conventional difference imaging reconstruction (Section \ref{subsec.diff.imaging}),
the global linearization of the EIT forward model is {\it not} needed here,
enabling the use of the two data realizations for quantitative imaging.
This of course necessitates that (possible) modeling errors are
handled properly, see Section \ref{sect.Introduction} and the references therein.
Moreover,
utilizing the information on the approximate position of the conductivity change
between the two measurement sets $V_1$ and $V_2$
is expected to improve the reconstructions,
especially if the ROI is relatively small in comparison with the volume of the target.
Note also that the two regularization terms
$p_{\sigma_1}(\sigma_1)$ and $p_{\delta\sigma_{\mathrm{ROI}}}(\delta\sigma_{\mathrm{ROI}})$ corresponding to
%conductivity 
$\sigma_1$ 
%at the initial time 
and 
%conductivity change 
$\delta\sigma_{\mathrm{ROI}}$, respectively,
may have different properties.
In the examples shown in the following sections,
$p_{\sigma_1}(\sigma_1)$ corresponds to 
the assumption of a smooth initial conductivity $\sigma_1$,
while $p_{\delta\sigma}(\delta\sigma)$ corresponds to a
total variation (TV) model for the conductivity change.

Finally, we note that
an alternative approach to non-stationary imaging is the state-estimation,
in which Kalman filter -type recursive estimators are used for reconstructing the
time-varying conductivity distribution.
The state-estimation approach has shown to be beneficial when
the measurements are obtained sequentially and
feasible models for the target evolution are available 
-- such as fluid dynamical models in industrial process imaging applications \cite{seppanen2001a}.
In principle, the region-of-interest information utilized in the present study
could be utilized in the state-estimation equivalently.
The implementation of the TV regularization used in this paper for the conductivity change,
however, would not be a straightforward task in the state-estimation framework.

%%%%%%%%%%%%%%%%%%%%%%%%%%%%%%%%%%%%%%%%%%%%%%%%%%%%%%%%%%
%%%%%%%%%%%%%%%%%%%%%%%%%%%%%%%%%%%%%%%%%%%%%%%%%%%%%%%%%%

\section{Experimental studies}
\label{sec.experimentalstudies}

The feasibility of the proposed reconstruction method
was studied experimentally.
All tests were carried out with targets that were translationally 
symmetric in the vertical direction,
and hence two-dimensional models were adequate
for modeling the measurements. 
The extension of the computational methods to a purely three-dimensional case
is straightforward.

%%%%%%%%%%%%%%%%%%%%%%%%%%%%%%%%%%%%%%%%%%%%%%%%%%%%%%%%%%%%
%%%%%%%%%%%%%%%%%%%%%%%%%%%%%%%%%%%%%%%%%%%%%%%%%%%%%%%%%%%%

\subsection{Experimental setup}

The experiments were carried out using a cylindrical tank shown in \cref{case13} in the top row.
The diameter of the tank was 28 cm.
Sixteen equally spaced metallic electrodes (width 2.5 cm, height 7.0 cm) 
were attached to the inner surface of the tank.
In \cref{case13},
the electrode positions are indicated with red stripes on the tank boundary.
The rightmost electrode was identified with electrode index $\ell=1$,
and the electrode indices increased in counter clockwise direction.
The tank was filled with saline,
and plastic objects with different shapes were placed in the tank
to form inhomogeneities to the conductivity distribution.
%All objects were symmetric in height. 
%In all cases,
%the excess water was removed from the tank, 
%so that the height of the water level was 7 cm, i.e. same as the height of the electrodes.

The EIT measurements were carried out with KIT4 measurement system developed 
in the Department of Applied Physics, 
University of Eastern Finland \cite{kourunen2009suitability}.
Pairwise current injections were applied in the measurements.
The frequency of the current was 1 kHz and the amplitude was 1 mA.
The currents were injected such that one electrode was fixed as
the sink electrode and then applying pairwise currents sequentally between the sink electrode
and each one of the 15 remaining electrodes. This process was repeated using electrodes $\{1,5,9,13\}$
as the sink, leading to total of 54 current injections when reciprocal curret injections were
not taken.
%were applied:
%between electrodes $e_1 \rightarrow e_i,  i=2,...,L$, 
%$e_5 \rightarrow e_i,  i=2,...,L \backslash 5$,
%$e_9 \rightarrow e_i, i=2,...,L \backslash 5,9$ and
%$e_{13} \rightarrow e_i, i=2,...,L \backslash 5,9,13$.
Corresponding to each current injection,
the potentials on 
all the 15 remaining electrodes were measured 
against the sink electrode, 
which was connected to the common ground.
With this measurement protocol, one measurement frame consists of
810 voltage readings (i.e.,  $V\in \mathbb{R}^{810}$). 

Three different experimental test cases were considered.
In all cases,
two realizations of EIT measurements were collected:
$V_1$ corresponding to an initial conductivity $\sigma_1$
and $V_2$ corresponding to conductivity $\sigma_2$ after a change.
%Between the two measurement sets, 
%an additional object was placed into the tank
%and, again, the excess water was removed from the tank
%to keep the water height in the original level.
%
The first test case is illustrated in \cref{case13}.
In the initial state $\sigma_1$ (top left),
a plastic circular cylinder (diameter 6.2 cm) was placed
in the tank.
In the second state $\sigma_2$
(top middle), a plastic triangular prism was added to the tank.
The top face of the prism was an equilateral triangle,
and the edges of the triangle were 8.5 cm long.
%In test cases 2 and 3, the conductivity distributions $\sigma_1$
%at the initial state were more complex than in case 1:
In test cases 2 and 3,
three plastic cylindrical objects were placed in the tank
in the initial state $\sigma_1$ (\cref{case59,case73}, top left).
In both cases, 
a tetragonal prism was added to the tank in the second state $\sigma_2$.
The difference between cases 2 and 3 was that
sizes of the inserted tetragonal prisms were different:
in case 2, the dimensions of the prism face were
8.2 cm $\times$ 3.5 cm (\cref{case59}, top middle)
and in case 3,
6.0 cm $\times$ 1.0 cm (\cref{case73}, top middle).

To estimate the noise level, we carried repeated measurement
(100 realizations) from the tank filled with saline. 
The noise covariance matrix $\Gamma_e$ was computed as a sample covariance
based on these realizations.

%%%%%%%%%%%%%%%%%%%%%%%%%%%%%%%%%%%%%%%%%%%%%%%%%%%%%%%%%%%%
%%%%%%%%%%%%%%%%%%%%%%%%%%%%%%%%%%%%%%%%%%%%%%%%%%%%%%%%%%%%

\subsection{Estimates}
\label{estsec}
%The computational code including also the forward solver was implemented for MatLab, and is
%an adaptation of the implementations described in
%\cite{Vauhkonen1999}

In each of the three test cases, the following estimates were computed:
\begin{itemize}
\item[(E1)] Conventional (linear) difference reconstruction by solving
\[
  \widehat{\delta\sigma}= {\rm arg} \min_{\delta\sigma}\{\Vert L_{\delta e}(\delta V-J\delta\sigma)\Vert^2
+ \Vert L_{\delta \sigma}{\delta\sigma} \Vert^2\}
\label{eq.difference.estimate2}
\]
where $L_{\delta \sigma}  \transp  L_{\delta \sigma} = \Gamma_{\delta \sigma}^{-1}$, and
\begin{equation}
\Gamma_{f}(i,j) = a \exp \left\{ - \frac{\Vert x_i - x_j \Vert_2^2}{2 b^2} \right\} + c \delta_{ij}.
\label{eq.sm.pr.covmat}
\end{equation}
is covariance matrix corresponding to a generic smoothness prior model for the unknown distributed
parameter $f$ \cite{lieberman2010}. Here $x_i$ and $x_j$ denote the coordinate points of nodes $i$ an $j$ 
in the parameterization of $f$, respectively.
%Reconstruction (E1) is shown top right in Figs \ref{case13}-\ref{case73}.

\item[(E2)] Absolute reconstructions of $\sigma_1$ and $\sigma_2$ by
solving
\[
  \hat\sigma_i = {\rm arg} \min_{\sigma_i > 0}\{\Vert L_{e}(V_i -U(\sigma_i))\Vert^2
+ \Vert L_{\sigma_i} (\sigma_i - \sigma^\ast) \Vert^2 \}. 
\]
%From these reconstructions,
%the estimate for the conductivity change was obtained simply by subtracting
%$\hat{\delta \sigma} = \hat\sigma_2 - \hat\sigma_1$. 
The regularization matrix 
$L_{\sigma_i}  \transp  L_{\sigma_i} = \Gamma_{\sigma_i}^{-1}$ was constructed by equation (\ref{eq.sm.pr.covmat}).
% (E2) is shown on the second row in Figs \ref{case13}-\ref{case73})

\item[(E3)] Absolute reconstructions of $\sigma_1$ and $\sigma_2$ by
solving
\[
  \hat\sigma_i = {\rm arg} \min_{\sigma_i >0}\{\Vert L_{e}(V_i -U(\sigma_i))\Vert^2
+ \alpha {\rm TV}(\sigma_i) \}, 
\]
where 
\[
{\rm TV}(f) = \sum_{k=1}^{N_e} \vert \Omega_k \vert \sqrt{ \| (\nabla f) \vert_{\Omega_k} \|^2 + \beta}
\]
is a differentiable approximation of the isotropic total variation functional \cite{RudinOsherFatemi} and
$(\nabla f )\vert_{\Omega_k}$ is the (constant) gradient of the (piecewise linear) $f$ at element $\Omega_k$, and $N_e$ is the number of elements.
%(E3) is shown on the third row in Figs \ref{case13}-\ref{case73}.

\item[(E4)] Reconstruction of 
$\bar{\sigma}=(\sigma_1\transp,\delta \sigma_{{\rm ROI}}\transp)\transp$ 
with 
%the ROI-based reconstruction method proposed in Section \ref{lad}
the proposed method
$$
  \hat{\bar{\sigma}}
  = {\rm arg} \min_{\bar\sigma} \left\{ \Vert L_{\bar e}(\bar V-\bar U(\bar \sigma))\Vert^2
 + p_{\bar{\sigma}} (\bar{\sigma}) \right\}  %\quad 
$$
$$
\sigma_1 > 0,\ \sigma_1+\mathcal{M}\delta\sigma_{\mathrm{ROI}} > 0
$$
with the choice
$$
p_{\bar{\sigma}} (\bar{\sigma})
=  \Vert L_{\sigma} (\sigma_i - \sigma^\ast) \Vert^2 
+ \alpha {\rm TV}(\delta \sigma_{\rm ROI})
$$
%(E4) is shown on the fourth row in Figs \ref{case13}-\ref{case73}.
In all three test cases, the ROI in estimate (E4) was selected to be a 
circular area with a diameter 7.0 cm and a center point at (0 cm, 6 cm).
%
%The ROI is marked with black lines in the images corresponding to estimates (E4)
%in Figs \ref{case13} - \ref{case73}
%(fourth row).
\end{itemize}
%

%The penalty functional $p_{\sigma}(\sigma_1)$ corresponding to the initial conductivity $\sigma_1$
%was constructed as
%$p_{\sigma}(\sigma_1) = \Vert L_{\sigma_{1}}(\sigma_1-\sigma_1^*) \Vert^2$
%where
%the matrix $L_{\sigma_{1}}$ is defined as
%$L_{\sigma_{1}}\transp  L_{\sigma_{1}}=\Gamma_{\sigma_{1}}^{-1}$
%and
%% 
%\begin{equation}
%\Gamma_{\sigma_{1}}(i,j) = a \exp \left\{ - \frac{\Vert x_i - x_j \Vert_2^2}{2 b^2} \right\} + c \delta_{ij}.
%\label{eq.sm.pr.covmat}
%\end{equation}
%%
%Here,
%$x_i$ and $x_j$ denote the spatial coordinates of the nodes in the FE mesh
%corresponding to pointal conductivity values $\sigma_1^{(i)}$ and $\sigma_1^{(j)}$
%in the finite dimensional approximation of the conductivity $\sigma_1$.
%It can be shown that the above form
%for the penalty functional corresponds to a certain type of a smoothness prior model,
%where 
%$\sigma_1^*$ and $\Gamma_{\sigma_{1}}$
%are, respectively, 
%the expectation and the covariance matrix
%of $\sigma_1$
%\cite{lieberman2010}.
%

%%%%%%%%%%%%%%%%%%%%%%%%%%%%%%%%%%%%%%%%%%%%%%%%%%%%%%%%%%%%
%%%%%%%%%%%%%%%%%%%%%%%%%%%%%%%%%%%%%%%%%%%%%%%%%%%%%%%%%%%%

\subsection{Parameters in the reconstructions}

The contact impedances of the electrodes were estimated using data from 
the tank filled solely with water.
Based on these measurements,
we computed an estimate for the water conductivity $\sigma_0 \in \mathbb{R}$
and a constant contact impedance $z_0 \in \mathbb{R}$ 
%by minimization of
%\[
%(\hat \sigma_0,\hat z_0) = \arg \min_{\sigma_0,z_0}\left\{ \| L_e (V - U(\sigma_0,z_0))\|^2\right\}
%\]
as a solution of a two parameter
non-linear least squares fitting problem.
The estimated water conductivity was $\sigma_0 = 3.7569 {\mathrm{m}\rm S}$
and the estimated contact impedance was $z_0 = 1.08\times 10^{-4} \Omega {\rm m}$.
Note that the measurements with a homogeneous object
are not a necessity for the reconstruction;
for computational methods for recovering from 
unknown contact impedances without the additional calibration measurements, see
\cite{vilhunen2002a,nissinen2009,kolehmainen2008electrical}.

In the reconstructions (E1)-(E4), the conductivity (or the change of conductivity) 
was approximated in a piece-wise linear first order nodal FE basis 
with $N_n = 1330$ nodes and $N_e = 2532$ 
triangular elements. 
Thus, the unknown conductivity vector was $\sigma \in \mathbb{R}^{1330}$. 
The electrical potential was approximated in a separate second order basis 
with $9109$ nodes and $4466$ elements. 

The parameters $a$, $b$ and $c$ 
in the construction of the prior covariance (\ref{eq.sm.pr.covmat}) 
used in (E1), (E2) and (E4)  were selected as
$a= 4.53 \mathrm{m}\mathrm{S}^2$,
$b = 4.61\mathrm{m}\mathrm{S}^2$
and
$c= 4.5\times10^{-3} \mathrm{cm}$.
The expectation $\sigma^\ast$ in (E2) and (E4) was set
equivalent to the estimated saline conductivity.
For interpretation of the prior covariance parameters, 
see e.g. \cite{Darde13b}.
The regularization parameter $\alpha$ for the total variation functional
was selected as $\alpha=1.30$. 
For the parameter $\beta$ in the TV functional we used 
value $\beta = 1\times10^{-3}$.

The estimates (E2)-(E4) were computed with the Gauss-Newton optimization method
which was equipped with an explicit line search algorithm. 
The positivity constraints were
taken into account by using an interior point method 
where the constrained minimization problem is approximated by a sequence 
of unconstrained problems which use a logarithmic barrier functional 
for enforcing positivity of the parameters, see \cite{fiacco,nocedalbook}

%%%%%%%%%%%%%%%%%%%%%%%%%%%%%%%%%%%%%%%%%%%%%%%%%%%%%%%%%%%%
%%%%%%%%%%%%%%%%%%%%%%%%%%%%%%%%%%%%%%%%%%%%%%%%%%%%%%%%%%%%

\subsection{Results and discussion}

\begin{figure}  [h] 
%\caption{Stimuli Category Explanations} 
\centering
  \begin{tabular} {lccc}
   & $\sigma_1$ & $\sigma_2$ & $\delta\sigma$ \\
      (E1) & \parbox[c]{3cm}{\includegraphics[width=3cm]{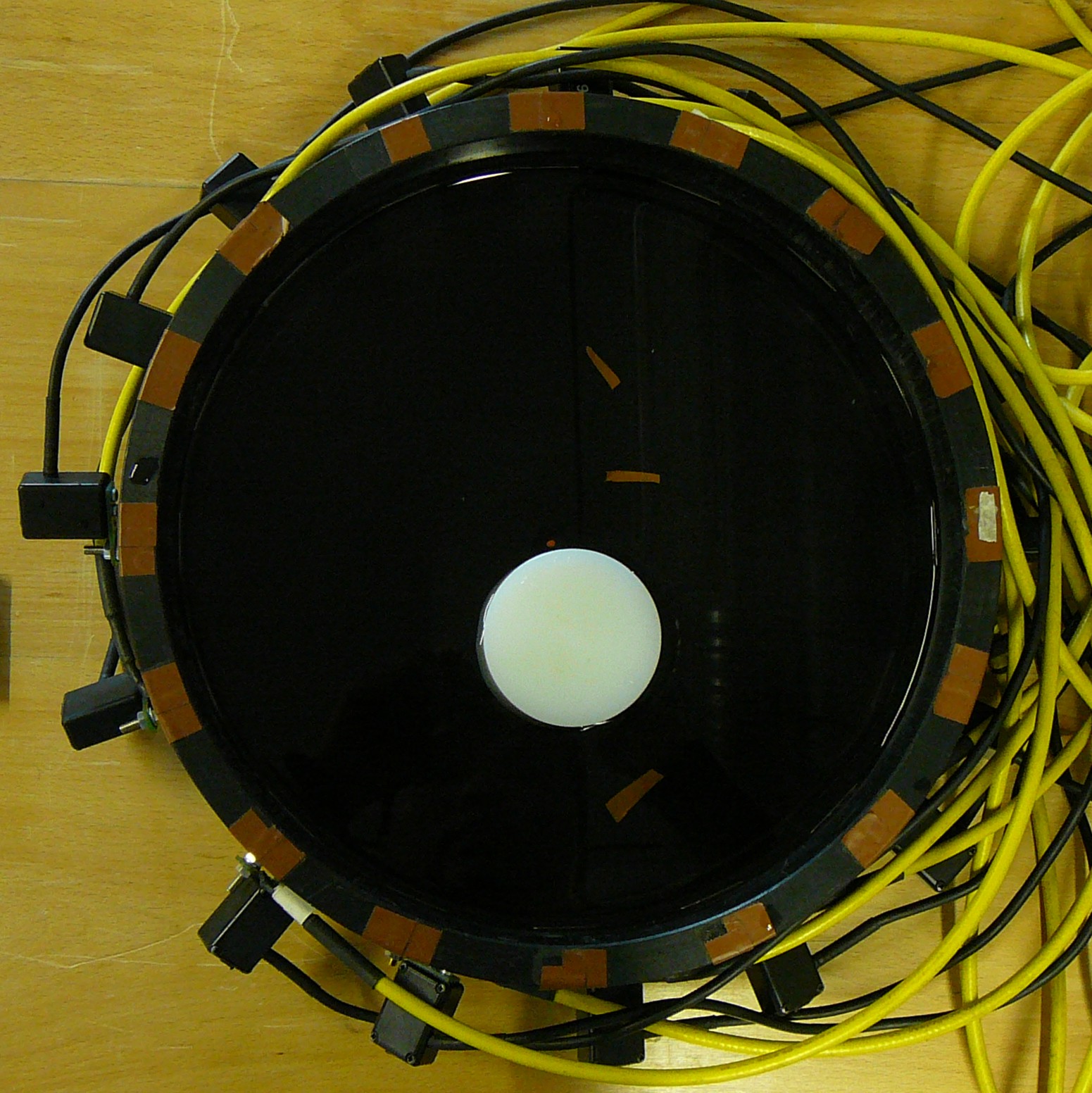}}
      & \parbox[c]{3cm}{\includegraphics[width=3cm]{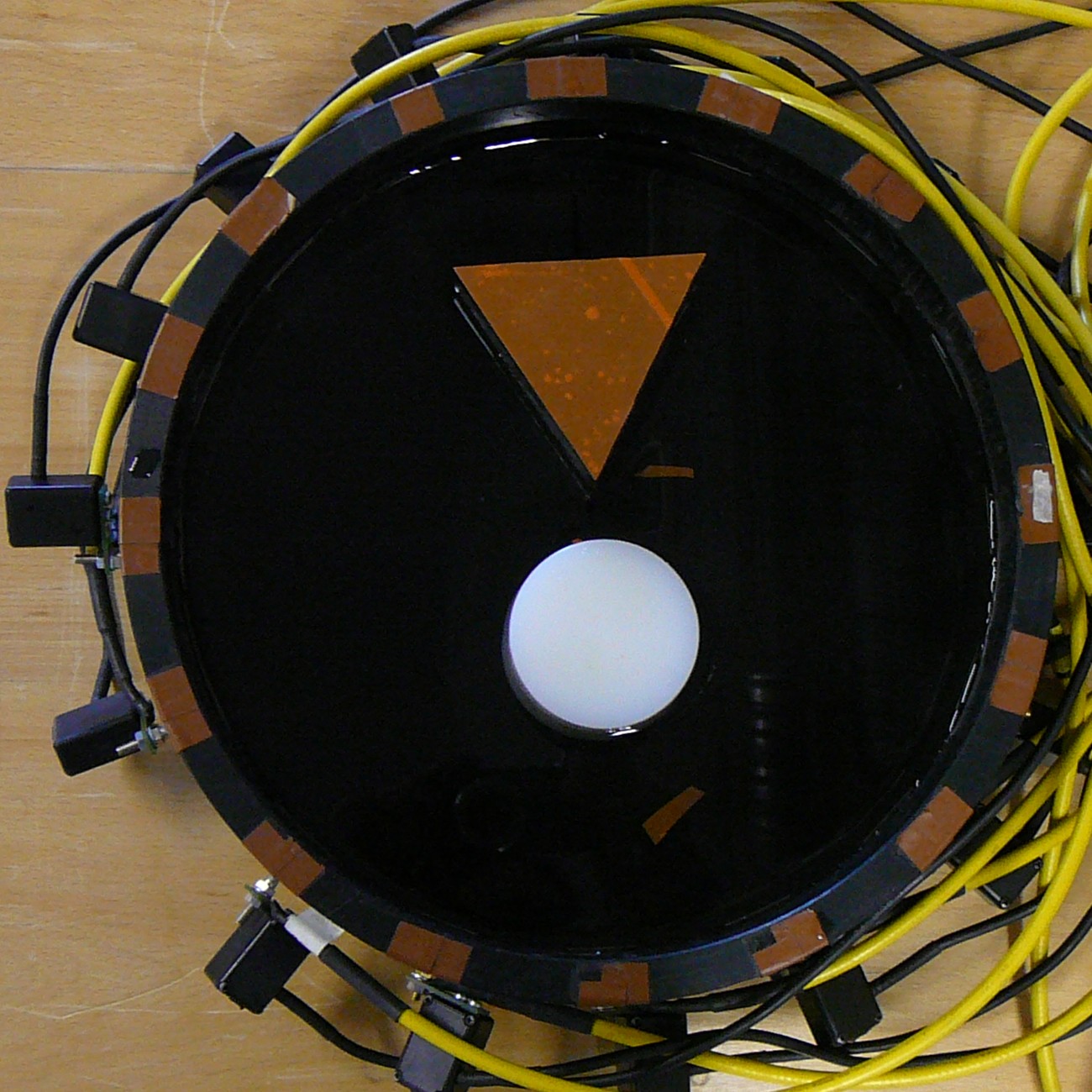}}
      &\parbox[c]{3cm}{\includegraphics[width=3cm]{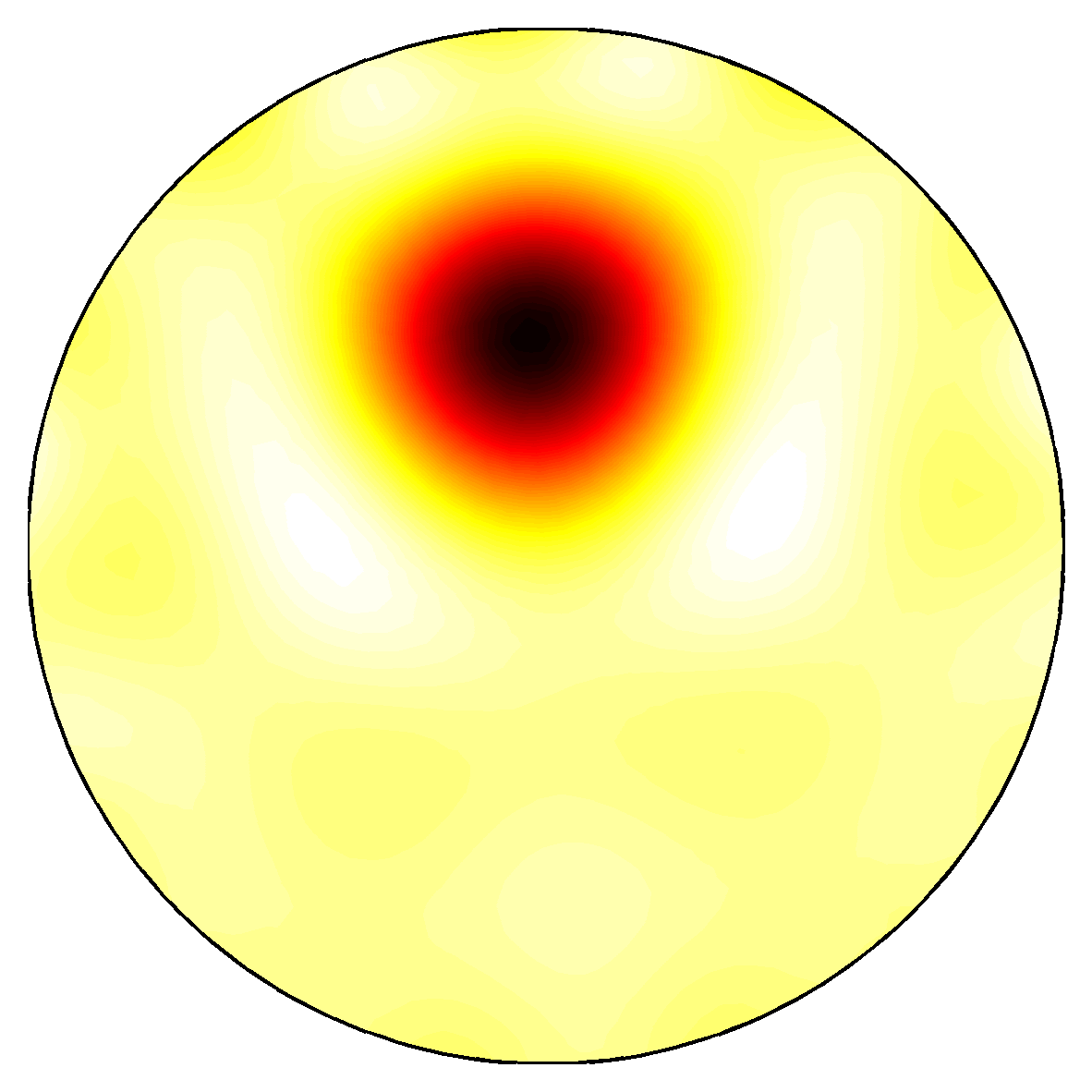}} \\[3 ex]
      & &  &\parbox[c]{3cm}{\includegraphics[width=3cm]{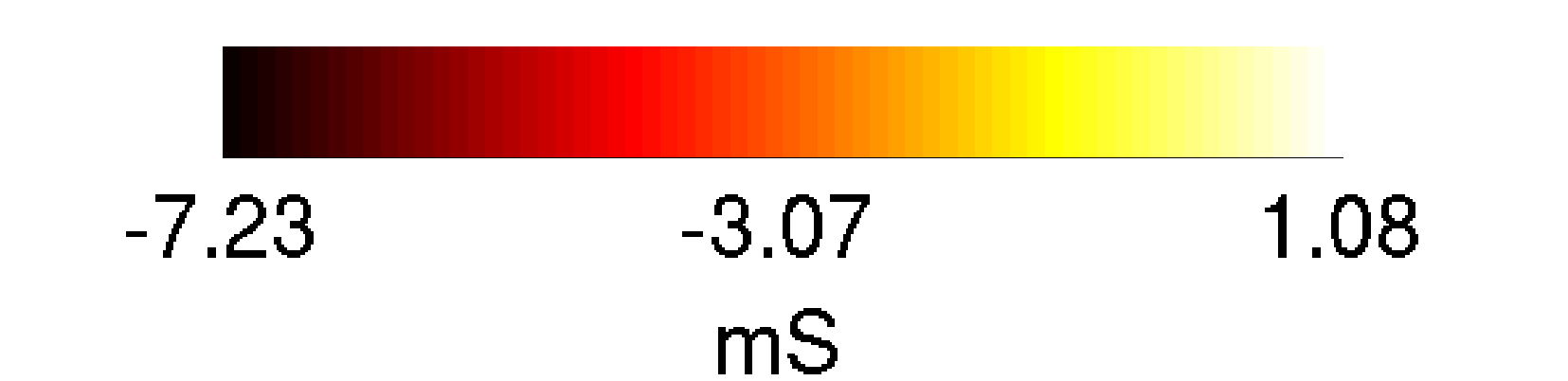}} \\[1 ex]
%%%%%%%%%%%%%%%%%%%%%%%
      (E2) & \parbox[c]{3cm}{\includegraphics[width=3cm]{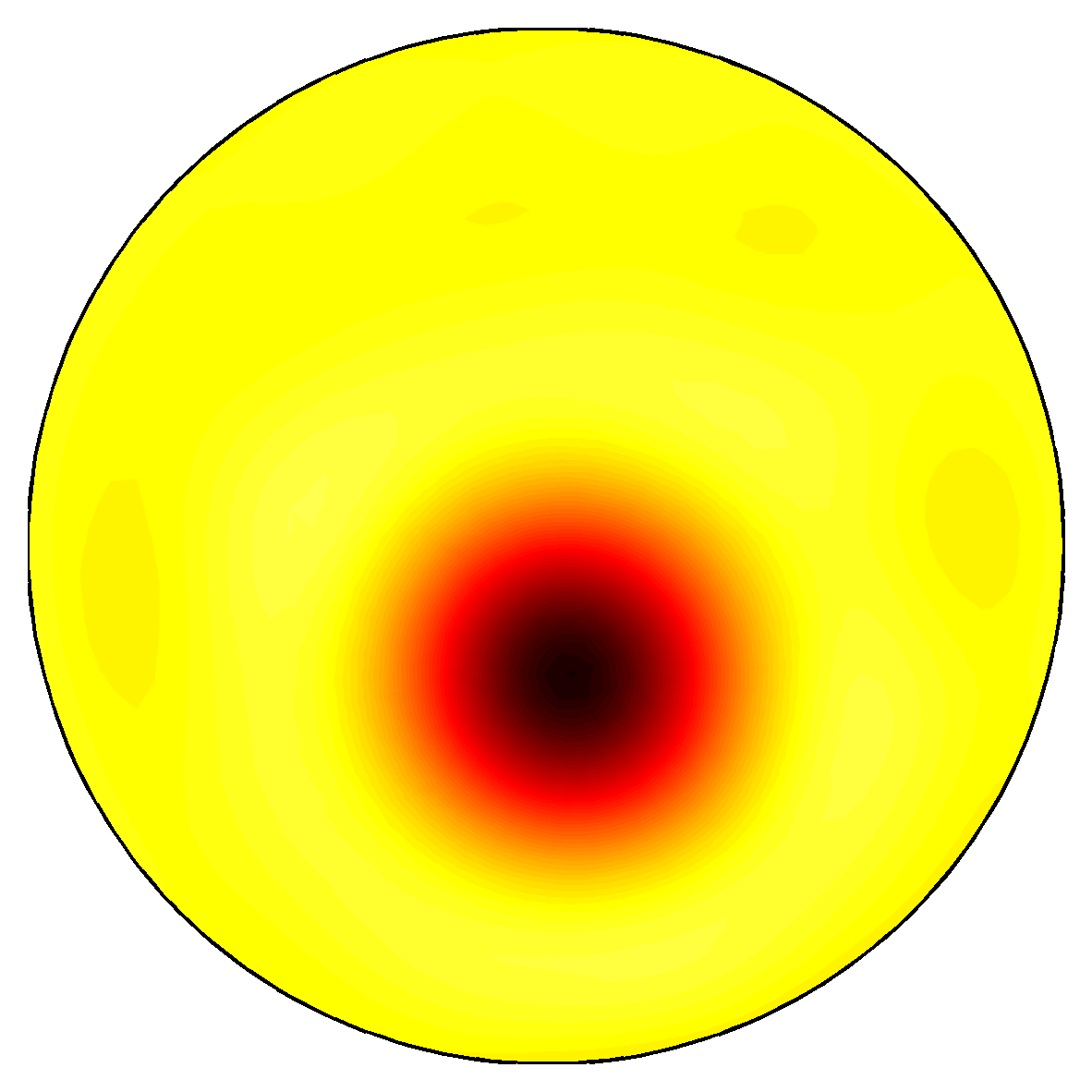}}
      & \parbox[c]{3cm}{\includegraphics[width=3cm]{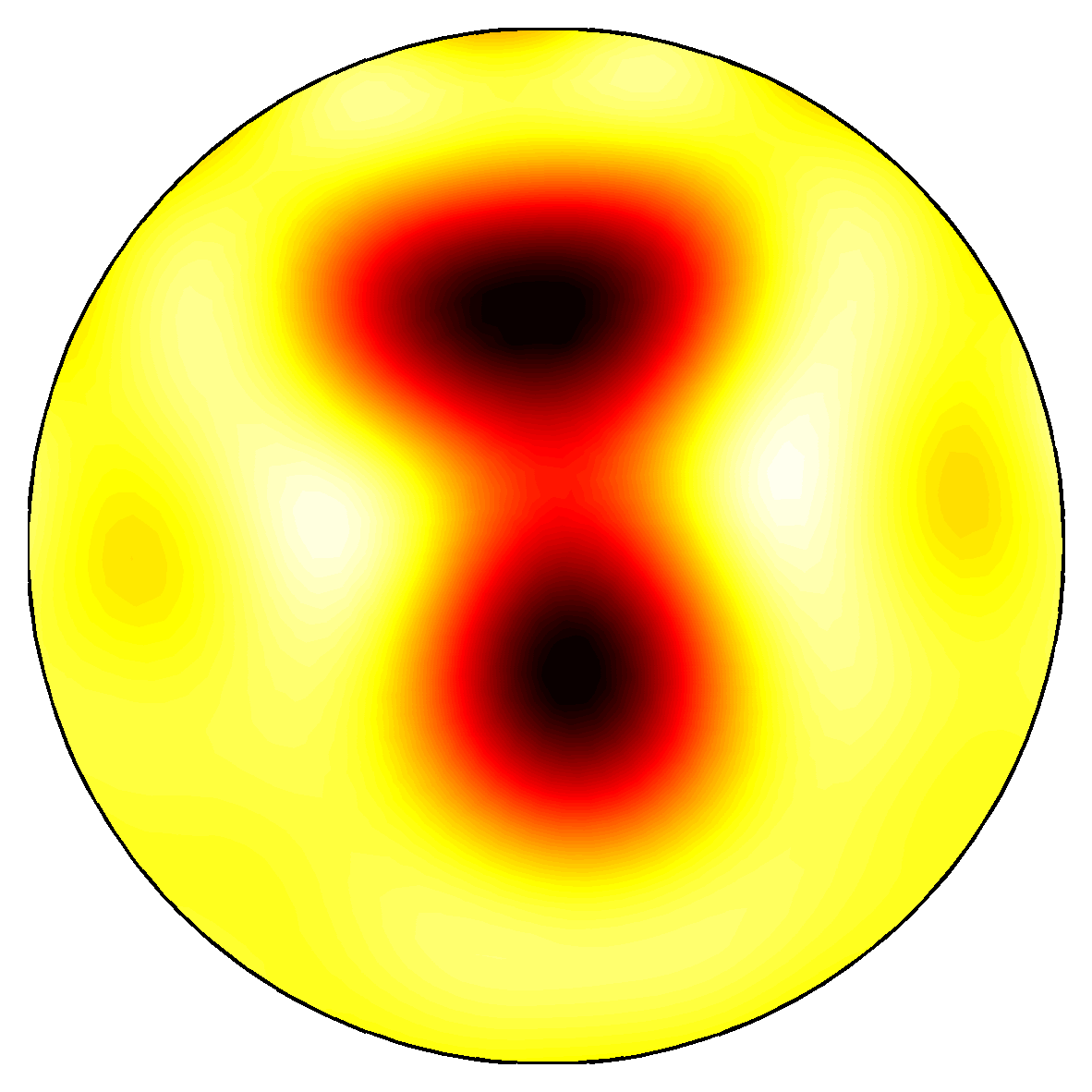}}
      & \parbox[c]{3cm}{\includegraphics[width=3cm]{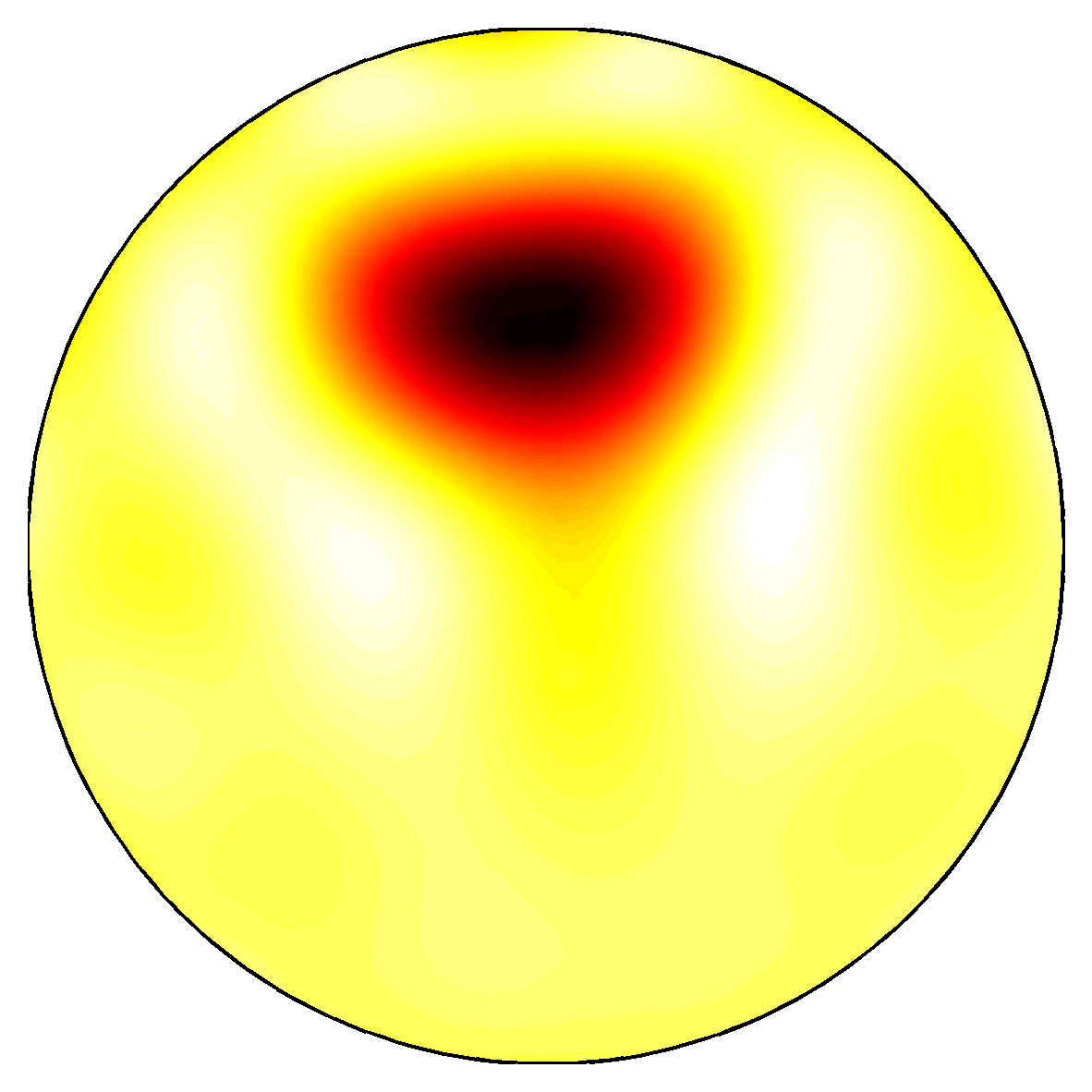}} \\[0 ex]
            %%%%%%%%%%%%%%%%%%%%%%%
      (E3) & \parbox[c]{3cm}{\includegraphics[width=3cm]{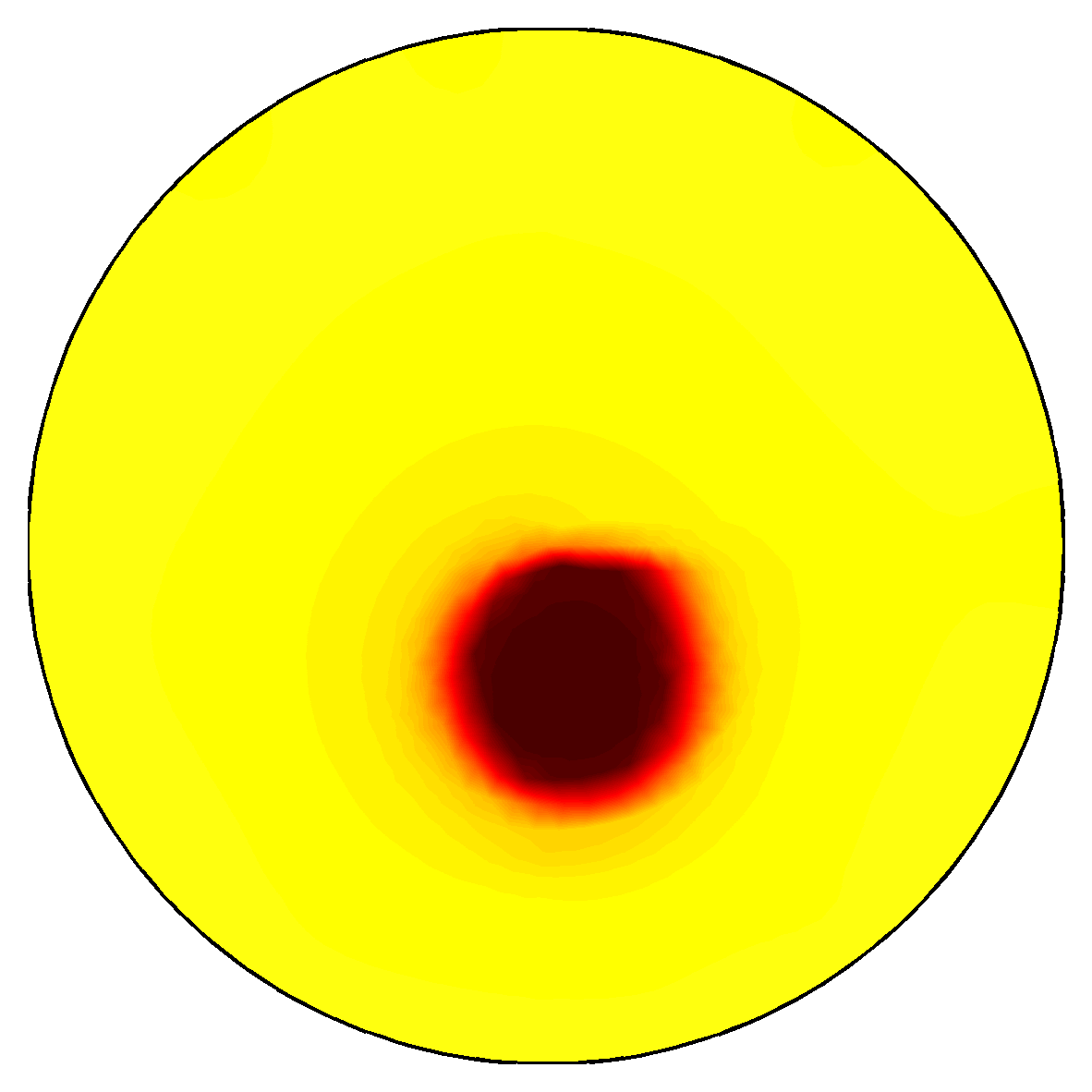}}
      & \parbox[c]{3cm}{\includegraphics[width=3cm]{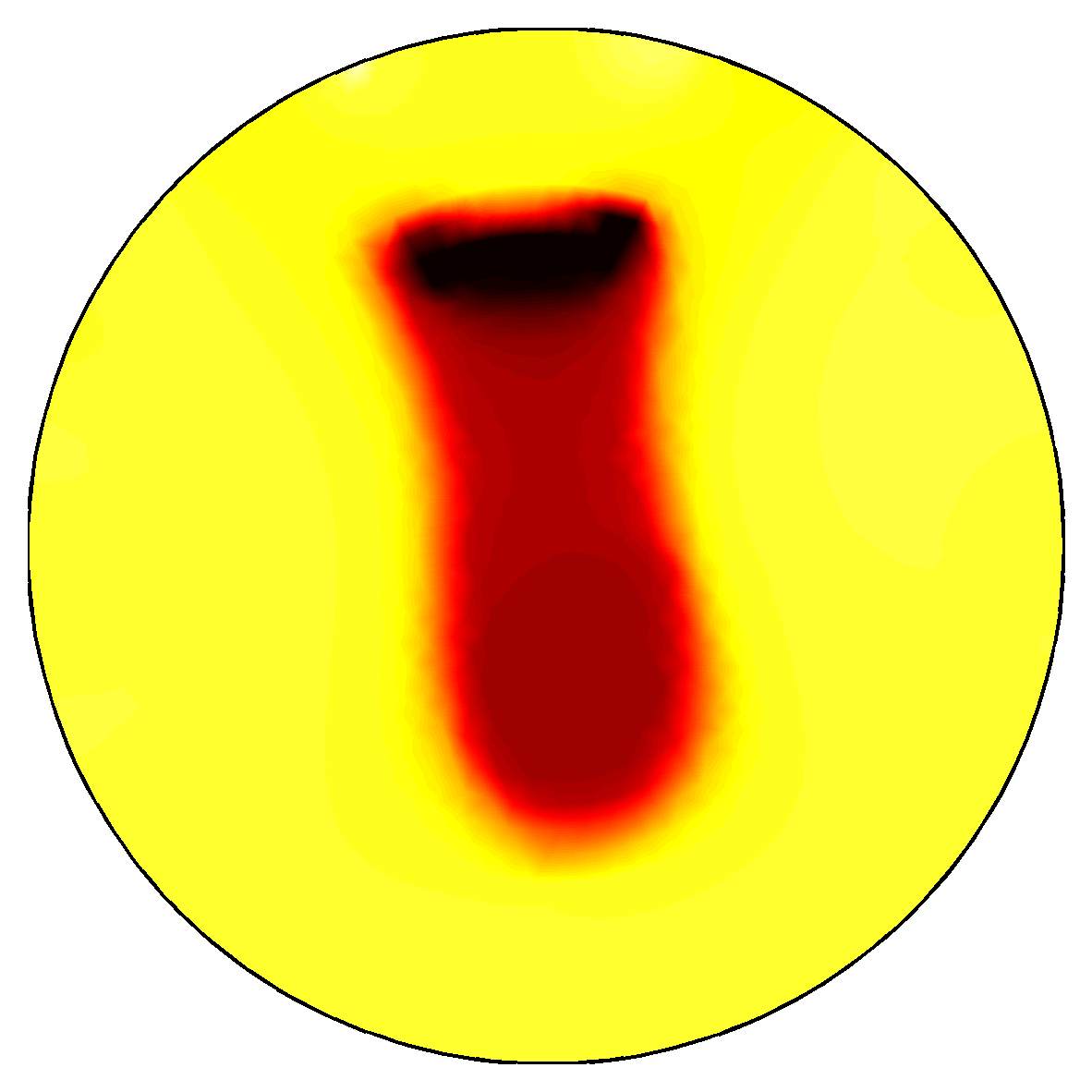}}
      & \parbox[c]{3cm}{\includegraphics[width=3cm]{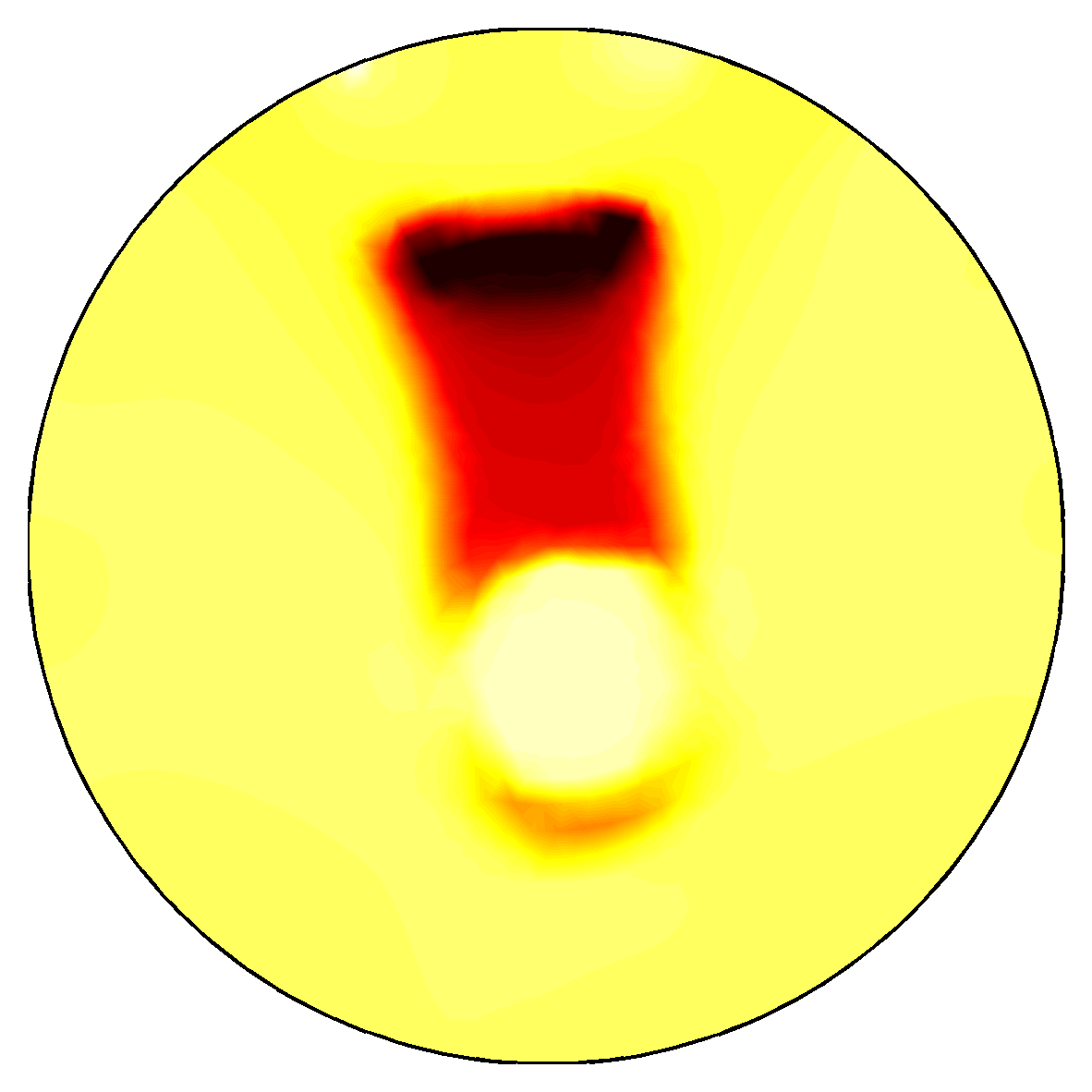}} \\[0 ex]
                  %%%%%%%%%%%%%%%%%%%%%%%
      (E4) & \parbox[c]{3cm}{\includegraphics[width=3cm]{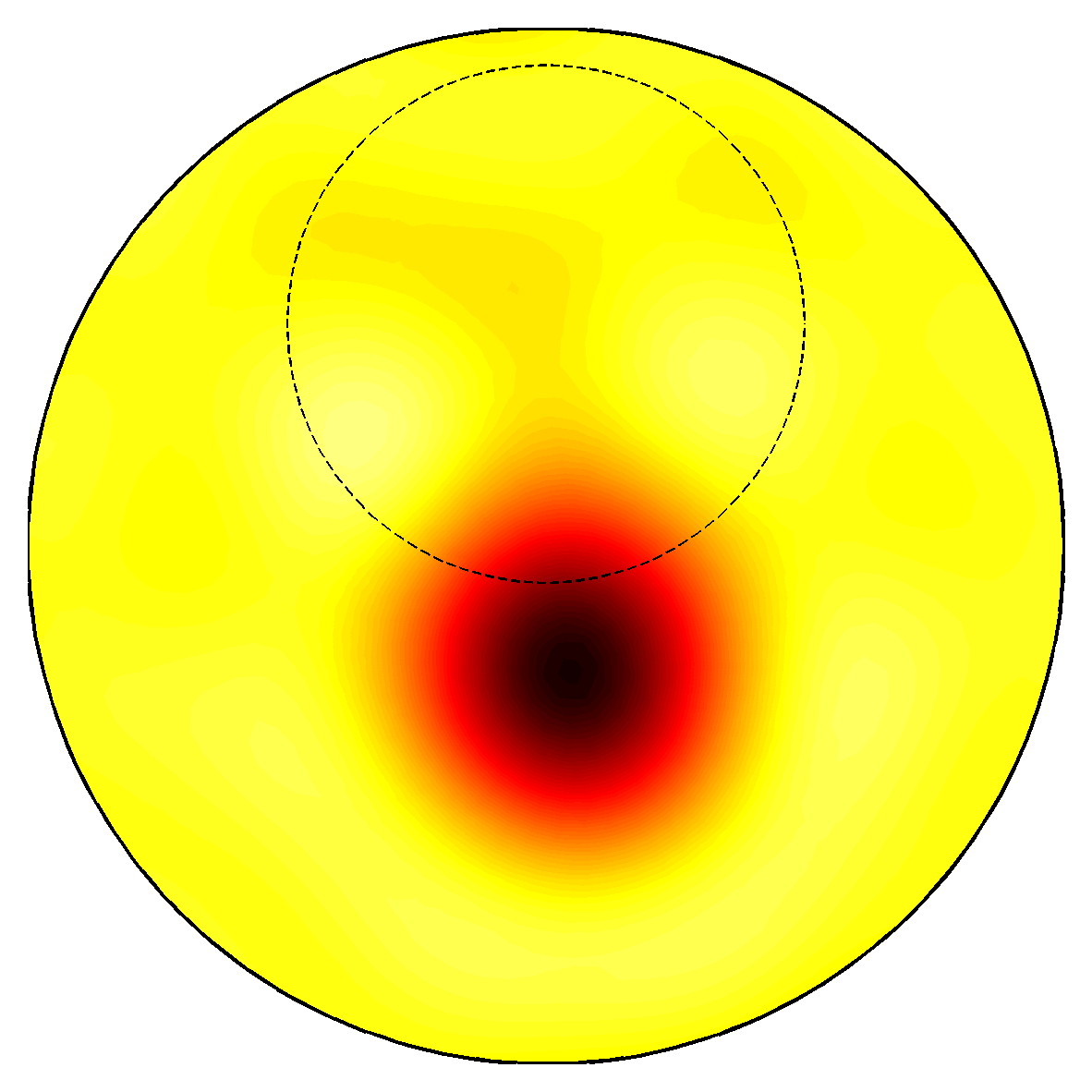}}
      & \parbox[c]{3cm}{\includegraphics[width=3cm]{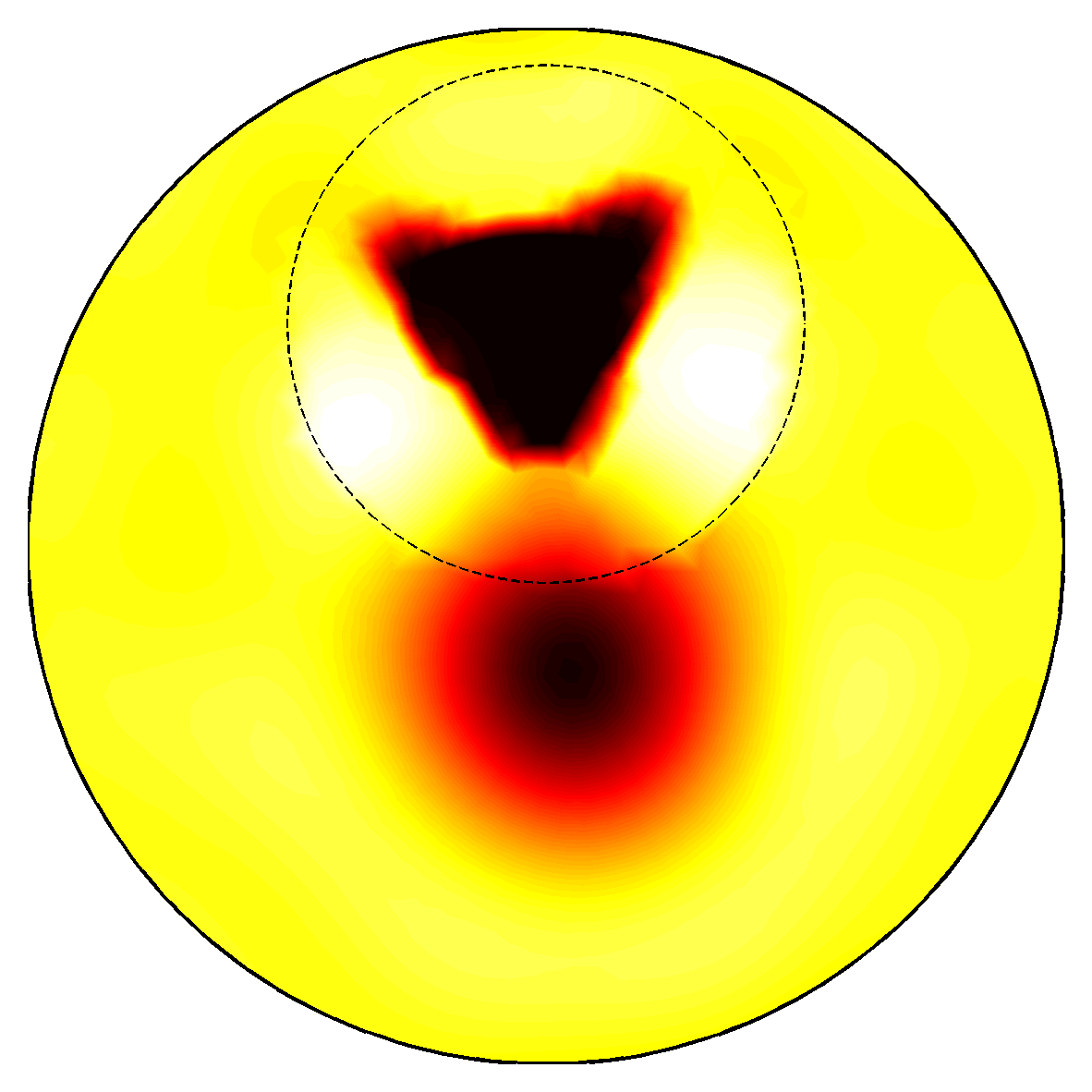}}
      & \parbox[c]{3cm}{\includegraphics[width=3cm]{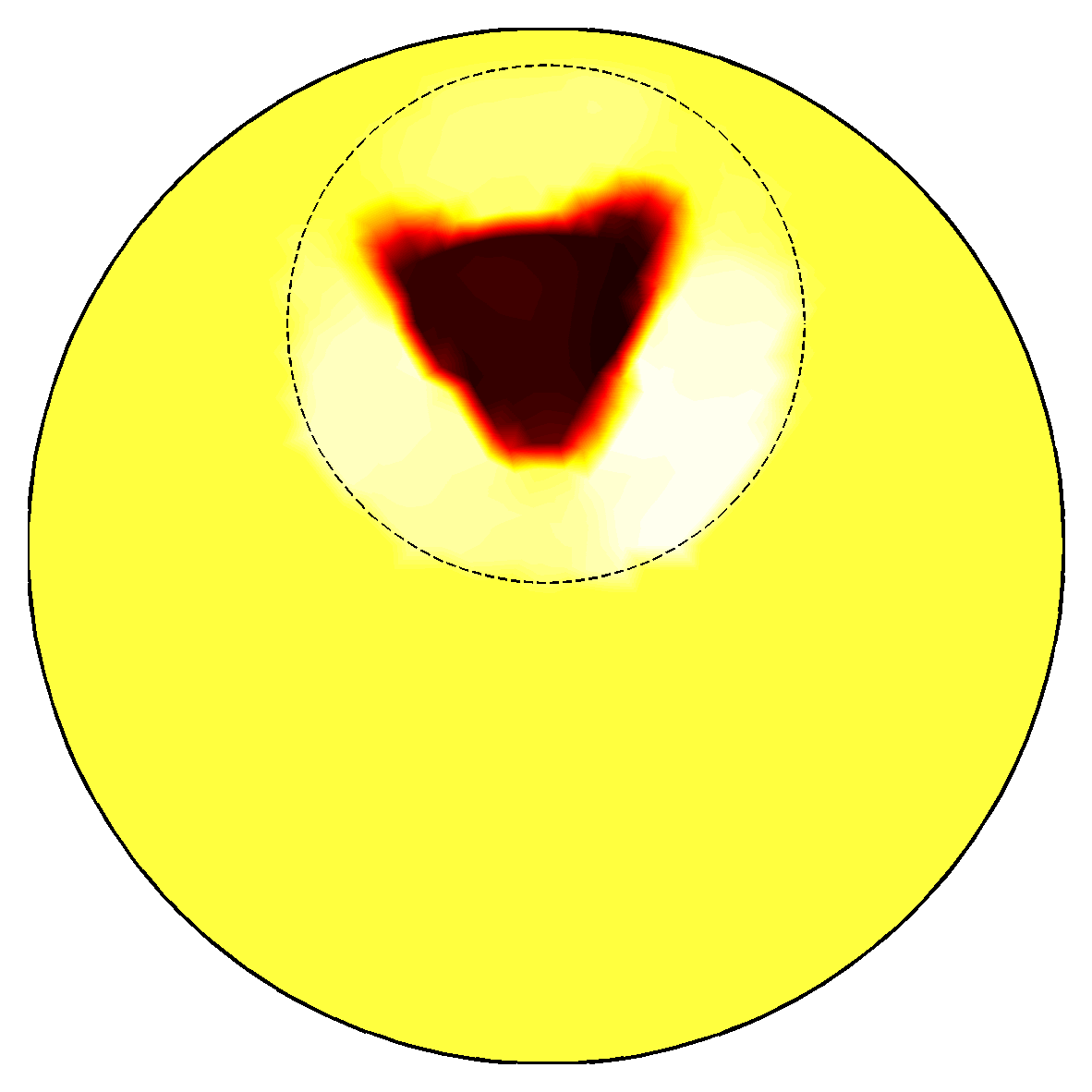}} \\[0 ex]
      %%%%%%%%%%%%%%%%%%%%%%%%%
       &\parbox[c]{3cm}{\includegraphics[width=3cm]{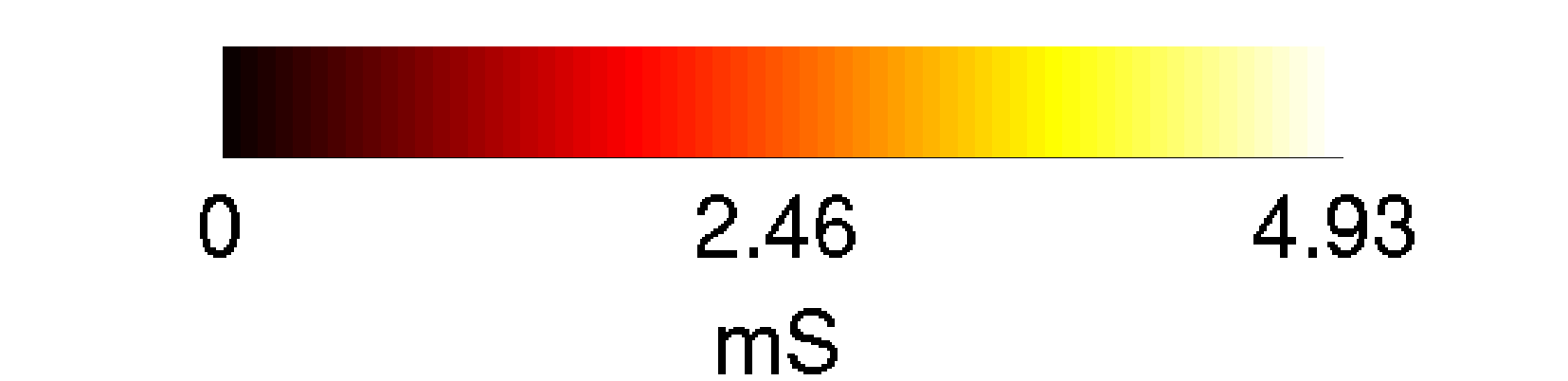}}
        &\parbox[c]{3cm}{\includegraphics[width=3cm]{figs/case13/cbs1.png}}
         &\parbox[c]{3cm}{\includegraphics[width=3cm]{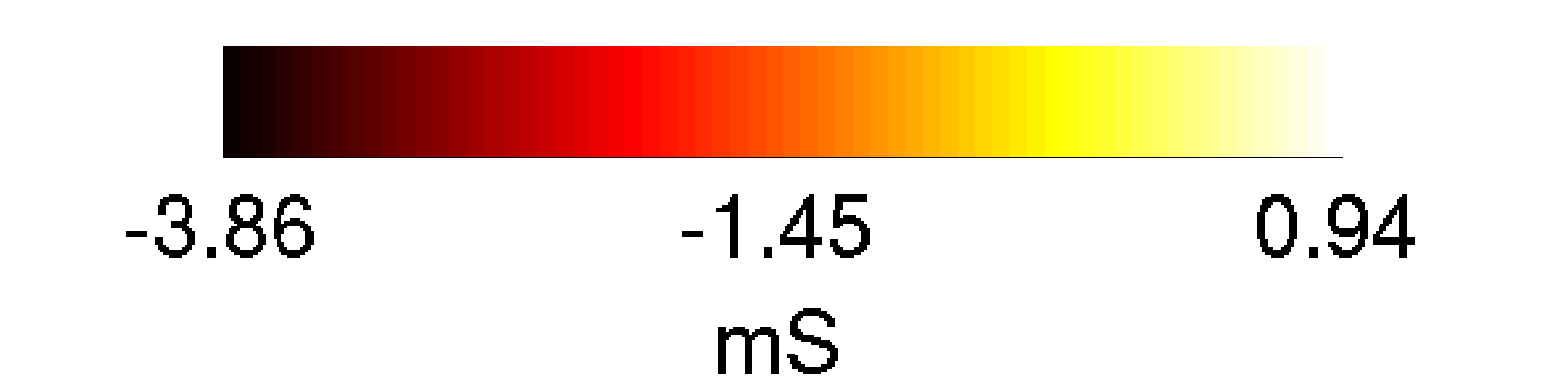}} \\[0 ex]
               %%%%%%%%%%%%%%%%%%%%%%%%%
    %   SD & & &\parbox[c]{3cm}{\includegraphics[width=3cm]{figs/case13/expdatadscase13Linearbeta1.png}} \\[3 ex]
       %%%%%%%%%%%%%%%
      %      & &  &\parbox[c]{3cm}{\includegraphics[width=3cm]{figs/case13/cbSD.png}} \\[0 ex]
    \end{tabular}
    \caption{Case 1: Reconstructions from real data. 
(E1)-(E4) refer to the estimates listed in
section \ref{estsec}.}
 \label{case13}
    \end{figure}

\begin{figure}  [h] 
%\caption{Stimuli Category Explanations} 
\centering
  \begin{tabular} {lccc}
   & $\sigma_1$ & $\sigma_2$ & $\delta\sigma$ \\
      (E1) & \parbox[c]{3cm}{\includegraphics[width=3cm]{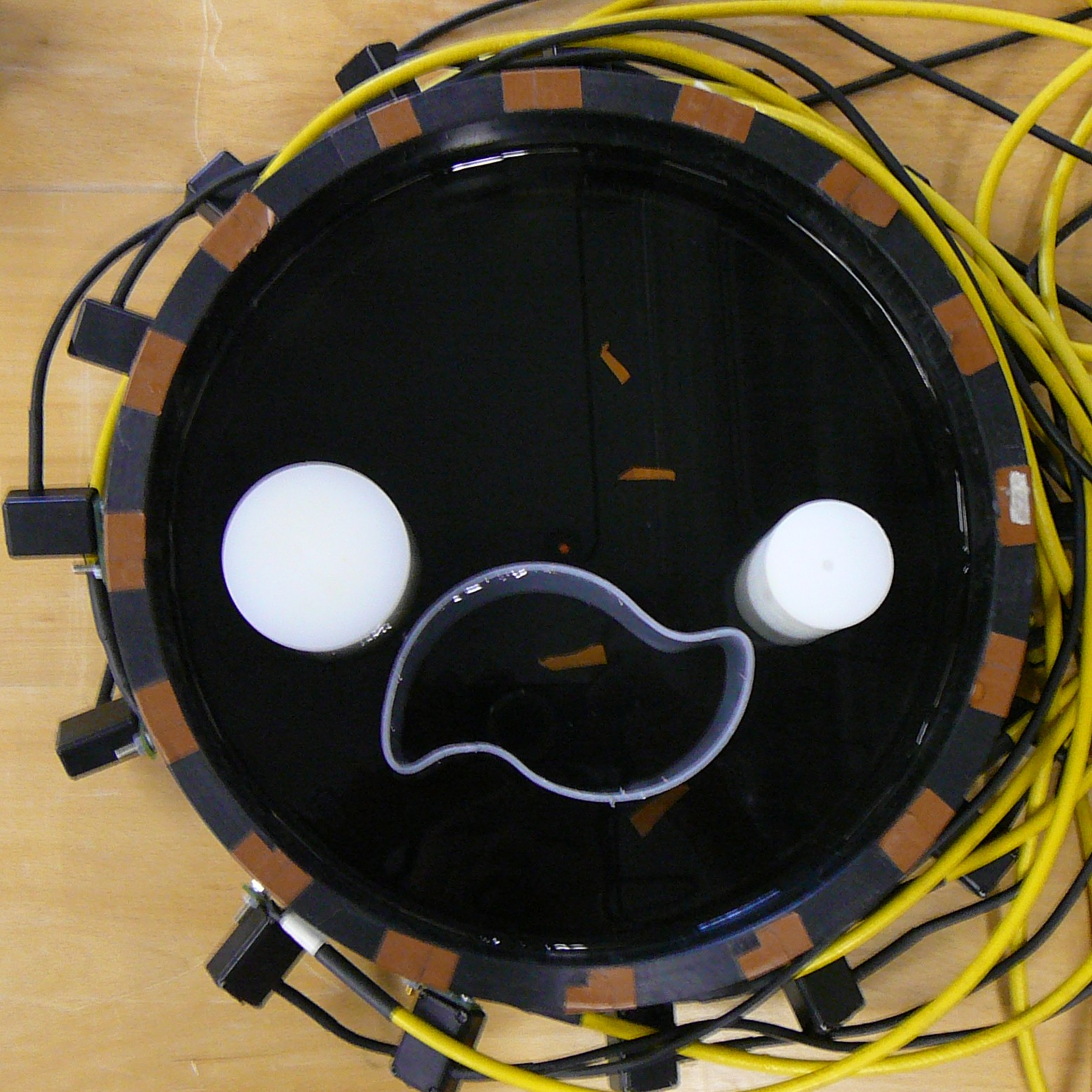}}
      & \parbox[c]{3cm}{\includegraphics[width=3cm]{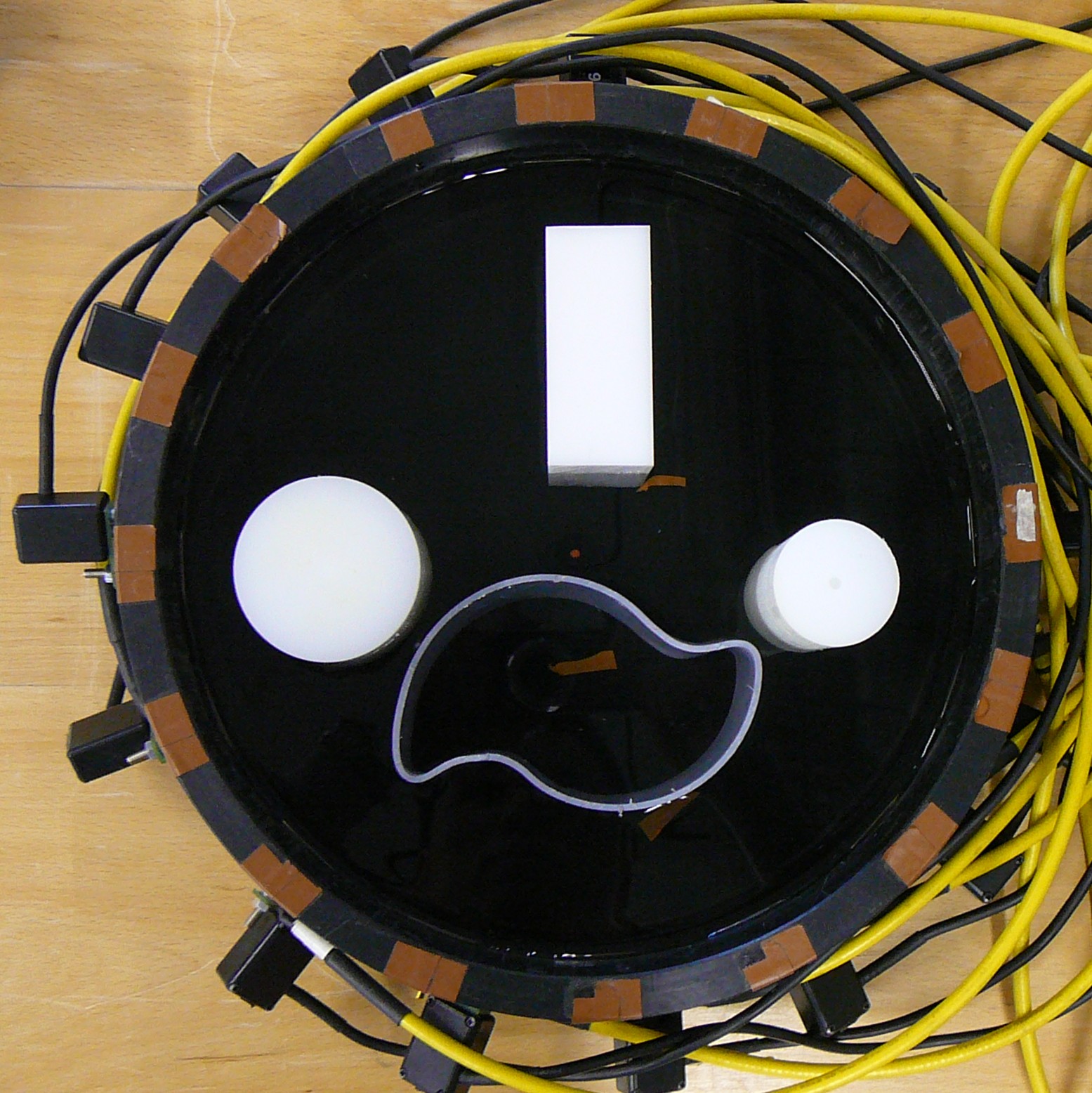}}
      &  \parbox[c]{3cm}{\includegraphics[width=3cm]{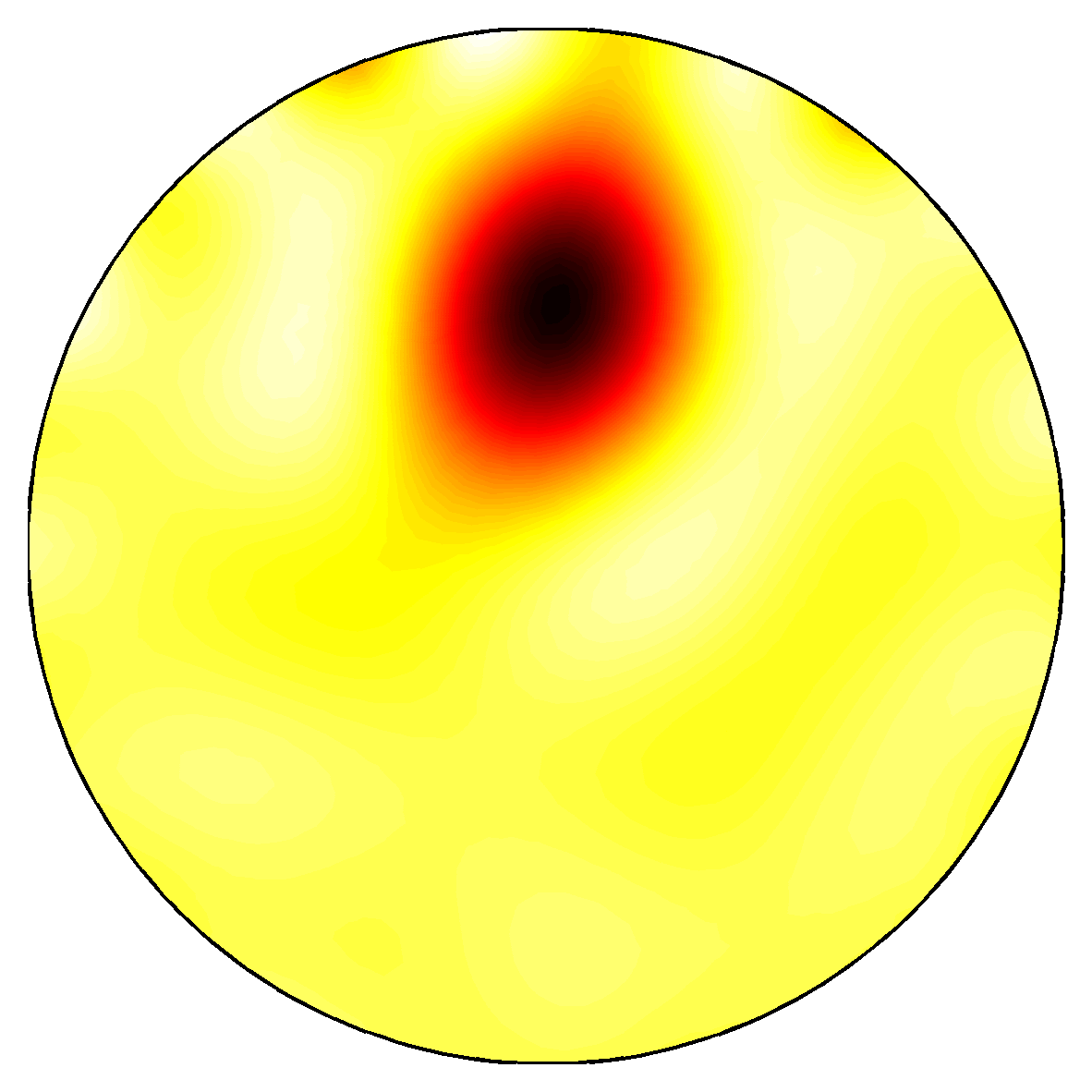}} \\[3 ex]
       %%%%%%%%%%%%%%%
            & &  &\parbox[c]{3cm}{\includegraphics[width=3cm]{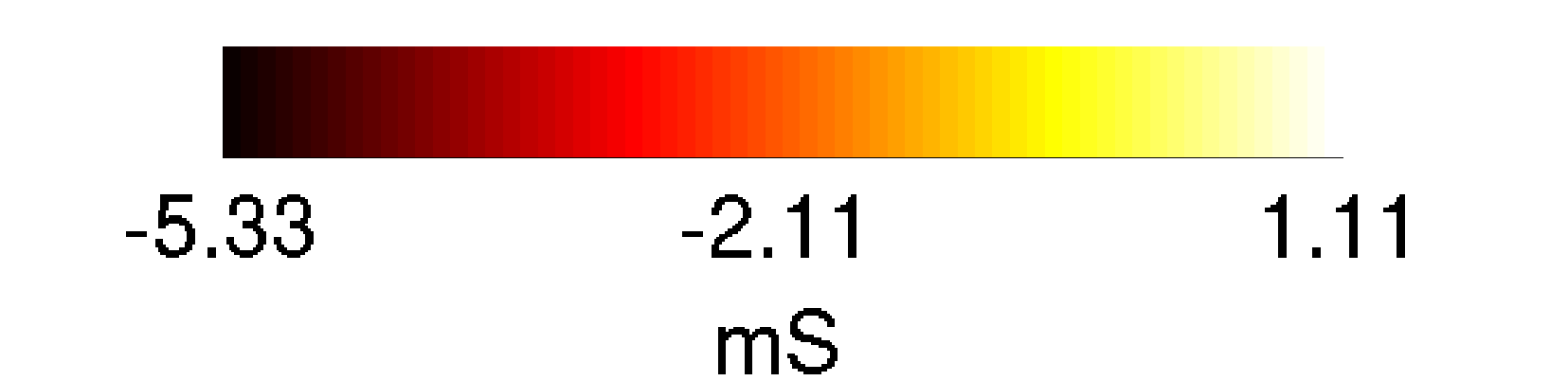}} \\[1 ex]  

 %%%%%%%%%%%%%%%%%%%%%%%
      (E2) & \parbox[c]{3cm}{\includegraphics[width=3cm]{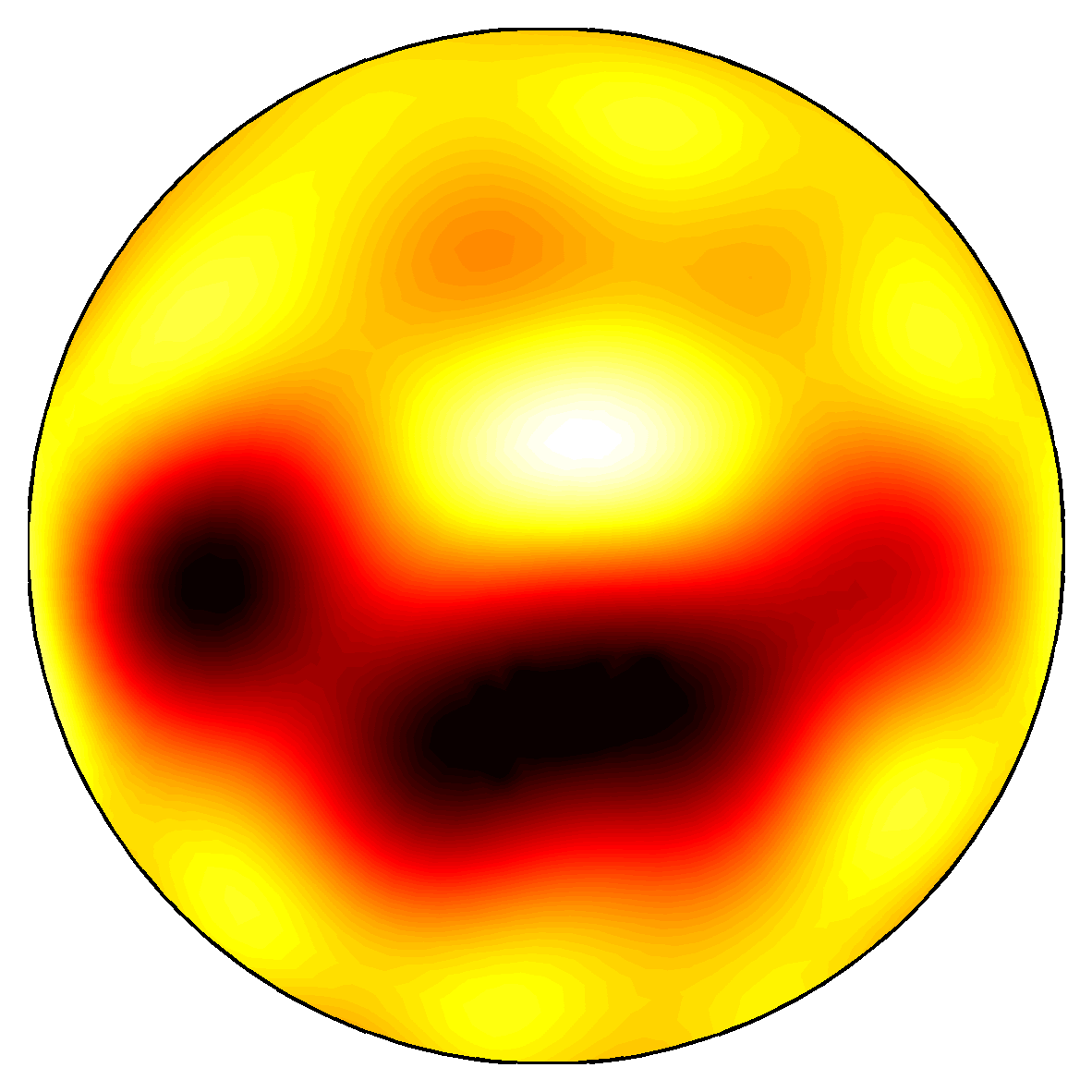}}
      & \parbox[c]{3cm}{\includegraphics[width=3cm]{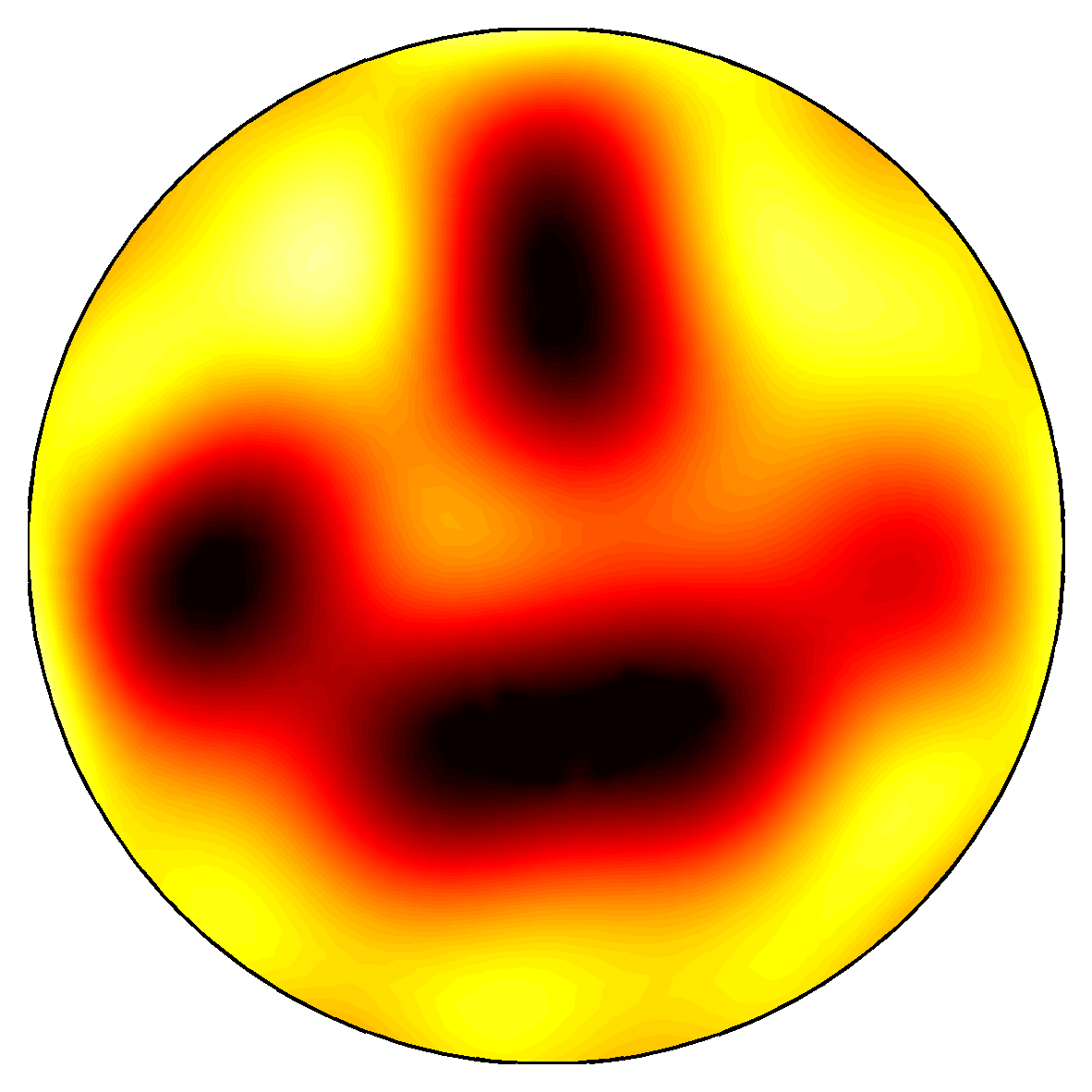}}
      & \parbox[c]{3cm}{\includegraphics[width=3cm]{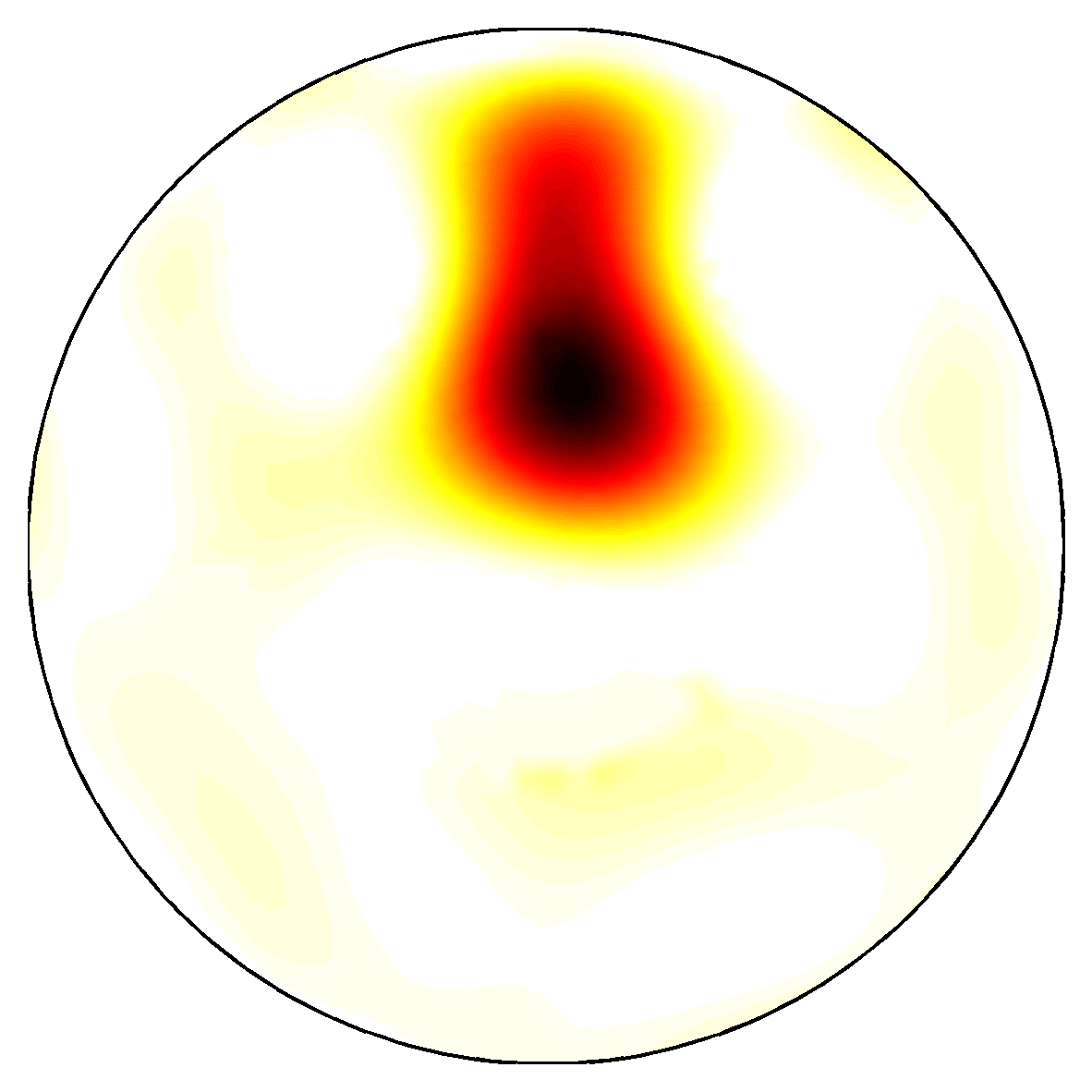}} \\[0 ex]
            %%%%%%%%%%%%%%%%%%%%%%%
      (E3) & \parbox[c]{3cm}{\includegraphics[width=3cm]{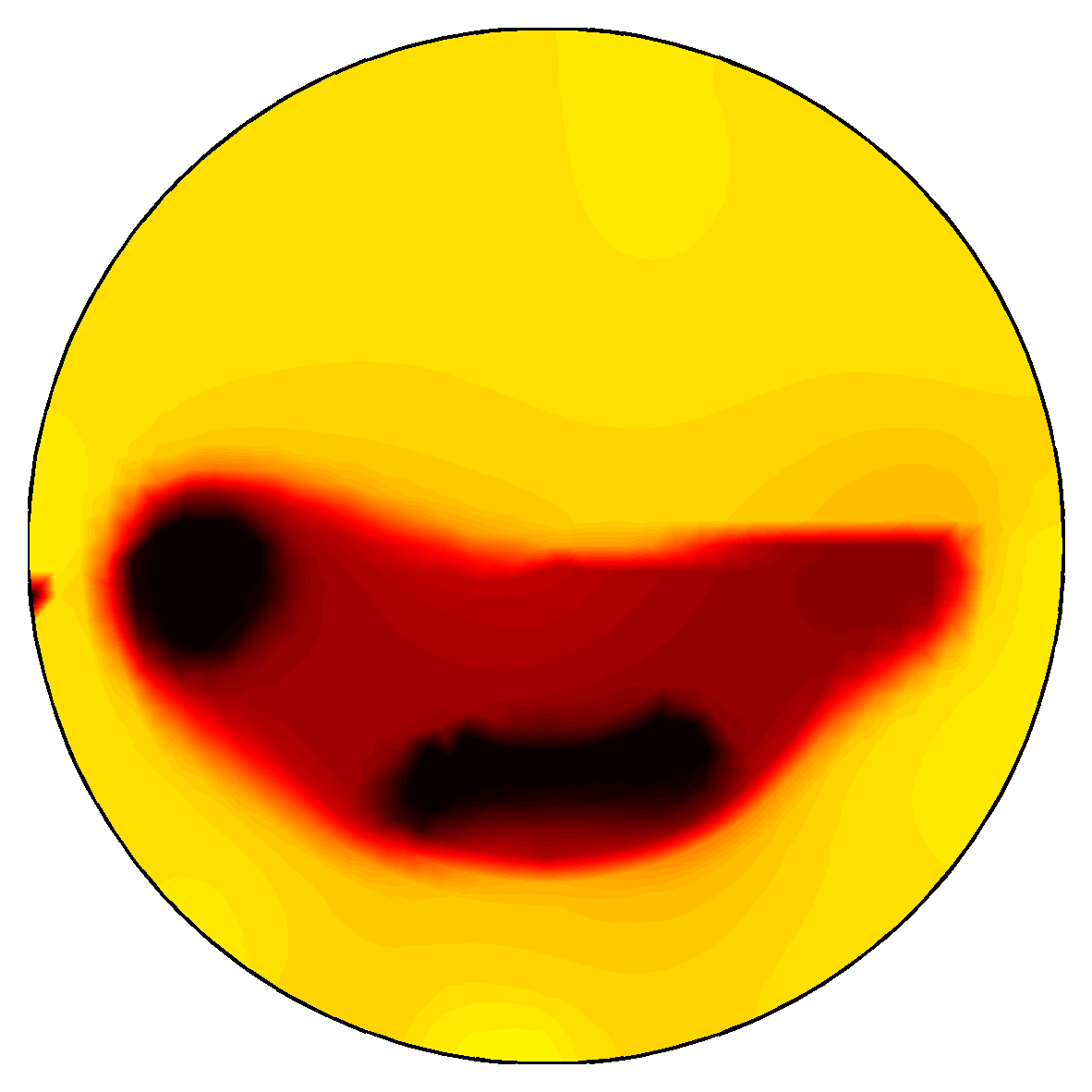}}
      & \parbox[c]{3cm}{\includegraphics[width=3cm]{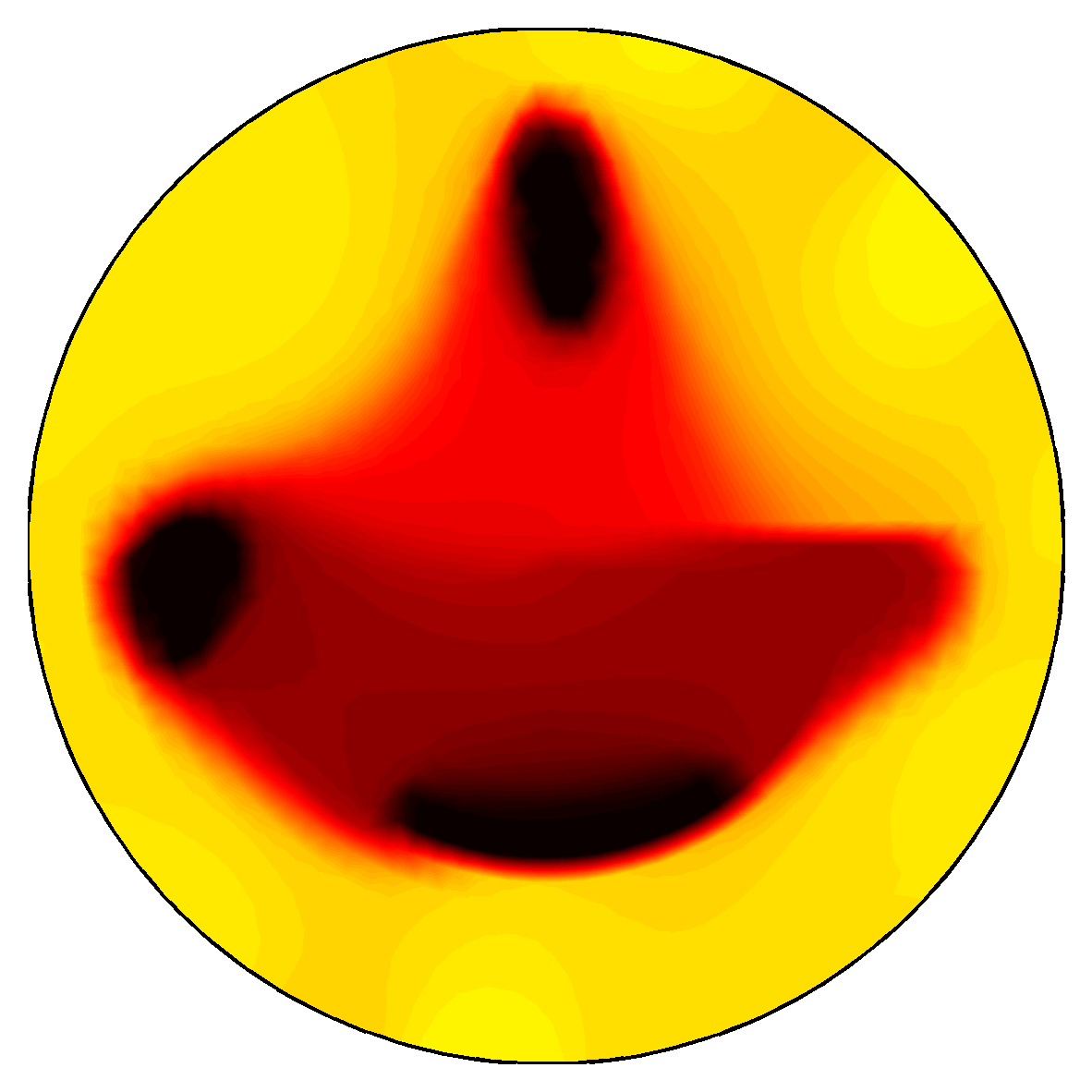}}
      & \parbox[c]{3cm}{\includegraphics[width=3cm]{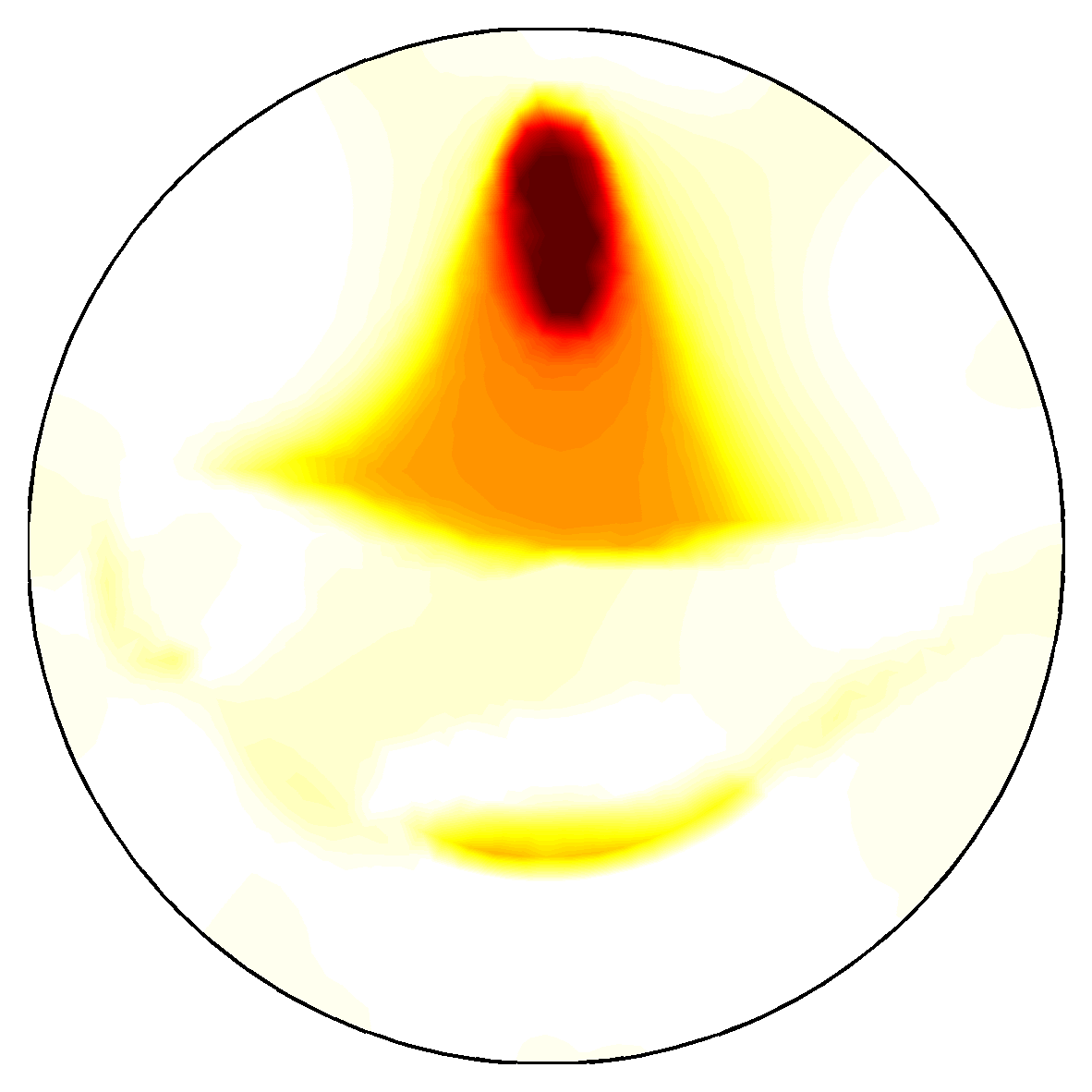}} \\[0 ex]
                  %%%%%%%%%%%%%%%%%%%%%%%
      (E4) & \parbox[c]{3cm}{\includegraphics[width=3cm]{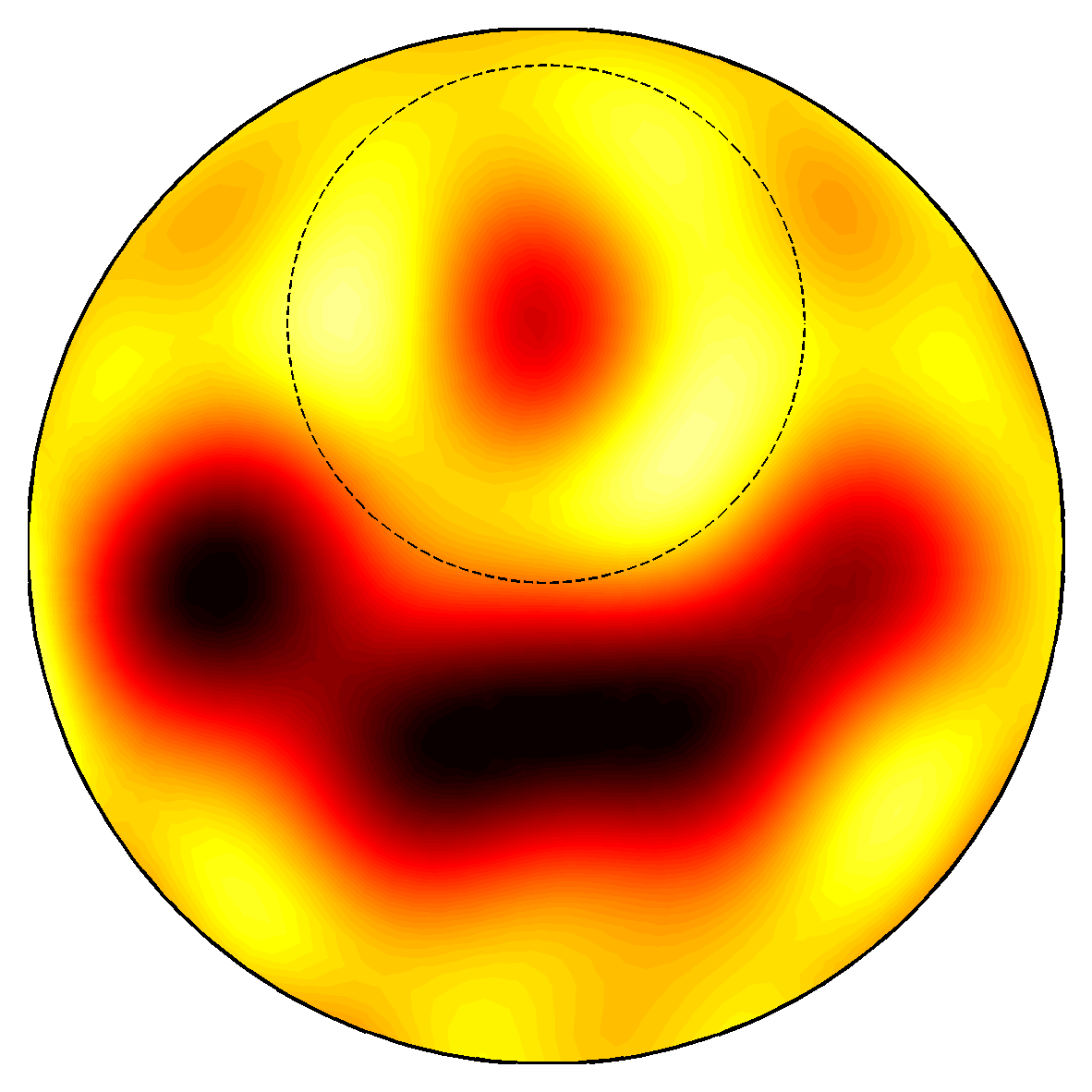}}
      & \parbox[c]{3cm}{\includegraphics[width=3cm]{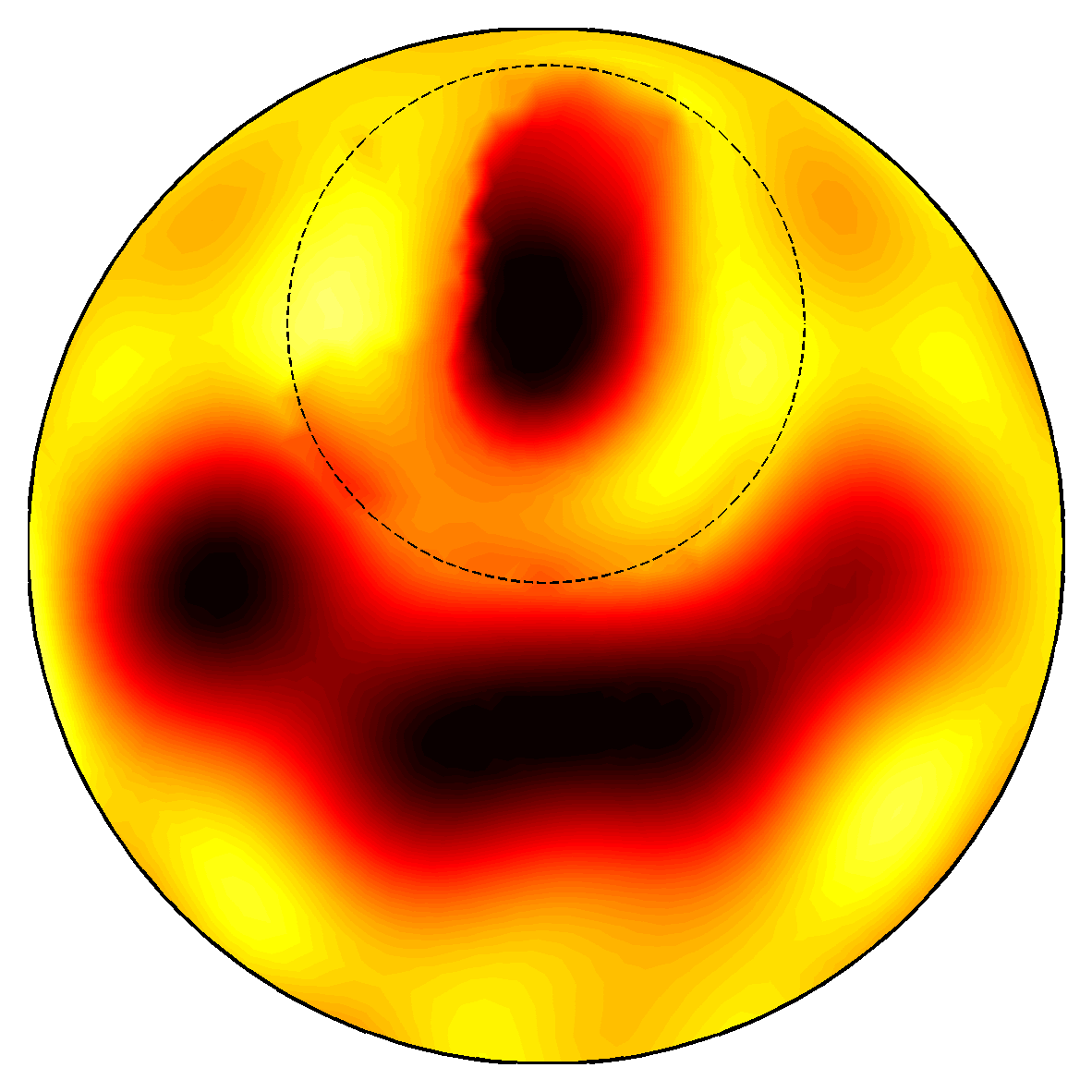}}
      & \parbox[c]{3cm}{\includegraphics[width=3cm]{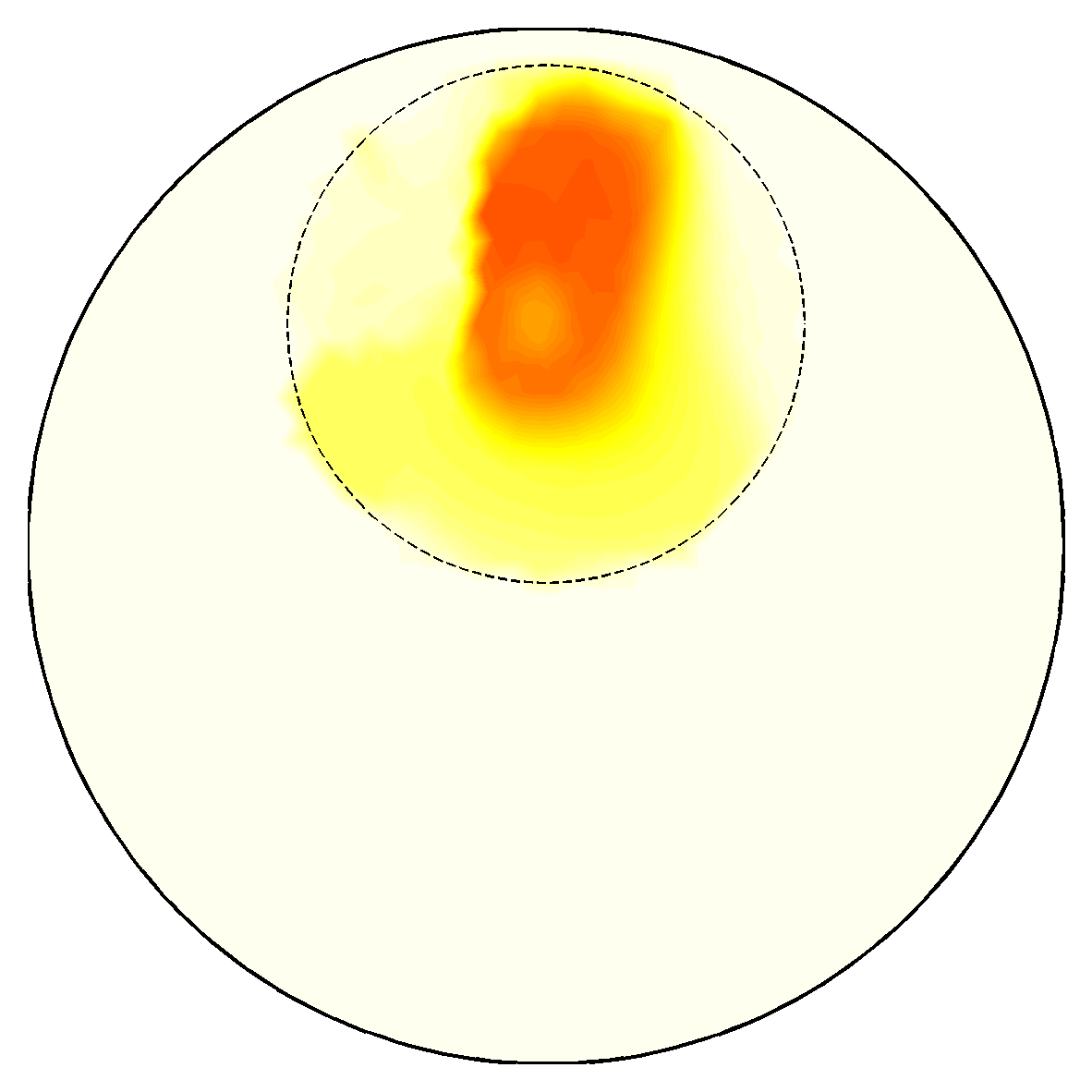}} \\[0 ex]
      %%%%%%%%%%%%%%%%%%%%%%%%%
       &\parbox[c]{3cm}{\includegraphics[width=3cm]{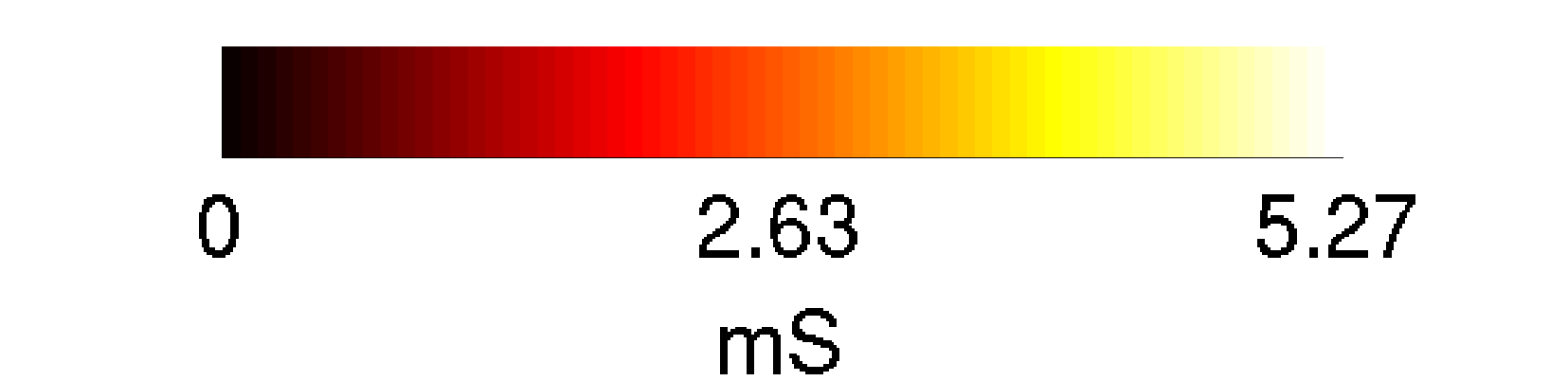}}
        &\parbox[c]{3cm}{\includegraphics[width=3cm]{figs/case59/cbs1.png}}
         &\parbox[c]{3cm}{\includegraphics[width=3cm]{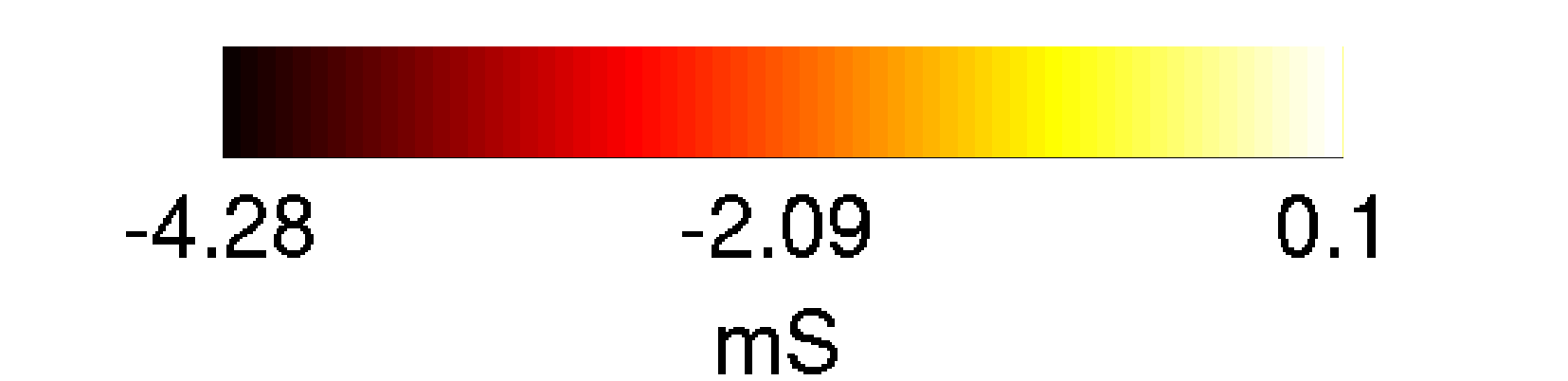}} \\[0 ex]
               %%%%%%%%%%%%%%%%%%%%%%%%%
%       SD & & &\parbox[c]{3cm}{\includegraphics[width=3cm]{figs/case59/expdatadscase59Linearbeta1.png}} \\[3 ex]
%       %%%%%%%%%%%%%%%
%            & &  &\parbox[c]{3cm}{\includegraphics[width=3cm]{figs/case59/cbSD.png}} \\[0 ex]
    \end{tabular}
    \caption{Case 2. Reconstructions from real data. 
(E1)-(E4) refer to the estimates listed in
section \ref{estsec}.} 
\label{case59}
    \end{figure}

\begin{figure}  [ht] 
\centering
  \begin{tabular} {lccc}
   & $\sigma_1$ & $\sigma_2$ & $\delta\sigma$ \\
      (E1) & \parbox[c]{3cm}{\includegraphics[width=3cm]{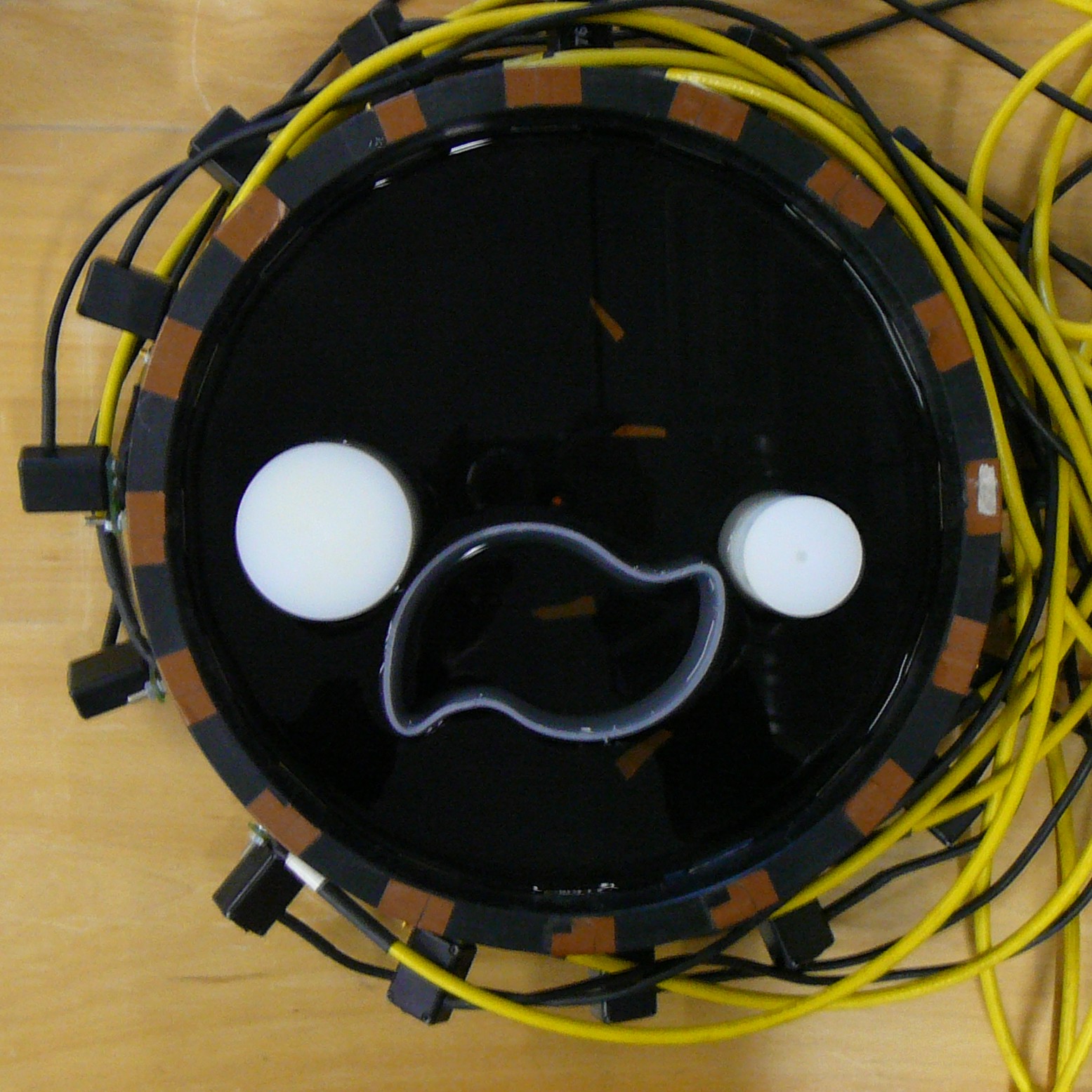}}
      & \parbox[c]{3cm}{\includegraphics[width=3cm]{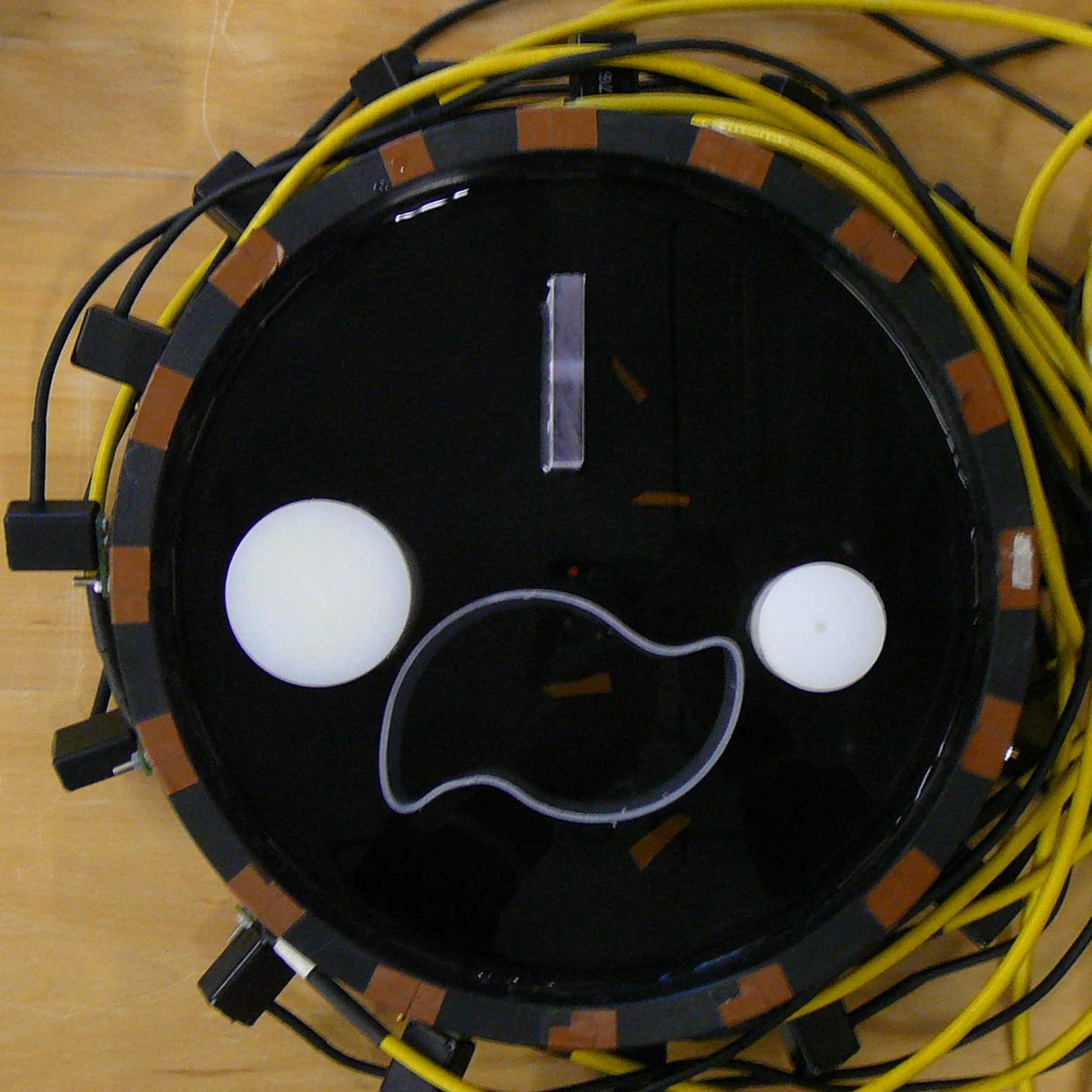}}
      % &  \\[5 ex]
       & \parbox[c]{3cm}{\includegraphics[width=3cm]{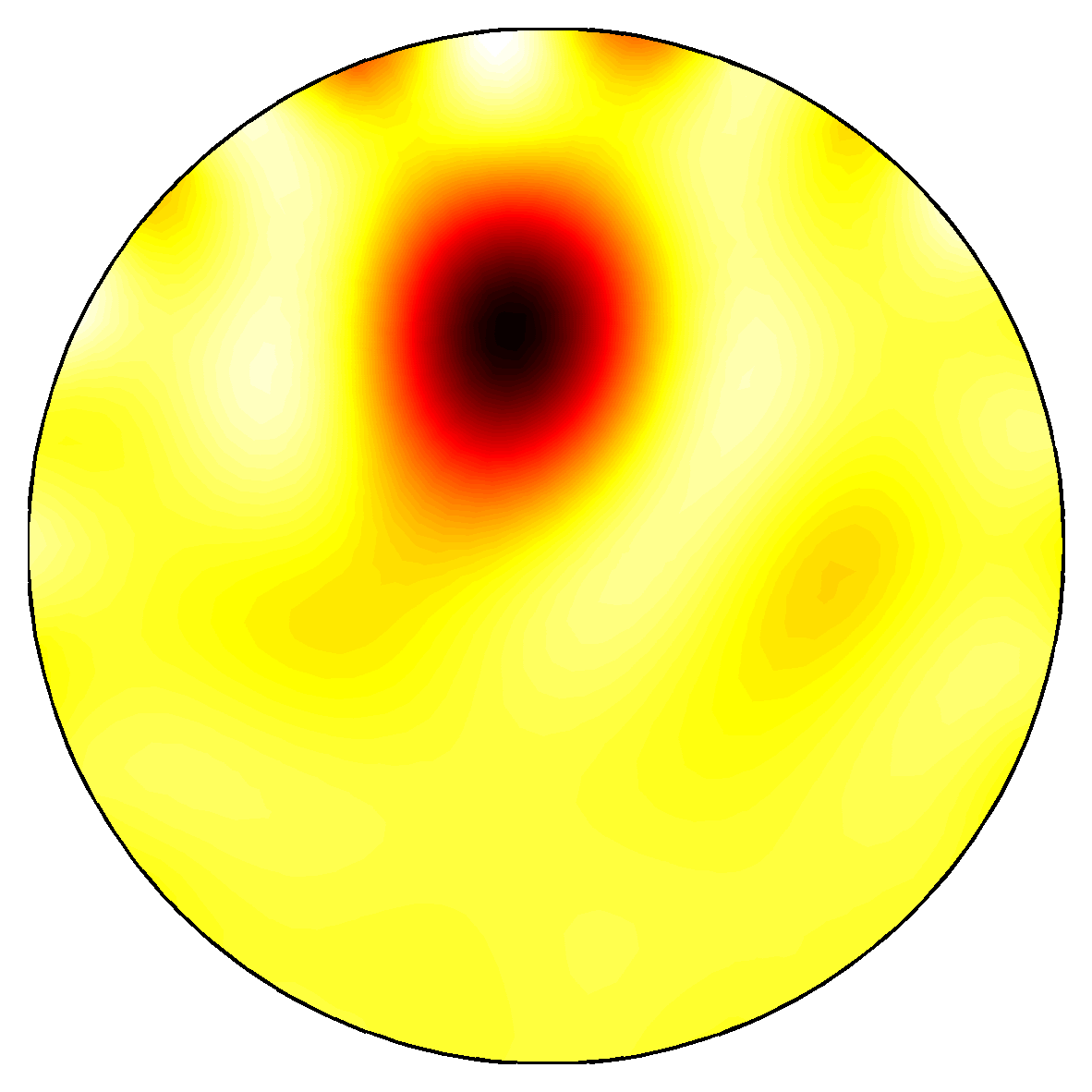}} \\[3 ex]
        & &  &\parbox[c]{3cm}{\includegraphics[width=3cm]{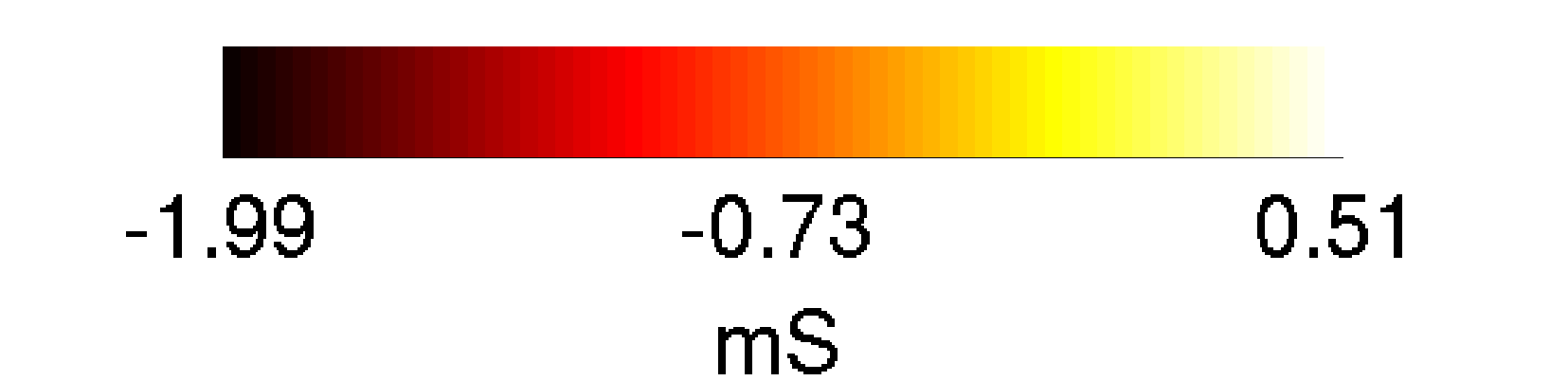}} \\[1 ex]
      %%%%%%%%%%%%%%%%%%%%%%%
      (E2) & \parbox[c]{3cm}{\includegraphics[width=3cm]{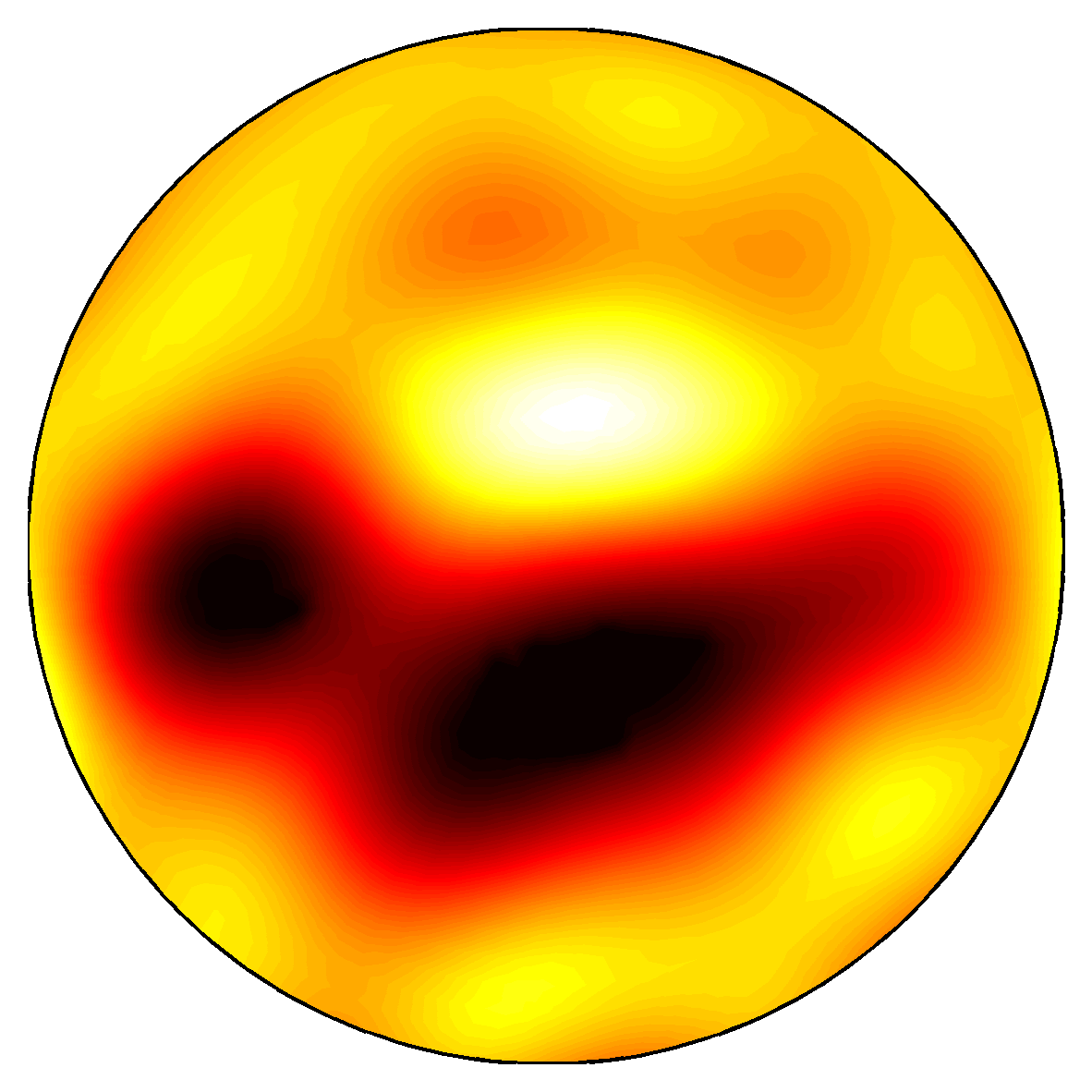}}
      & \parbox[c]{3cm}{\includegraphics[width=3cm]{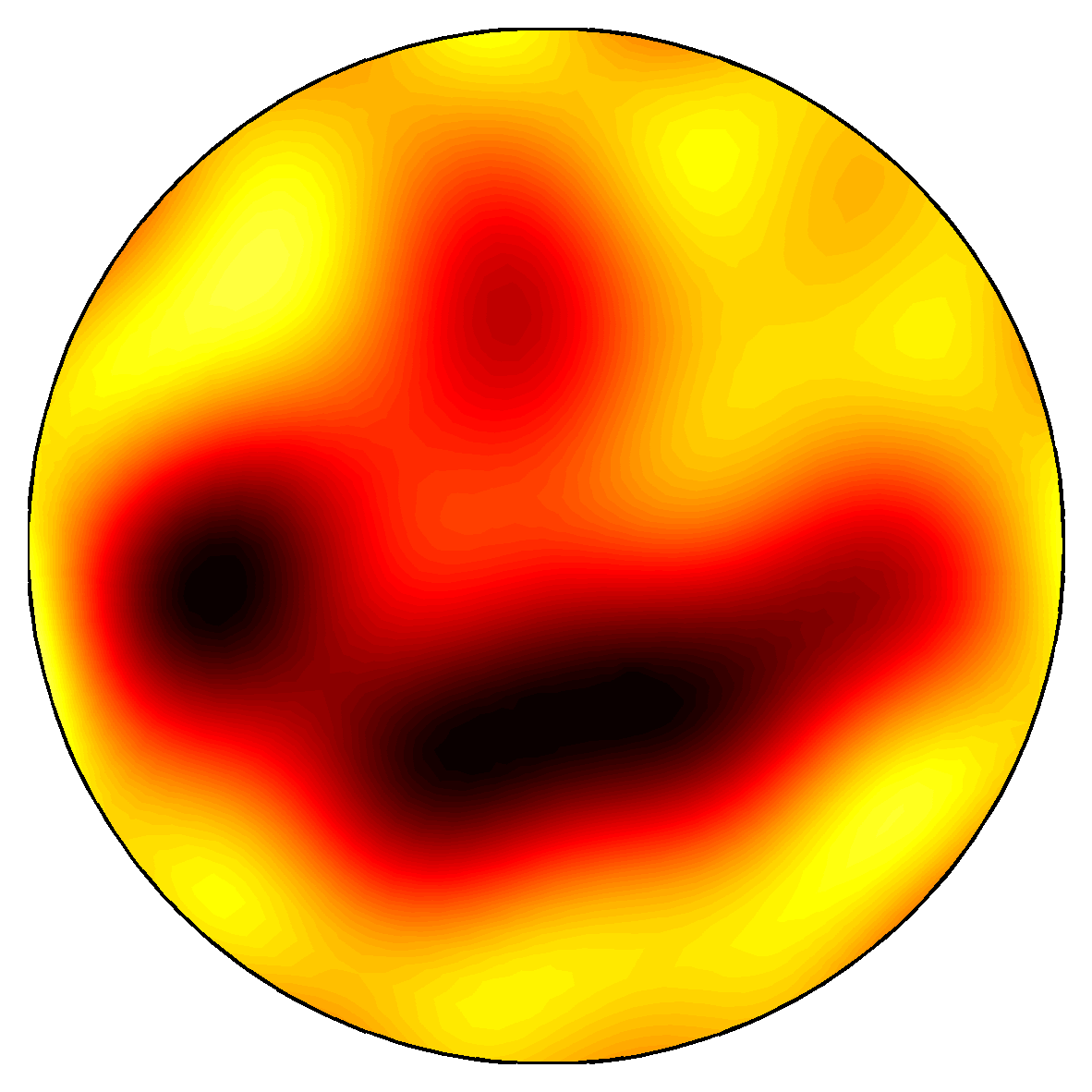}}
      & \parbox[c]{3cm}{\includegraphics[width=3cm]{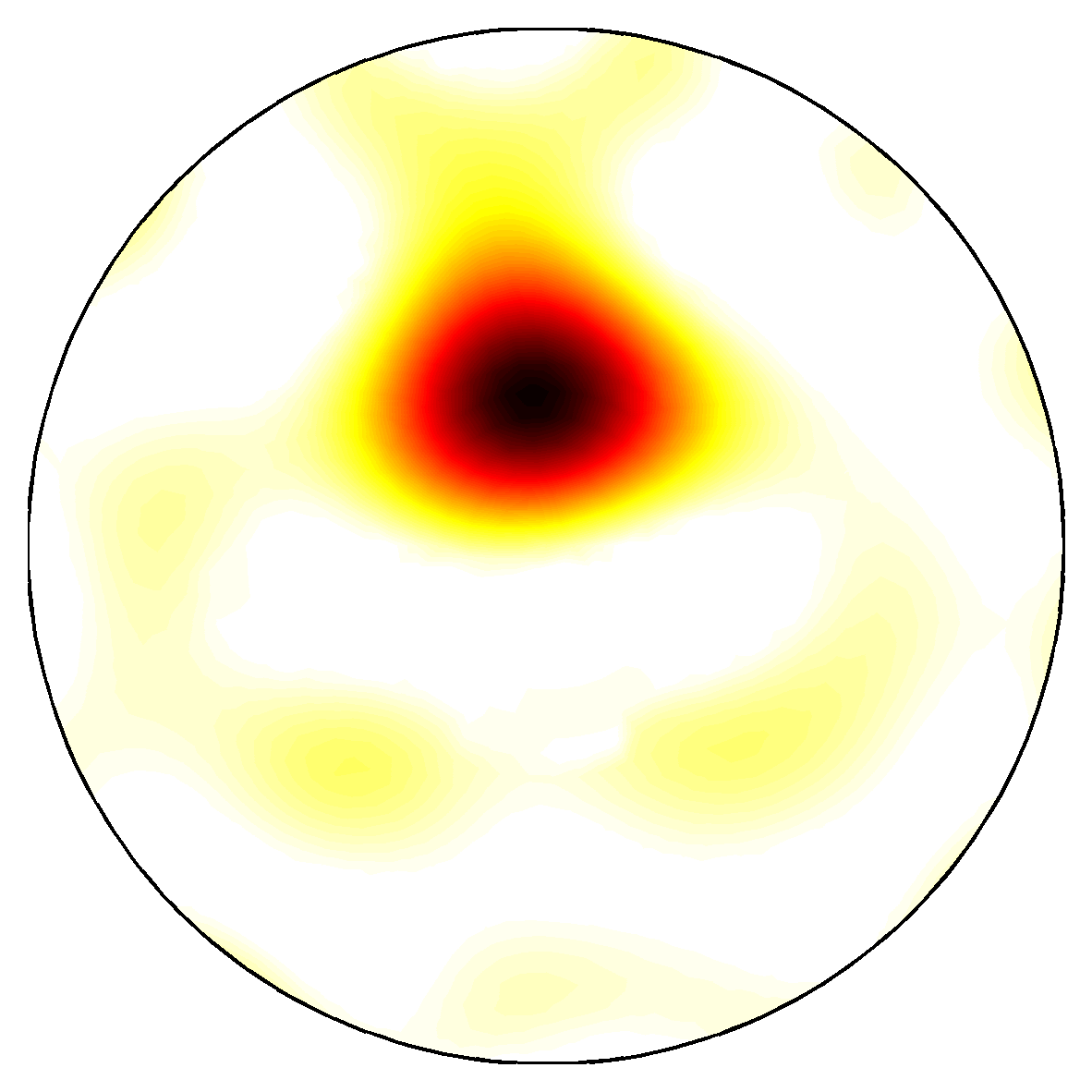}} \\[3 ex]
            %%%%%%%%%%%%%%%%%%%%%%%
      (E3) & \parbox[c]{3cm}{\includegraphics[width=3cm]{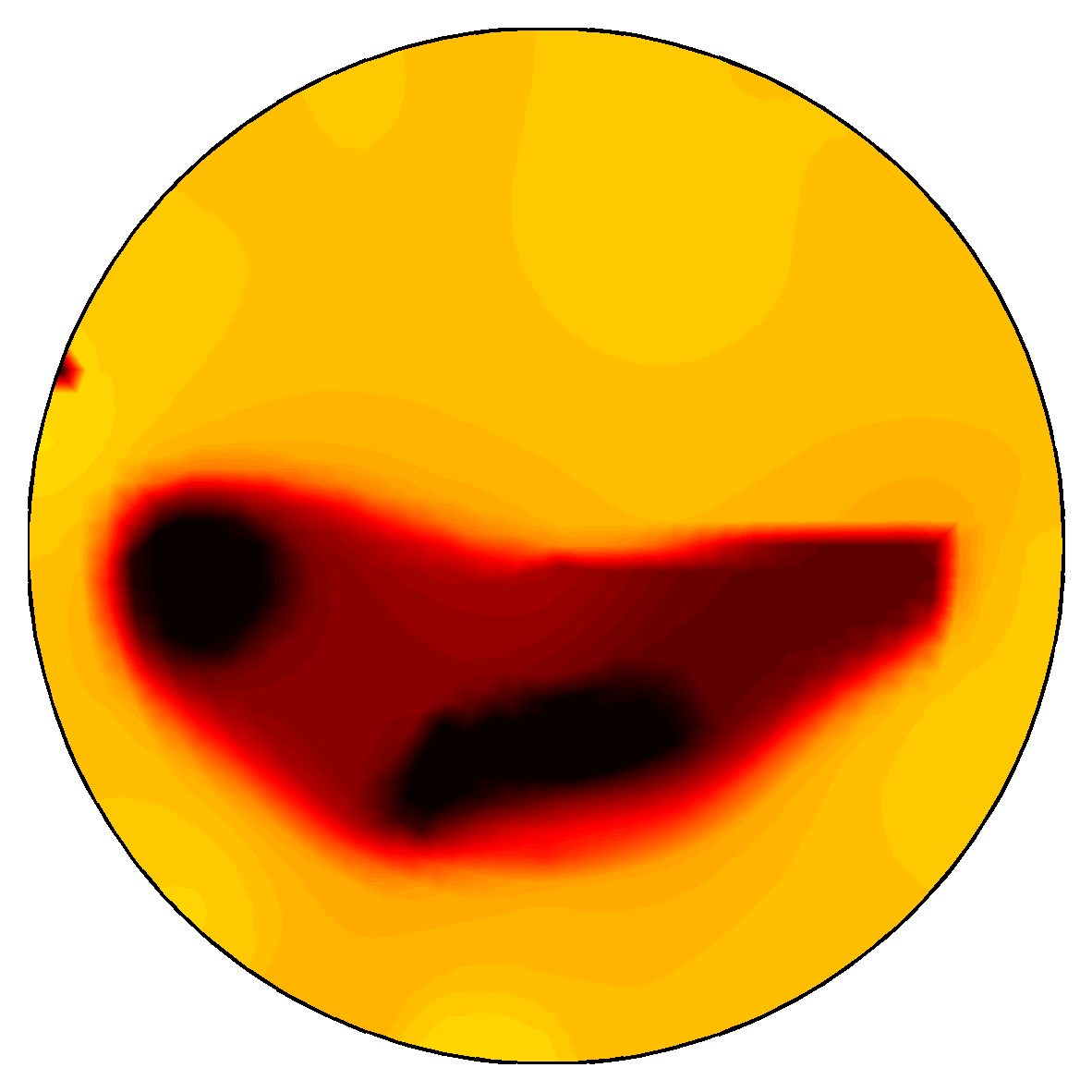}}
      & \parbox[c]{3cm}{\includegraphics[width=3cm]{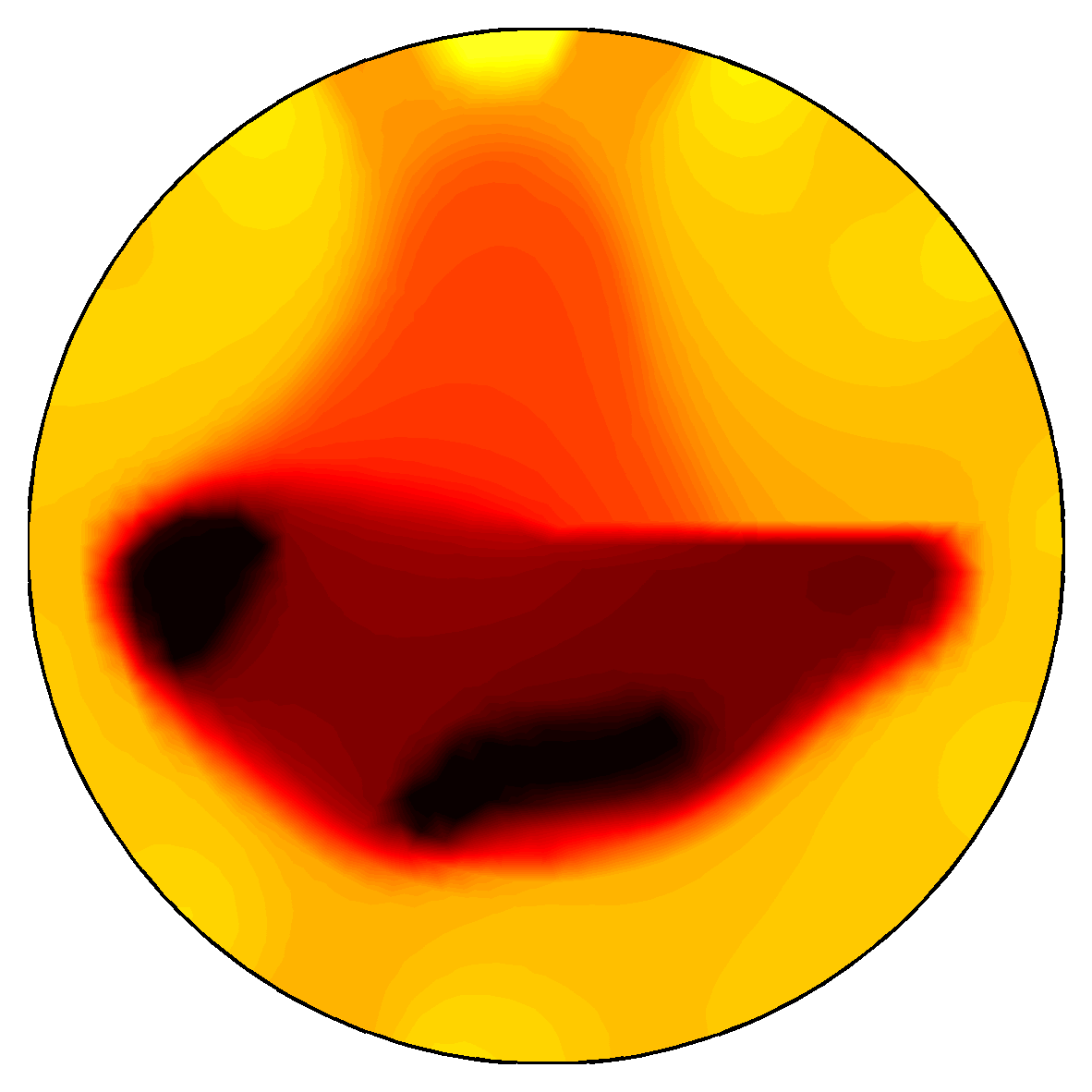}}
      & \parbox[c]{3cm}{\includegraphics[width=3cm]{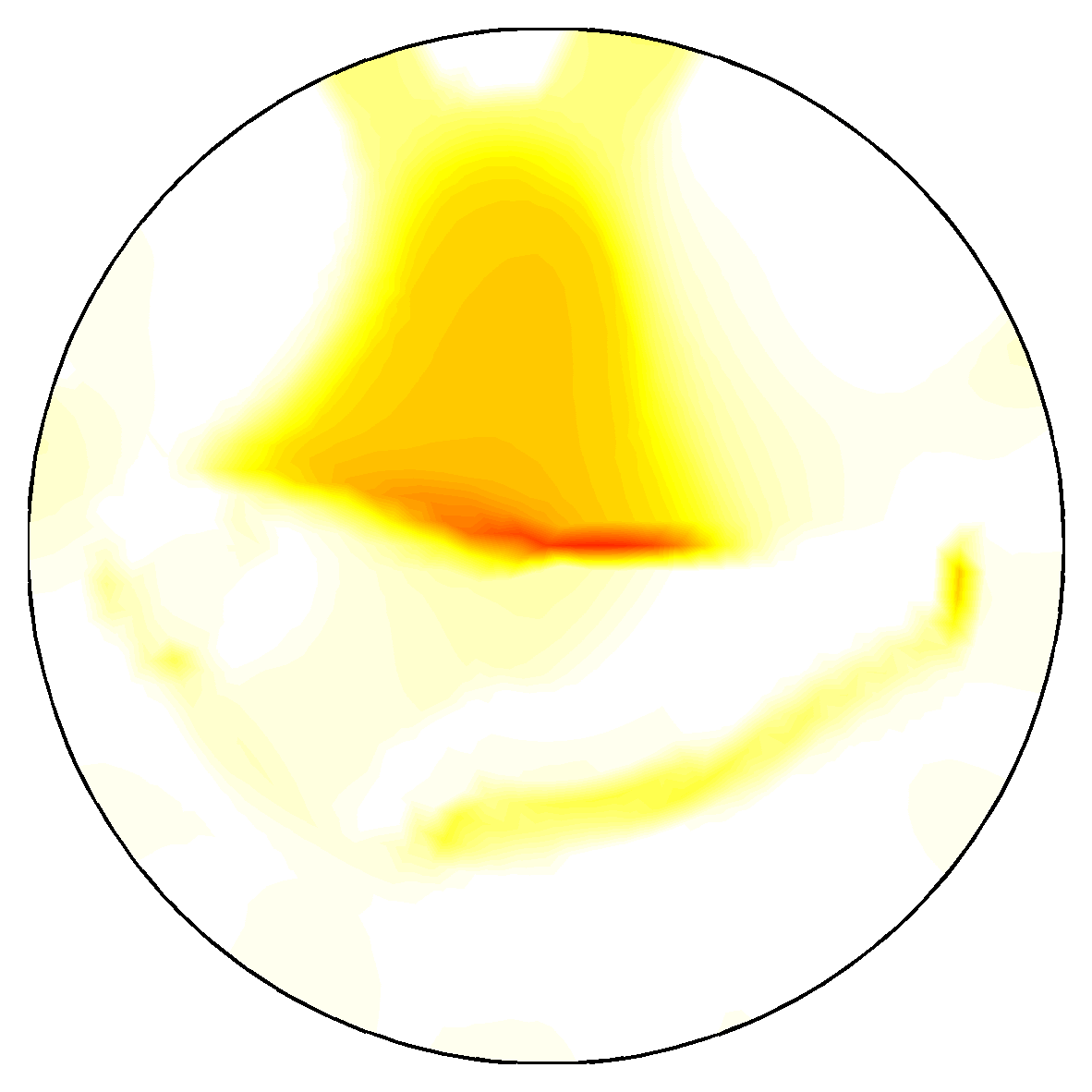}} \\[3 ex]
                  %%%%%%%%%%%%%%%%%%%%%%%
      (E4) & \parbox[c]{3cm}{\includegraphics[width=3cm]{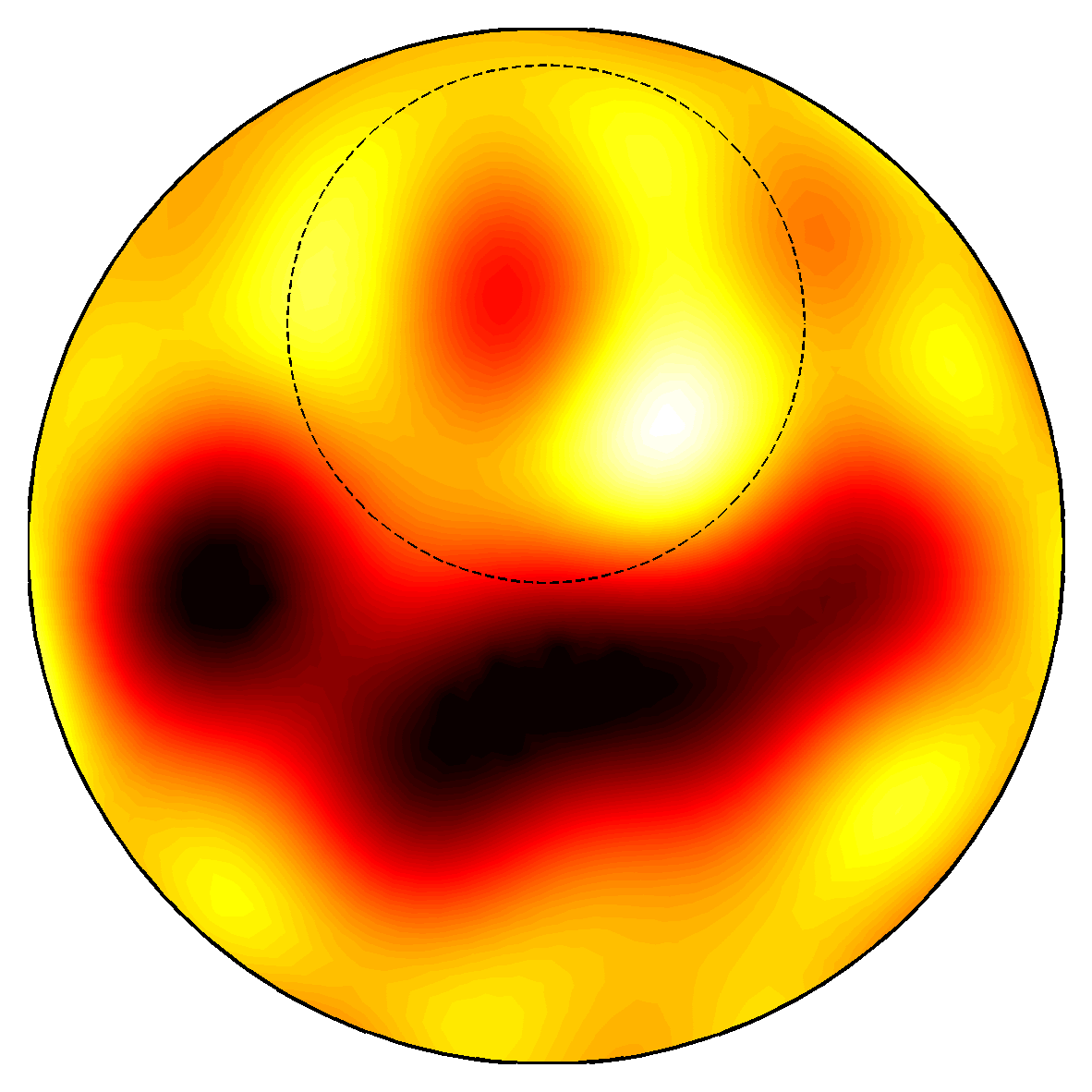}}
      & \parbox[c]{3cm}{\includegraphics[width=3cm]{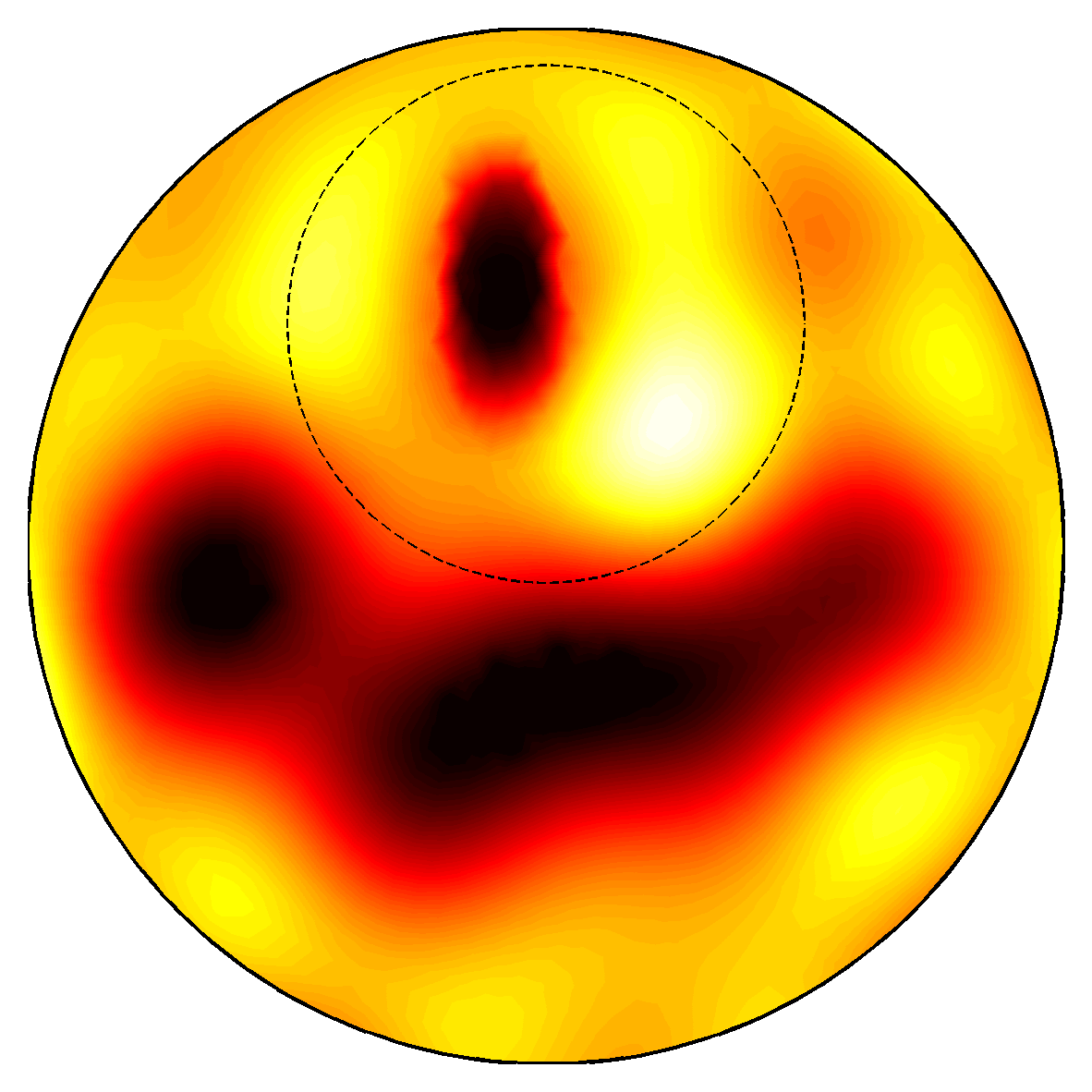}}
      & \parbox[c]{3cm}{\includegraphics[width=3cm]{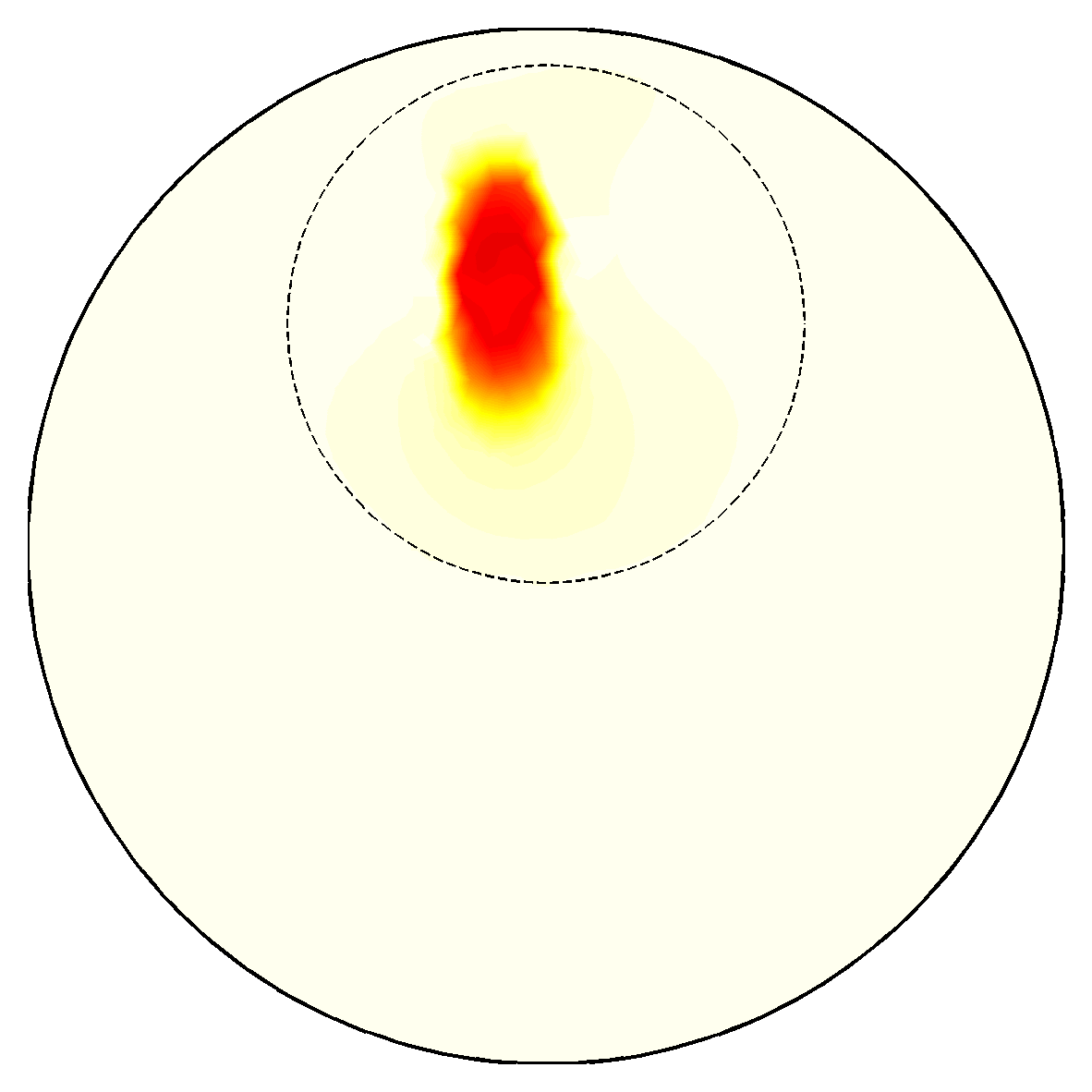}} \\[3 ex]
      %%%%%%%%%%%%%%%%%%%%%%%%%
       &\parbox[c]{3cm}{\includegraphics[width=3cm]{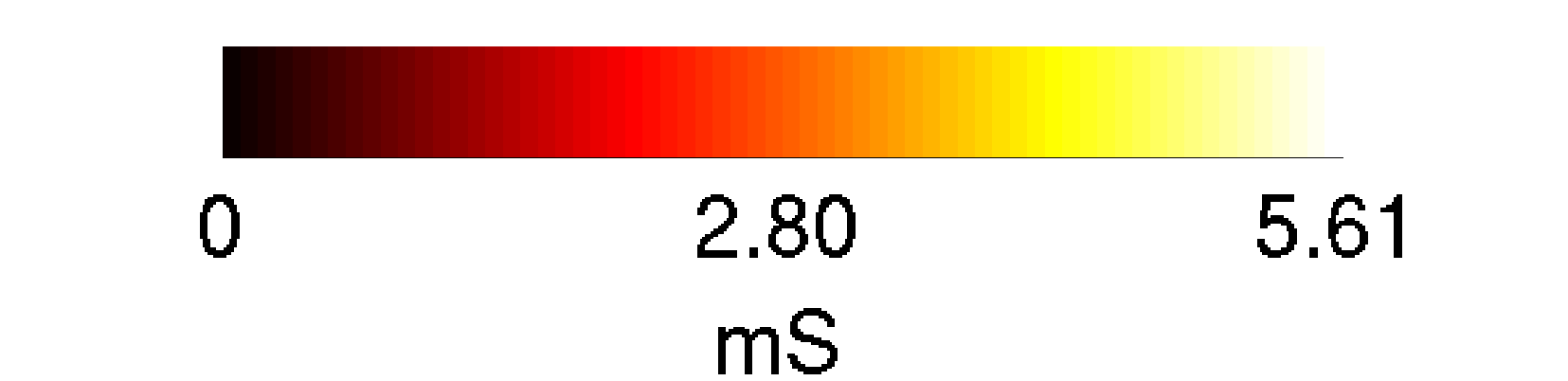}}
        &\parbox[c]{3cm}{\includegraphics[width=3cm]{figs/case73/cbs1.png}}
         &\parbox[c]{3cm}{\includegraphics[width=3cm]{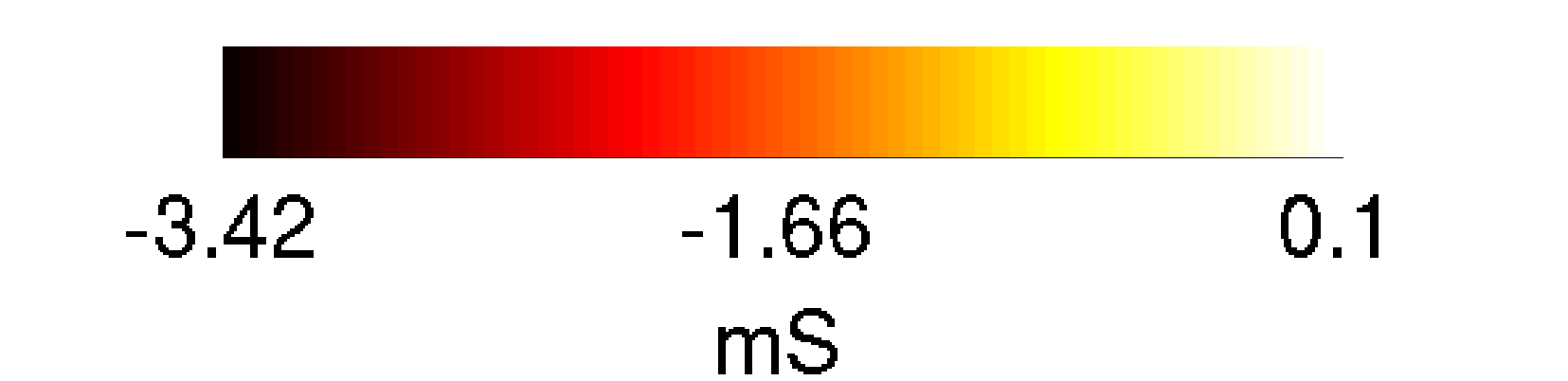}} \\[0 ex]
               %%%%%%%%%%%%%%%%%%%%%%%%%
%       SD & & &\parbox[c]{3cm}{\includegraphics[width=3cm]{figs/case73/expdatadscase73Linearbeta1.png}} \\[3 ex]
       %%%%%%%%%%%%%%%
           % & &  &\parbox[c]{3cm}{\includegraphics[width=3cm]{figs/case73/cbSD.png}} \\[0 ex]
    \end{tabular}
    \caption{Case 3: Reconstructions from real data. 
(E1)-(E4) refer to the estimates listed in
section \ref{estsec}.} 
\label{case73}
    \end{figure}

The results for the experimental test cases are shown in Figs \ref{case13}-\ref{case73}. 
Notice that since we are employing a two-dimensional model 
for vertically translationally symmetric 3D objects,
the conductivity values represent the product $\gamma h$, 
where $\gamma$ is the (cylindrically symmetric) 
three-dimensional conductivity distribution 
and $h$ is the height of the cylinder. 
Accordingly, the contact impedance  $z_{\ell}$ represents $z_{\ell} = \xi_{\ell}/h$,
where $\xi_{\ell}$ is the contact impedance in a three-dimensional model.

In Figs \ref{case13}-\ref{case73},
the first row shows photographs of the measurement
tank at the initial state (conductivity $\sigma_1$) and after the change 
(conductivity $\sigma_2$) and the linear difference reconstruction (E1). 
The second row shows the reconstruction (E2), where $\hat\sigma_1$ and $\hat\sigma_2$ are
computed as separate absolute reconstructions with the smoothness regularization,
and the third row represents the reconstruction (E3), which is
based on computing separate absolute reconstruction using the total variation regularization.
In both (E2) and (E3), the estimate for the target change $\delta \sigma$ is obtained
simply as
$\widehat{\delta \sigma} = \hat\sigma_2 - \hat\sigma_1$.
The fourth row shows the reconstruction (E4) with the proposed method. 
The region of interest
is shown with black line in the reconstructed images. 
The number of nodes inside the region of interest
was $320$, and thus 
$\bar{\sigma} = (\sigma_1\transp,\delta \sigma_{{\rm ROI}}\transp)\transp \in \mathbb{R}^{1650}$.

For quantitating the results,
we computed size estimates for the inclusions in the reconstructed images
%corresponding to the 
of the conductivity change $\delta\sigma$.
First, we estimated the areas of the inclusions in (E1)-(E4);
as a threshold for the inclusion boundaries we used
the mean values of the respective estimates $\widehat{\delta\sigma}$.
Secondly,
for test cases 2 and 3, where the true changes of the conductivities
were due to inclusions with rectangular shapes,
we also determined the widths of the inclusions in the reconstructions
$\widehat{\delta\sigma}$.
The width estimates were calculated as half widths of the
inclusions along the horizontal line at distance of $5.5$ cm above the center of
the domain.
Both the area and the width estimates and the true values are listed in 
Table \ref{tab.widthsandareas}.

%As can be seen from Figs \ref{case13}-\ref{case73}, 
%the proposed approach 
%gives in all three test cases the best reconstruction of the conductivity change $\delta \sigma$,
%and also the reconstructions of $\sigma_1$ and $\sigma_2$ are better with (E4) than (E2) or (E3).  
%The proposed approach gives in all cases good reconstruction of the shape and size of the inclusion, 
%and especially, 
%in  Figs \ref{case59} and \ref{case73}  (E4) is the only reconstruction from which  
%one can differentiate between the case of wider versus narrow  
%rectangular prism as the change from the same (inhomogeneous) background. 

The results of Case 1
with triangular inclusion are illustrated in 
Fig \ref{case13}.
All reconstructions methods detect the inclusion at least roughly
(estimates for the conductivity change $\delta\sigma$, third column).
However, the quality of the estimate (E4) obtained with the proposed
method is superior to qualities of the other estimates;
indeed, the triangular shape of the inclusion is tracked notably well in (E4).
Moreover, Table \ref{tab.widthsandareas} shows that the area estimate based on (E4)
is closest to the true area of the triangle.

Also in Cases 2 and 3,
different reconstruction methods detected the inclusions with varying accuracies.
In Case 2 (Fig \ref{case59}), where the rectangular inclusion was larger,
the inclusion is clearly visible in all estimates for $\delta\sigma$.
The size of the inclusion, however, is significantly overestimated in all reconstructions:
While the true area and width of the rectangle were 
28.70 cm$^2$ and 3.5 cm, respectively,
the area estimates vary between 41.04 cm$^2$ and 73.69 cm$^2$
and the width estimates between 5.54 cm and 6.47 cm, see Table \ref{tab.widthsandareas}.
The width estimate 5.54 cm closest to the true value was obtained with the proposed ROI-based
reconstruction method (E4).
In Case 3 (Fig \ref{case73}), estimate (E4) is clearly the best one:
(E4) is the only estimate showing even approximatively the  	
elongated shape of the inclusion.
Also the area and width estimates corresponding to (E4) are clearly closest to the true values.
Most importantly,
{\it (E4) 
is the only estimate capable for showing a clear difference between
the size of the inclusion in Case 2 and that in Case 3.}
Indeed, with reconstructions (E1)-(E3) the estimated widths of the inclusion
in Case 3 are close to those in Case 2,
but with (E4), the width estimate 
drops from 5.54 cm in Case 2
to 2.76 cm in Case 3.


\begin{table}[tb]
\caption{True areas and widths of the inclusions 
related to change of the conductivity $\delta\sigma$ (third column),
and
area and width estimates for the inclusions based on reconstructions (E1)-(E4).}
   \label{tab.widthsandareas}
\begin{tabular}{|c|c|c|c|c|c|c|}
\hline
&&True&(E1)&(E2)&(E3)&(E4)
\\
\hline
Case 1
&area (cm$^2$)&31.29&40.44&50.37&61.56&35.44\\
\hline
\multirow{2}{12mm}{\centering Case 2}
&area (cm$^2$)&28.70&41.04&51.30&73.69&46.64\\
&width (cm)&3.5&6.42&5.85& 6.47&5.54\\
\hline
\multirow{2}{8mm}{\centering Case 3}
&area (cm$^2$)&6.00&35.91&42.44&100.74&14.45\\
&width (cm)&1.0&6.13&6.91&8.68&2.76\\
\hline
\multirow{2}{8mm}{\centering Case 4}
&area (cm$^2$)&0.64&47.93&6.00&2.40&0.75\\
&width (cm)&0.39&1.13&1.11&1.17&0.47\\
\hline
\end{tabular}
\end{table}

%%%%%%%%%%%%%%%%%%%%%%%%%%%%%%%%%%%%%%%%%%%
%%%%%%%%%%%%%%%%%%%%%%%%%%%%%%%%%%%%%%%%%%%

\section{Simulation study with a larynx model}
\label{sec.simulation}

%With the aid of EIT, it might also be possible to get information on human larynx: 

As the last test case (Case 4), 
we consider a simulation study of a possible new application of EIT;
imaging of the vocal folds. 
The conductivity changes detected by EIT 
could potentially be used for 
detecting the vocal folds movement during speech production, 
or estimating the physiological changes in the vocal fold tissue caused by vocal loading. 
This information could be utilized for detecting and quantifying vocal loading (i.e. getting estimates of stresses acting upon the tissue) and measuring the consequences of vocal loading (i.e. changes in the tissue). 
Indeed, EIT has a high potential for glottal diagnostics. 
Basically the data used in EIT consists of similar measurements that are also used in electroglottography (EGG) 
\cite{Fourcin1971,Rotternberg1992}, 
which is a widely used tool in the assessment of voice production. 
However, while in standard EGG two-channel impedance
measurement data is measured for producing a (scalar valued) time serie of changes in tissue impedance, 
EIT is based on multi-channel data and produces 
%advanced mathematical modeling for formation of 
tomography images of the impedance changes between the measurement frames.
%in the data processing. 
%This enables estimation of spatial properties of the larynx, in addition to temporal change information provided by EGG. 
% A dual-channel EGG has been used since the 1990’s \cite{Rotternberg1992}. 
Recently, in \cite{kob2009,Hezard}
multi-channel-EGG systems for improving the assessment of glottal opening and the laryngeal position
have been proposed.
In these papers, the measurement setup was similar to EIT, 
but the data was not used for 3D image
reconstruction. 
However, the results in \cite{kob2009} indicated that 
it is possible to track the location of glottis during a
swallowing manoeuvre.

\begin{figure}  [ht] 
%\caption{Stimuli Category Explanations} 
\centering
  \begin{tabular} {cccc}
   & $\sigma_1$ & $\sigma_2$ & $\delta\sigma$ \\
      True & \parbox[c]{3cm}{\includegraphics[width=3cm]{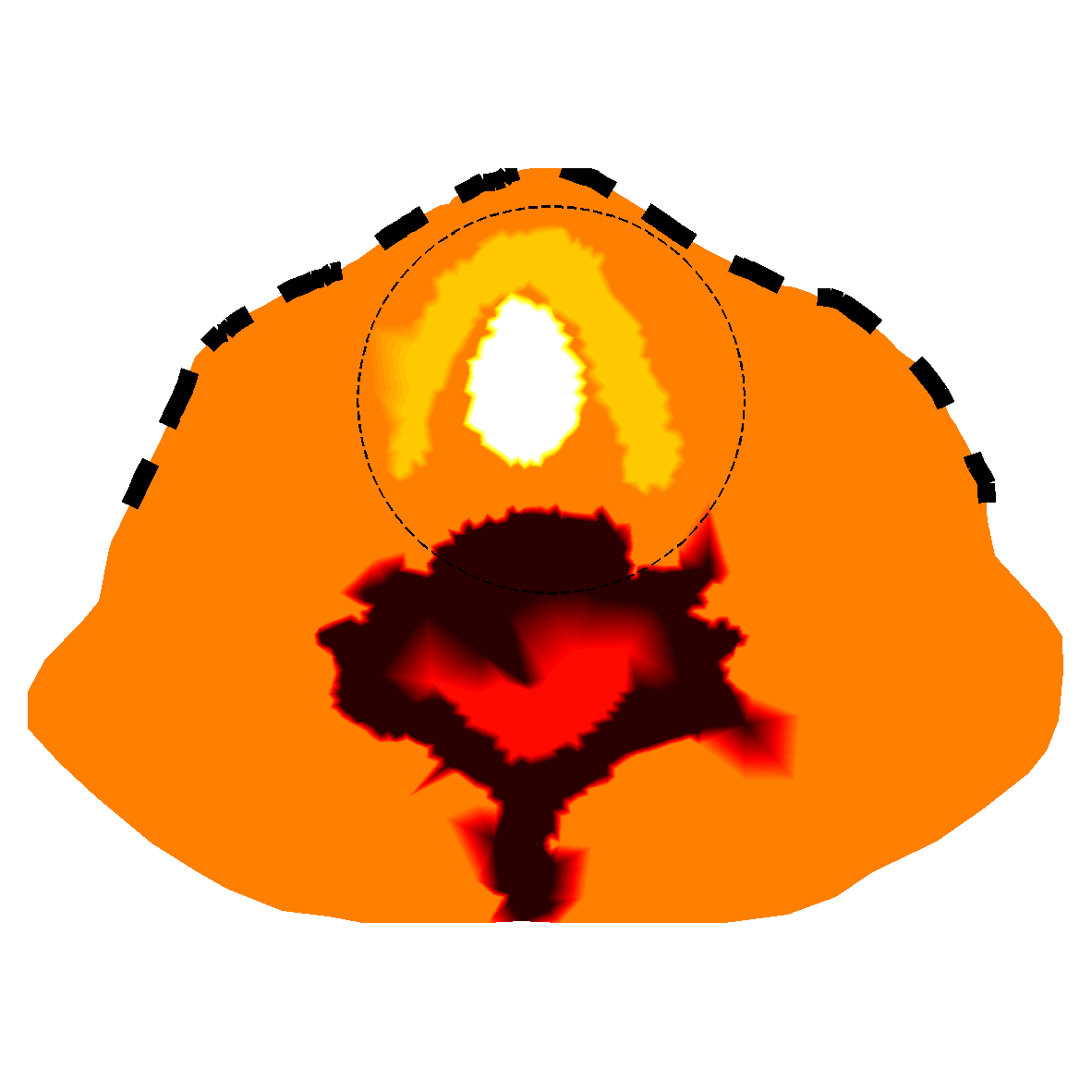}}
      & \parbox[c]{3cm}{\includegraphics[width=3cm]{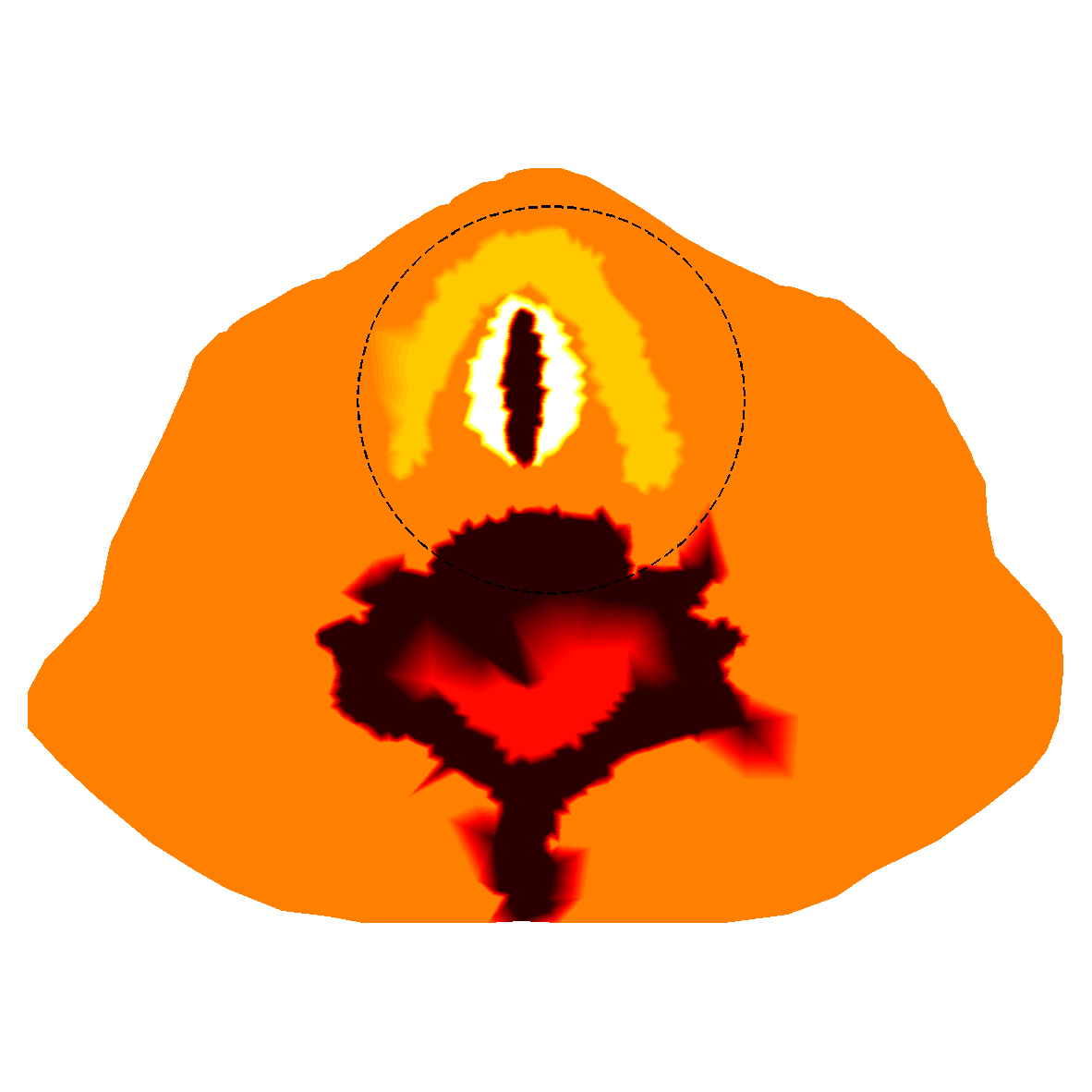}}
      & \parbox[c]{3cm}{\includegraphics[width=3cm]{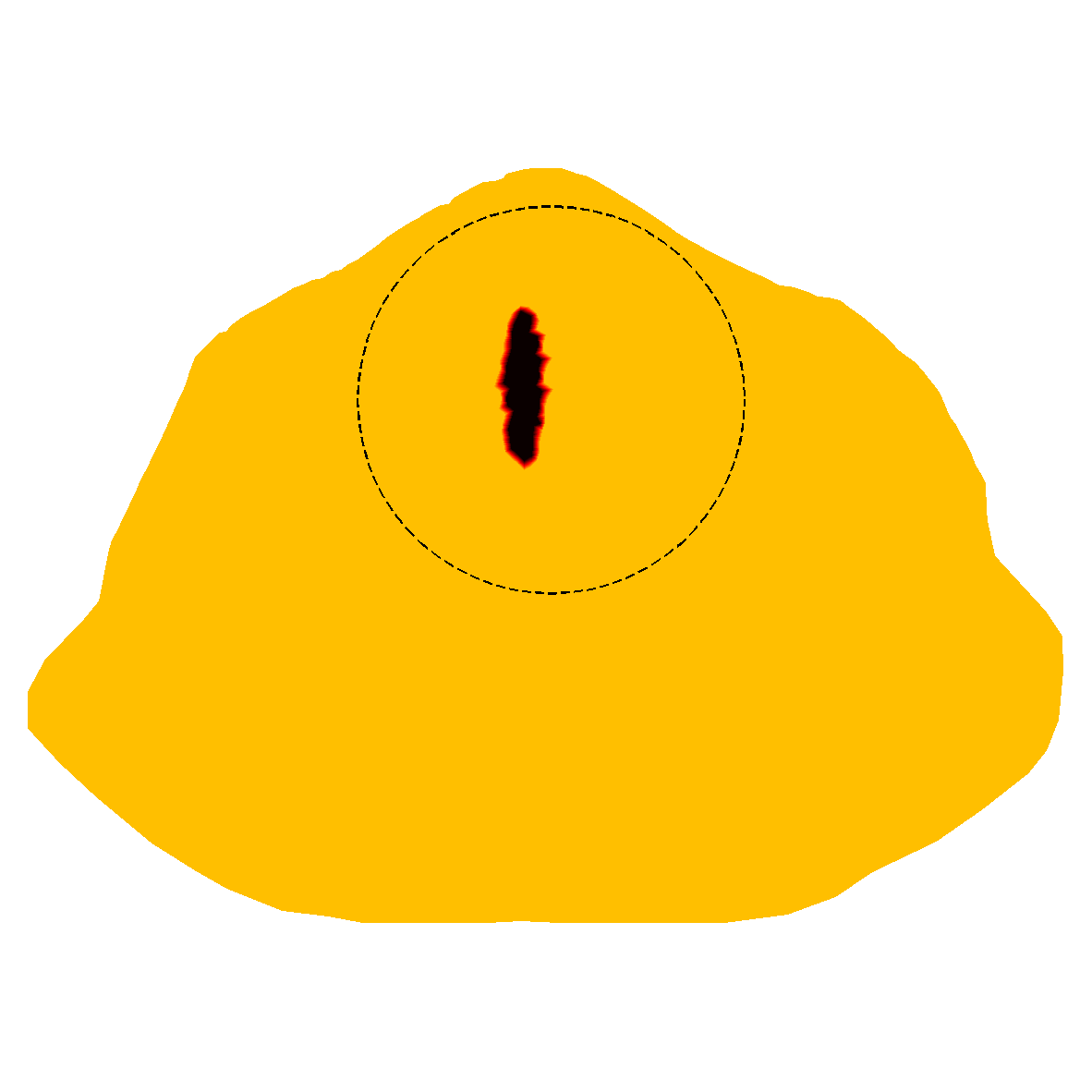}} \\[3 ex]
      %%%%%%%%%%%%%%%%%%%%%%%
      (E2) & \parbox[c]{3cm}{\includegraphics[width=3cm]{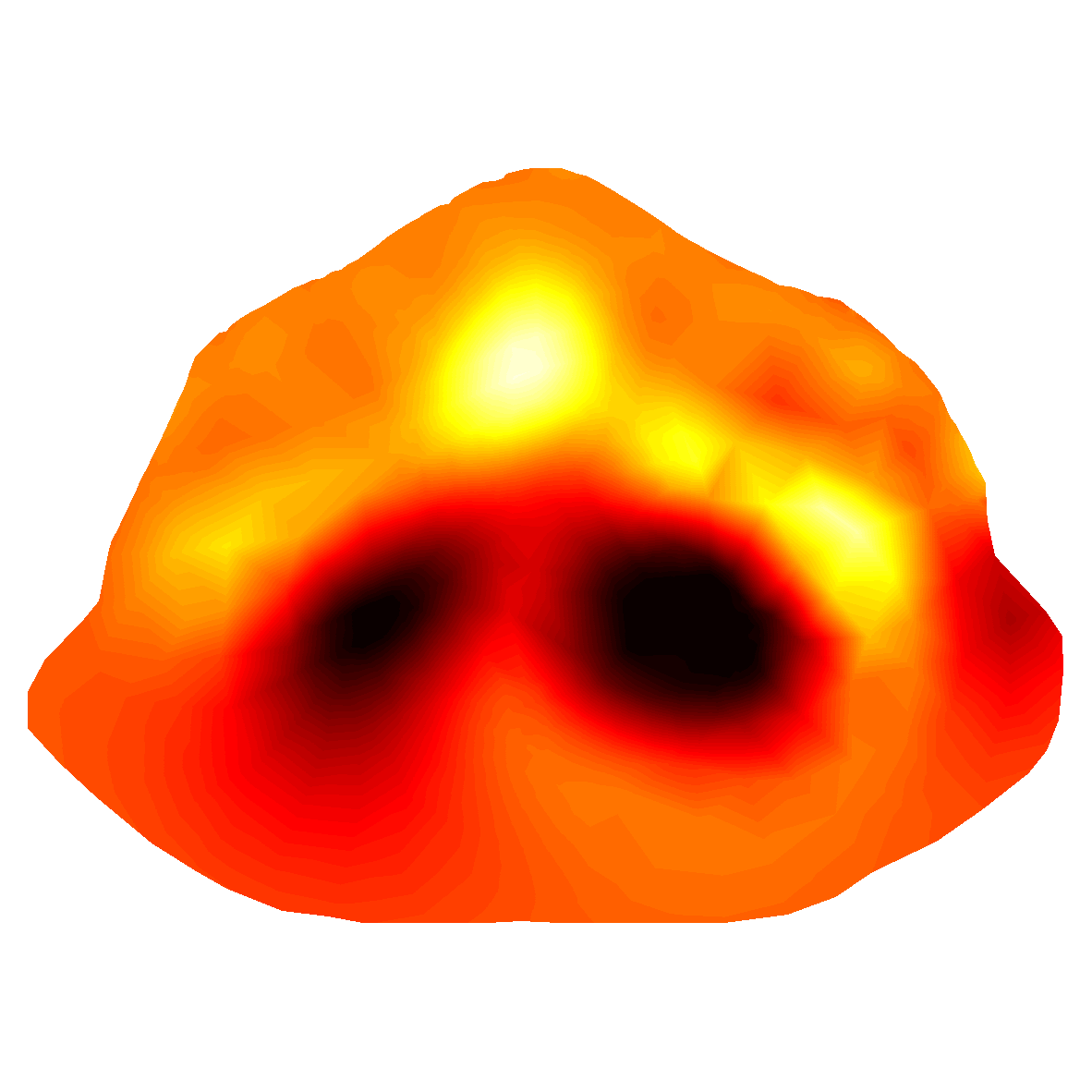}}
      & \parbox[c]{3cm}{\includegraphics[width=3cm]{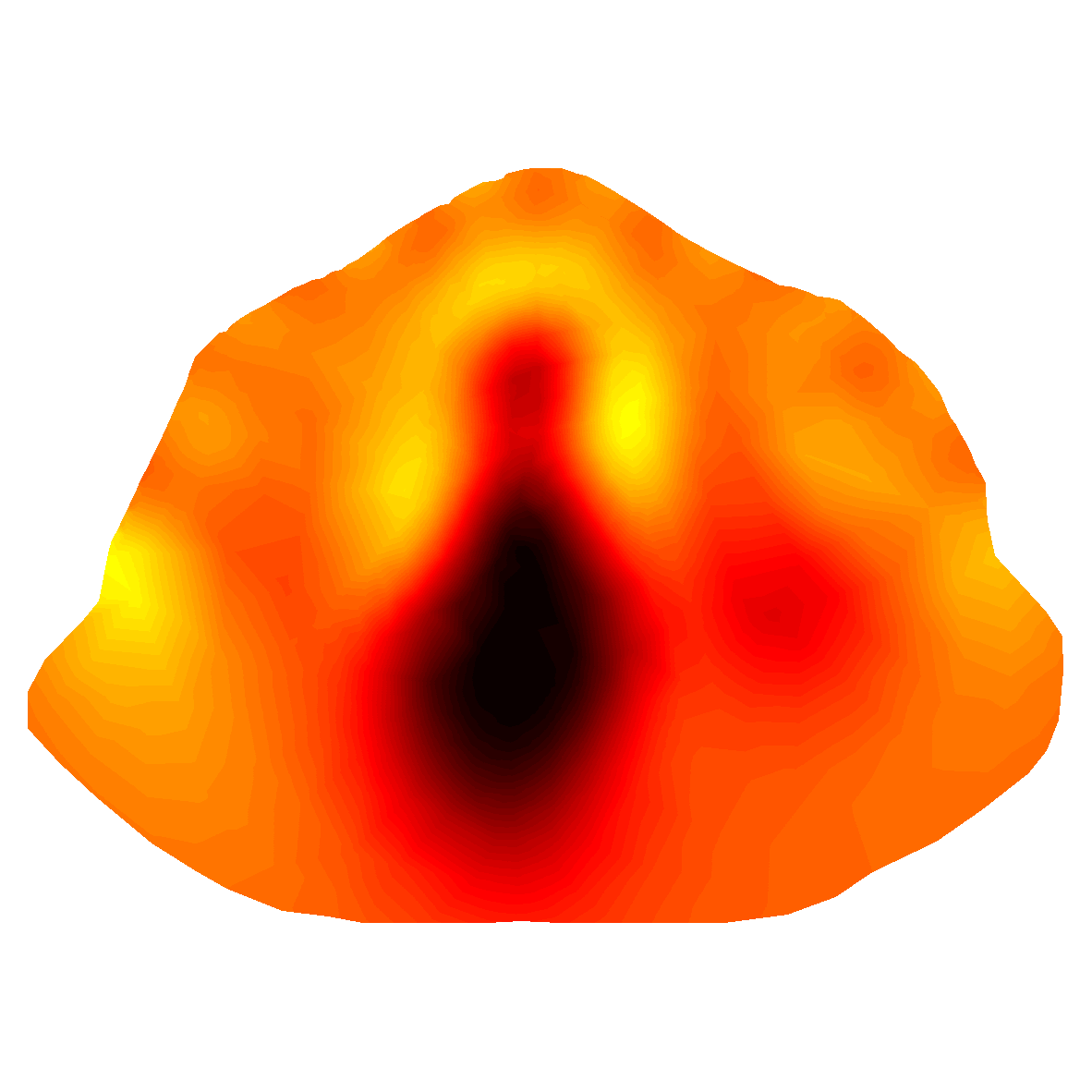}}
      & \parbox[c]{3cm}{\includegraphics[width=3cm]{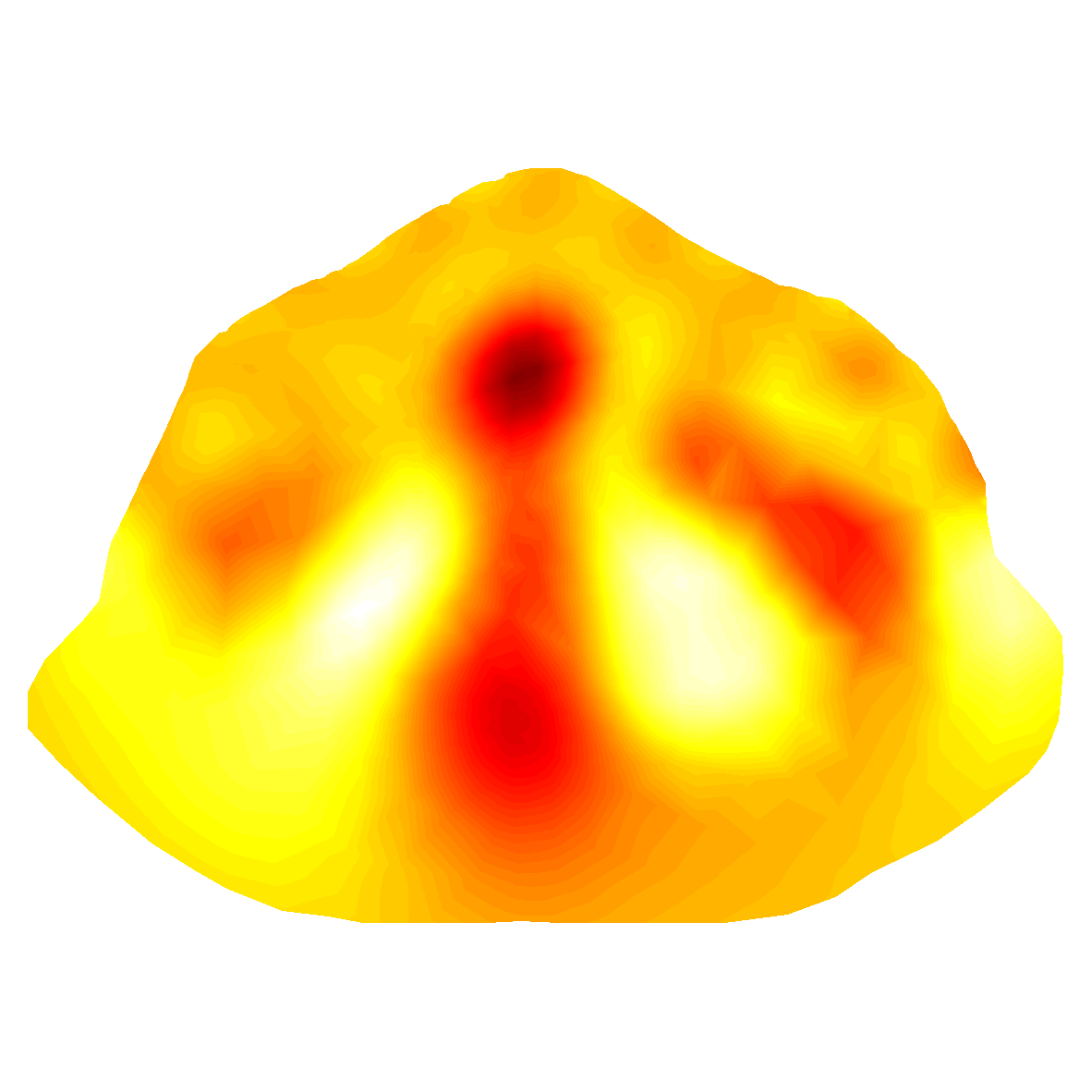}} \\[3 ex]
            %%%%%%%%%%%%%%%%%%%%%%%
      (E3) & \parbox[c]{3cm}{\includegraphics[width=3cm]{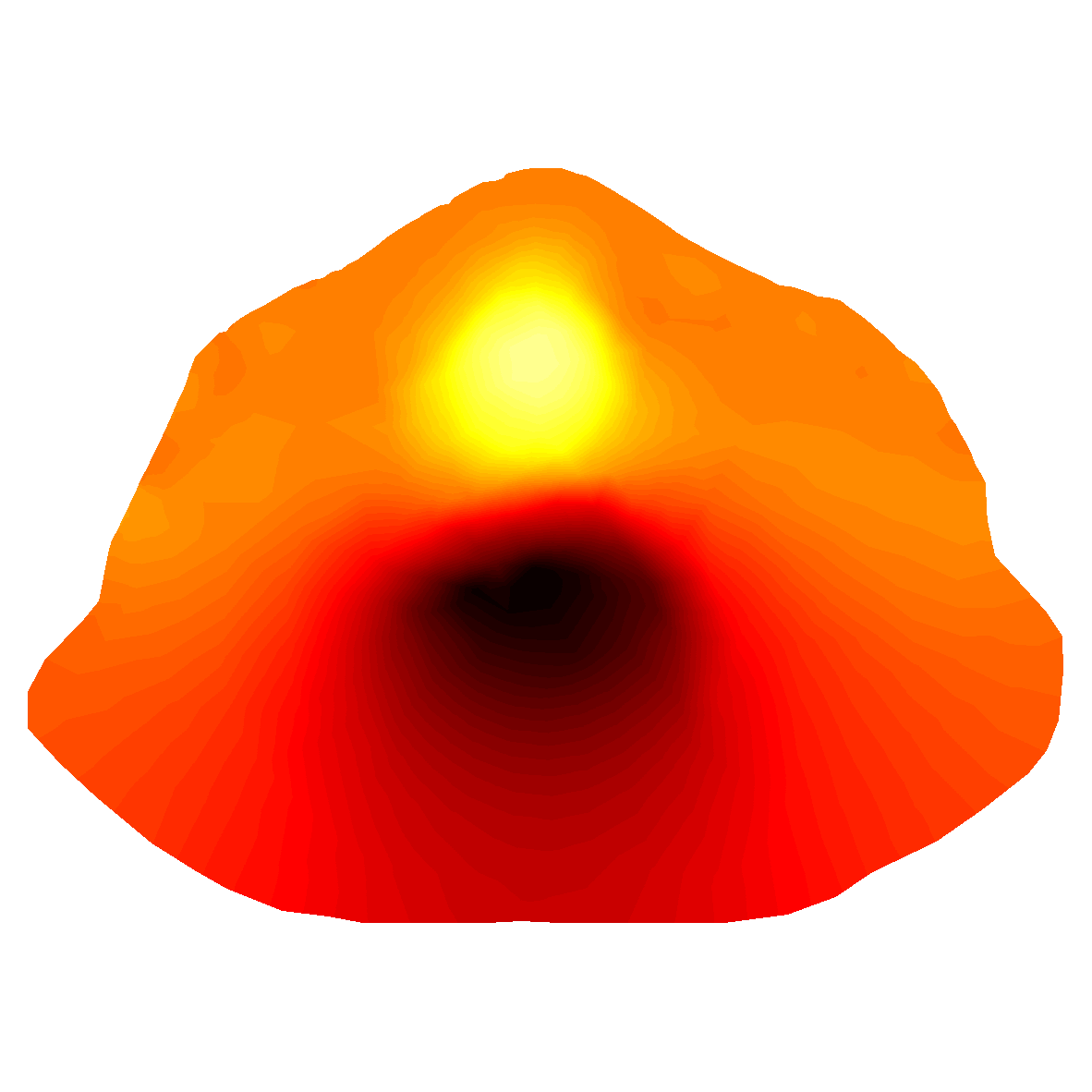}}
      & \parbox[c]{3cm}{\includegraphics[width=3cm]{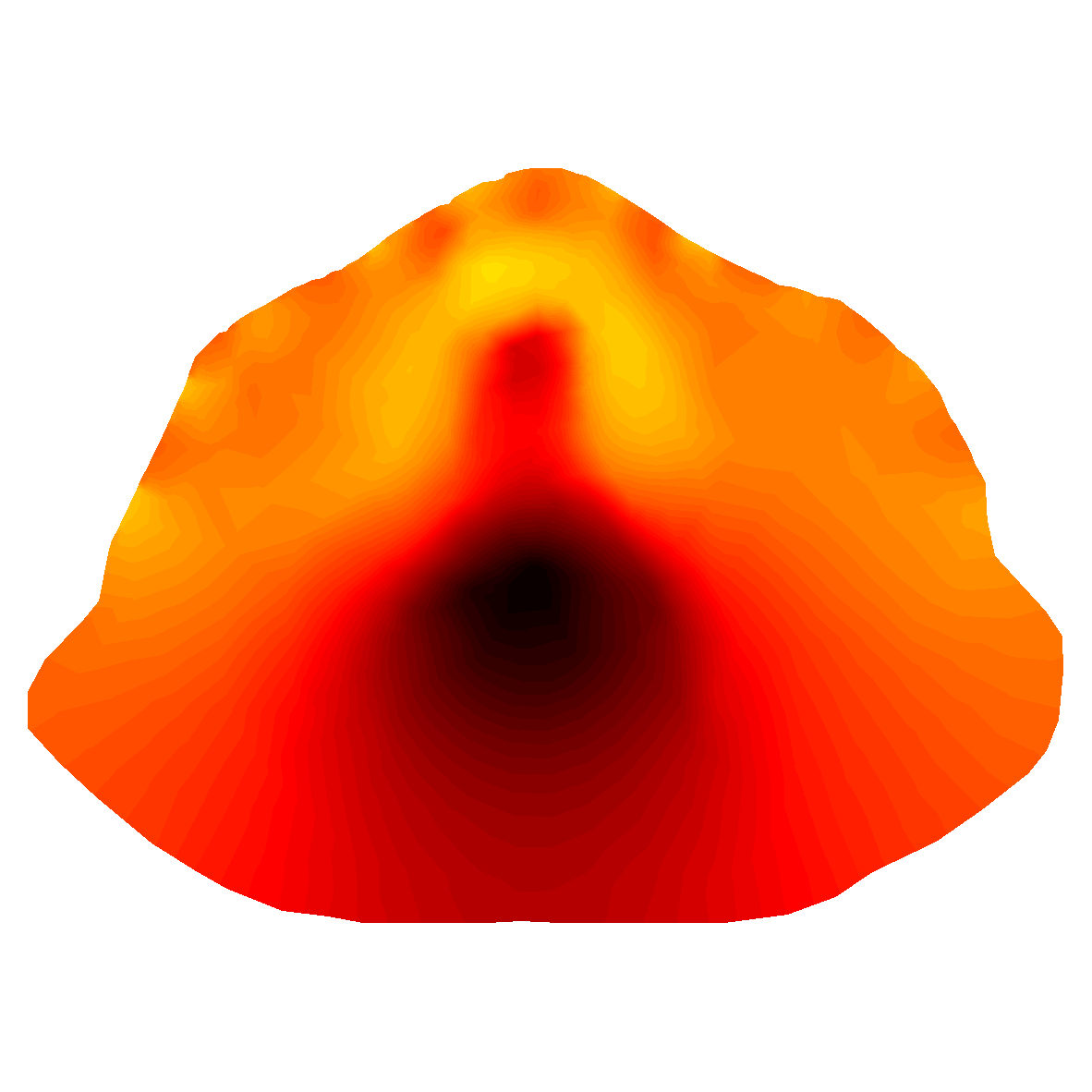}}
      & \parbox[c]{3cm}{\includegraphics[width=3cm]{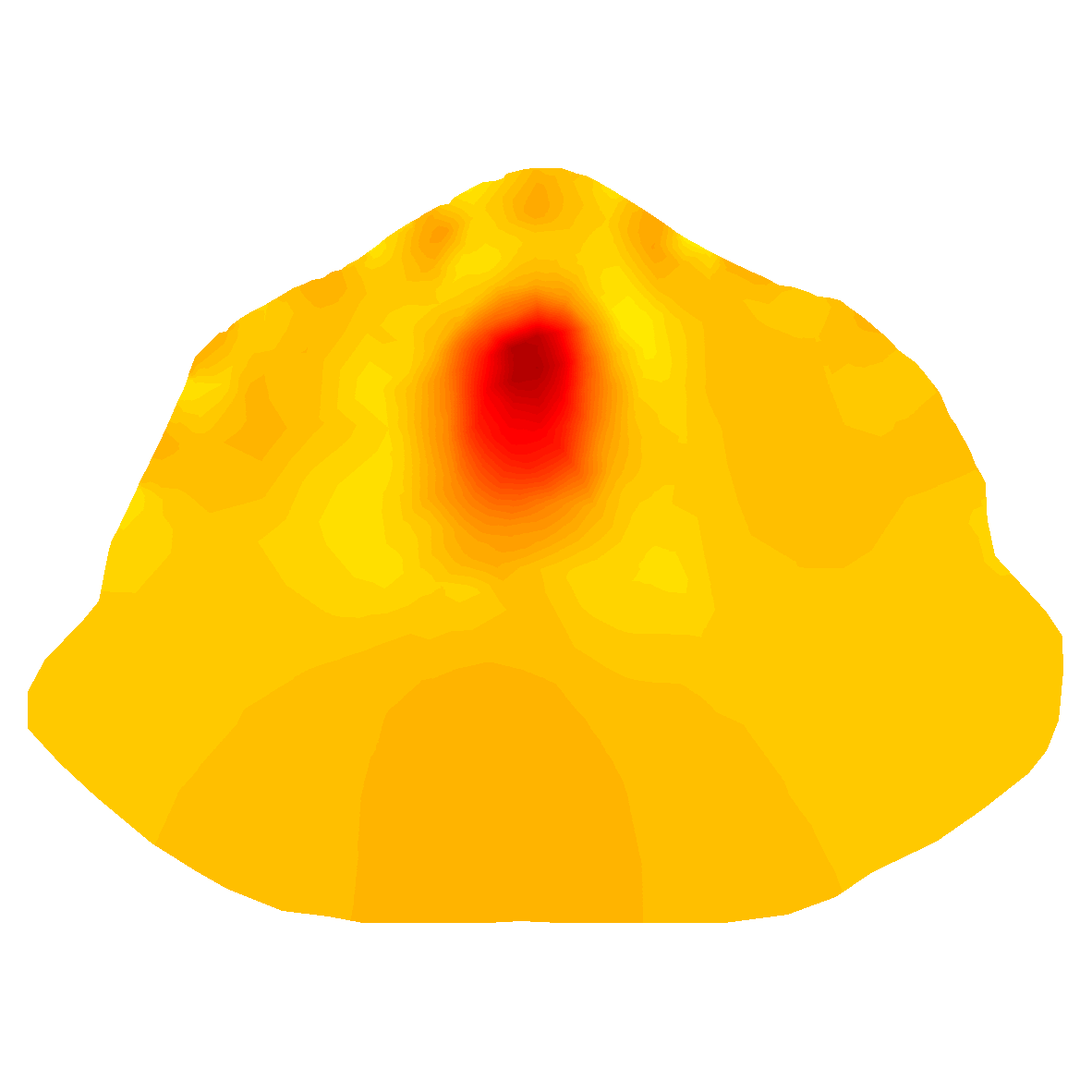}} \\[3 ex]
                  %%%%%%%%%%%%%%%%%%%%%%%
     (E4) & \parbox[c]{3cm}{\includegraphics[width=3cm]{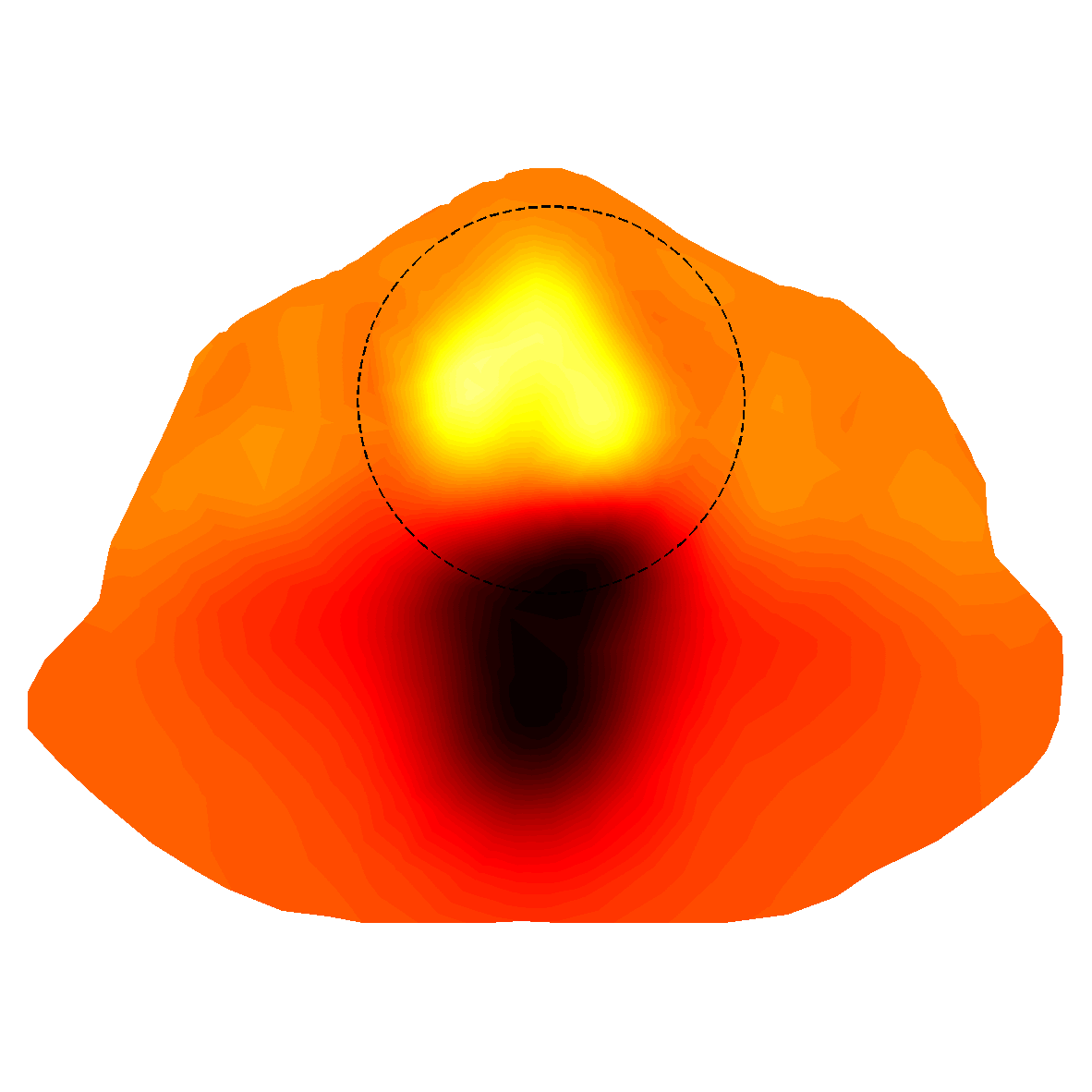}}
      & \parbox[c]{3cm}{\includegraphics[width=3cm]{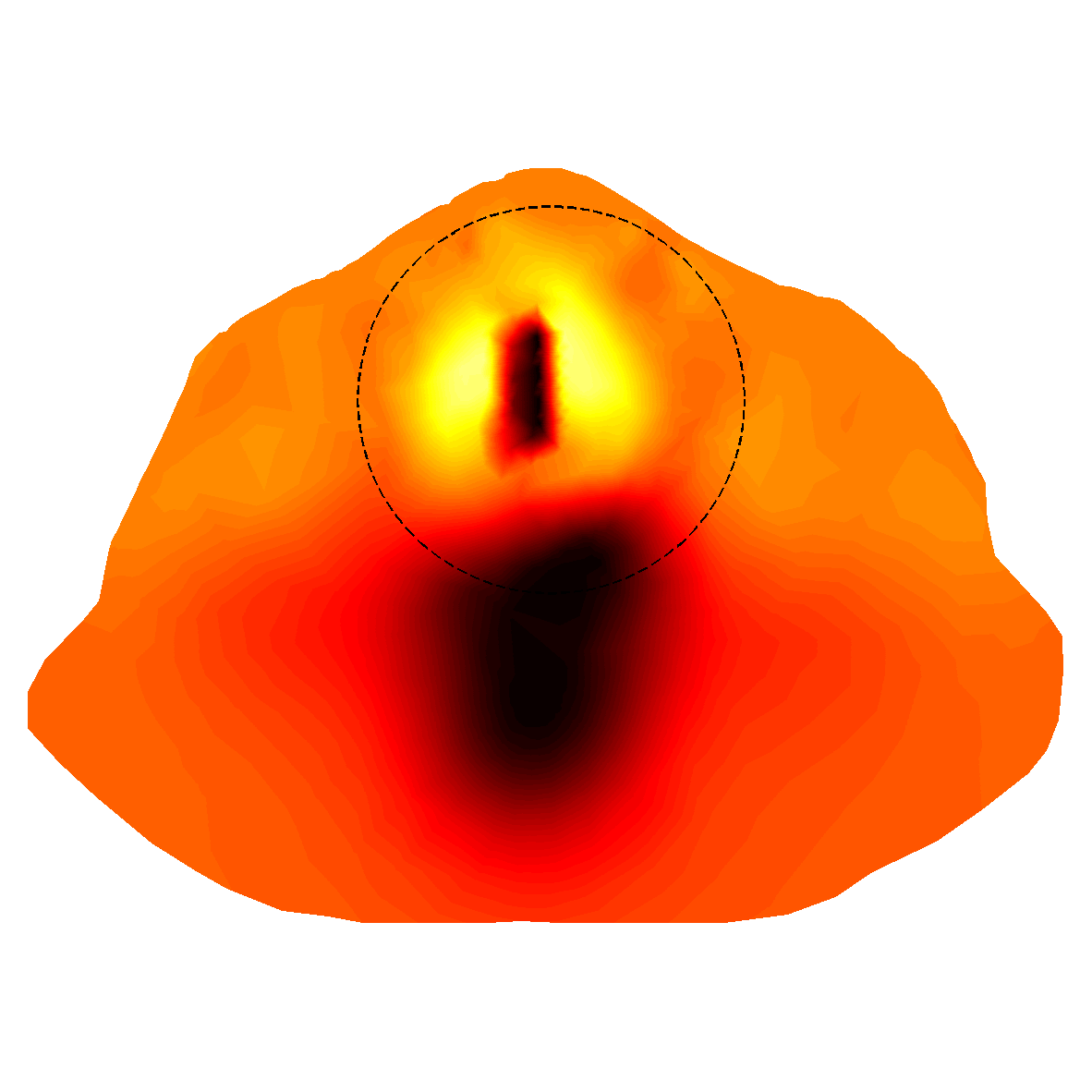}}
      & \parbox[c]{3cm}{\includegraphics[width=3cm]{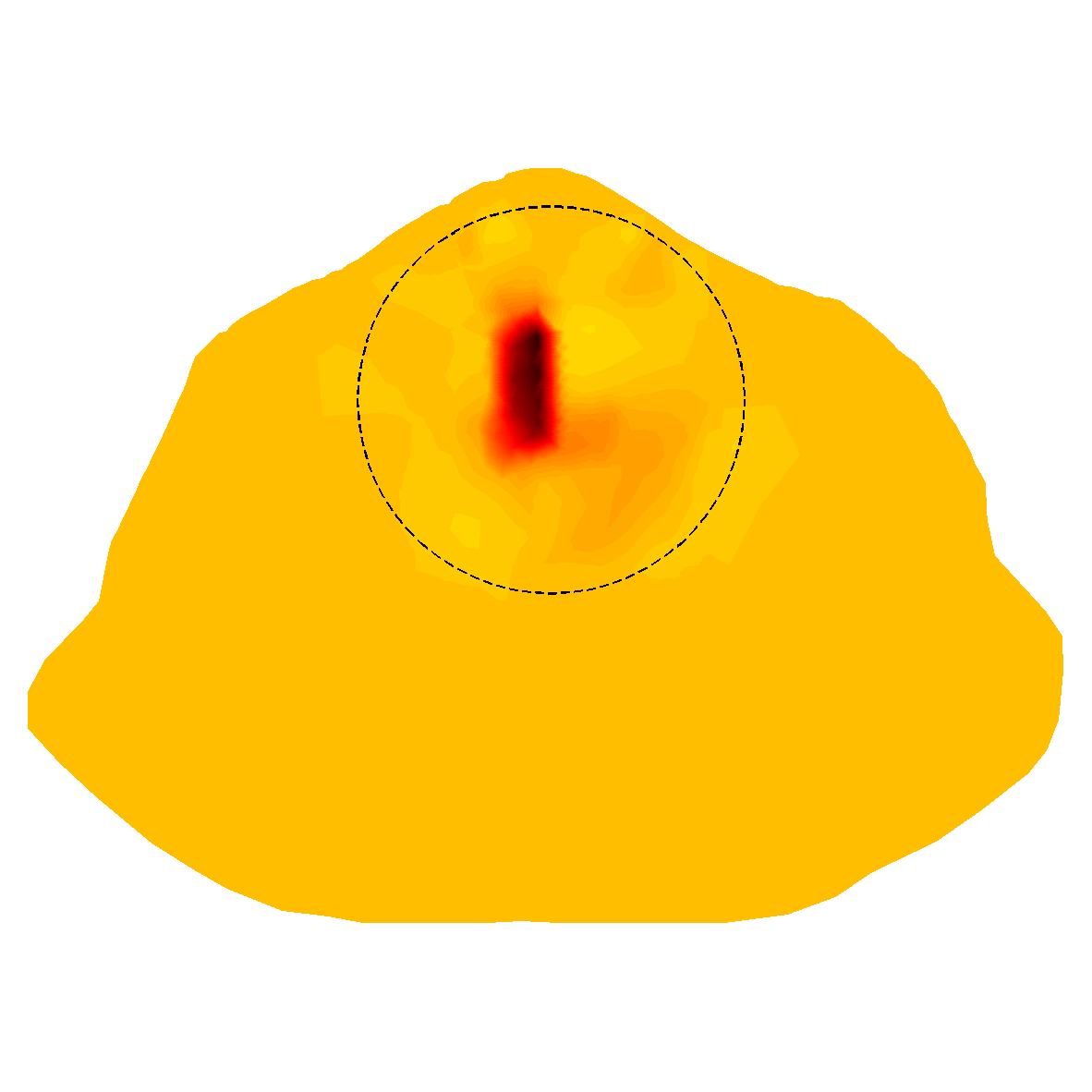}} \\[3 ex]
      %%%%%%%%%%%%%%%%%%%%%%%%%
       &\parbox[c]{3cm}{\includegraphics[width=3cm]{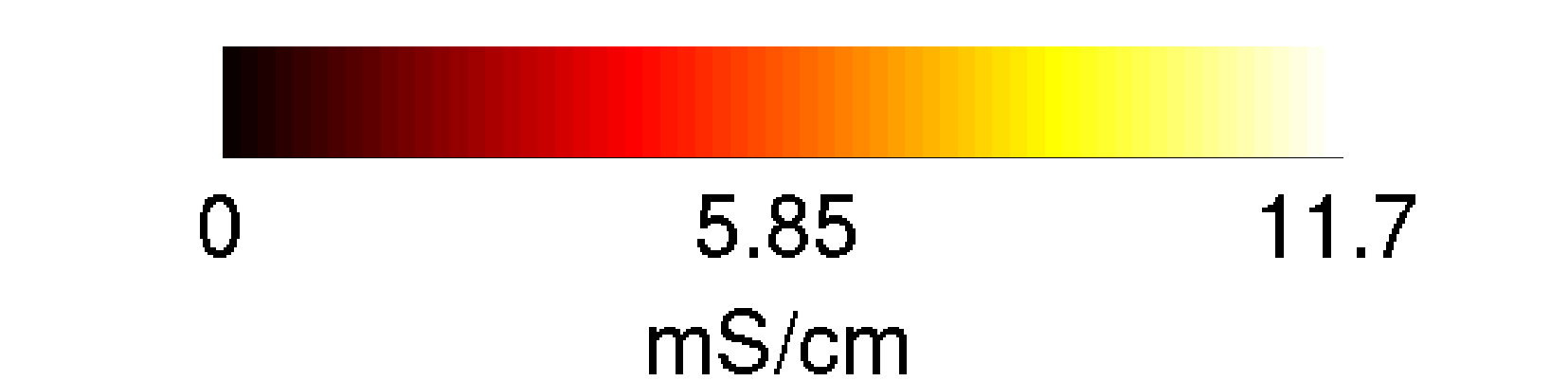}}
        &\parbox[c]{3cm}{\includegraphics[width=3cm]{figs/case99/cbs1.png}}
         &\parbox[c]{3cm}{\includegraphics[width=3cm]{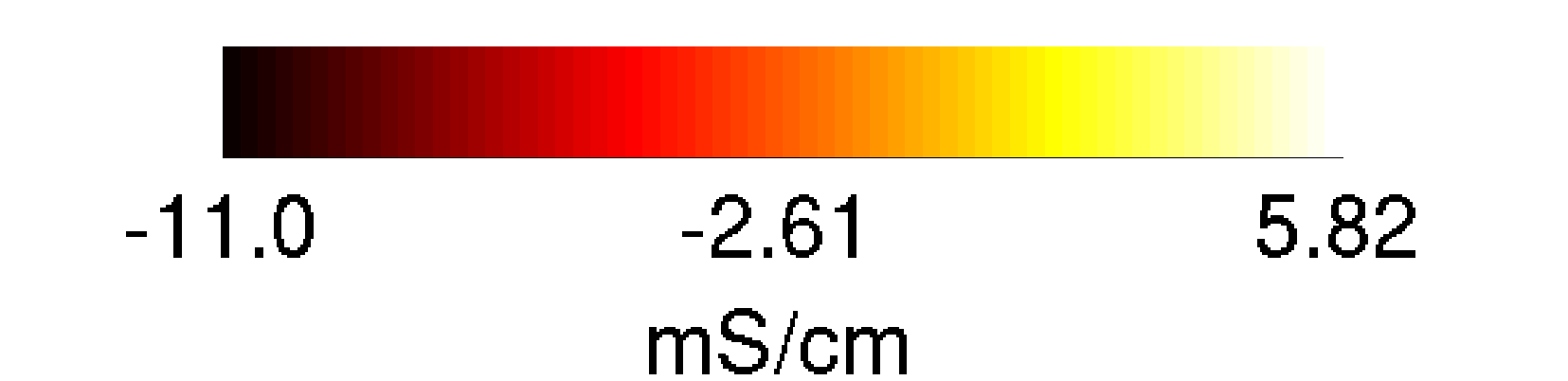}} \\[0 ex]
               %%%%%%%%%%%%%%%%%%%%%%%%%
       (E1) & & &\parbox[c]{3cm}{\includegraphics[width=3cm]{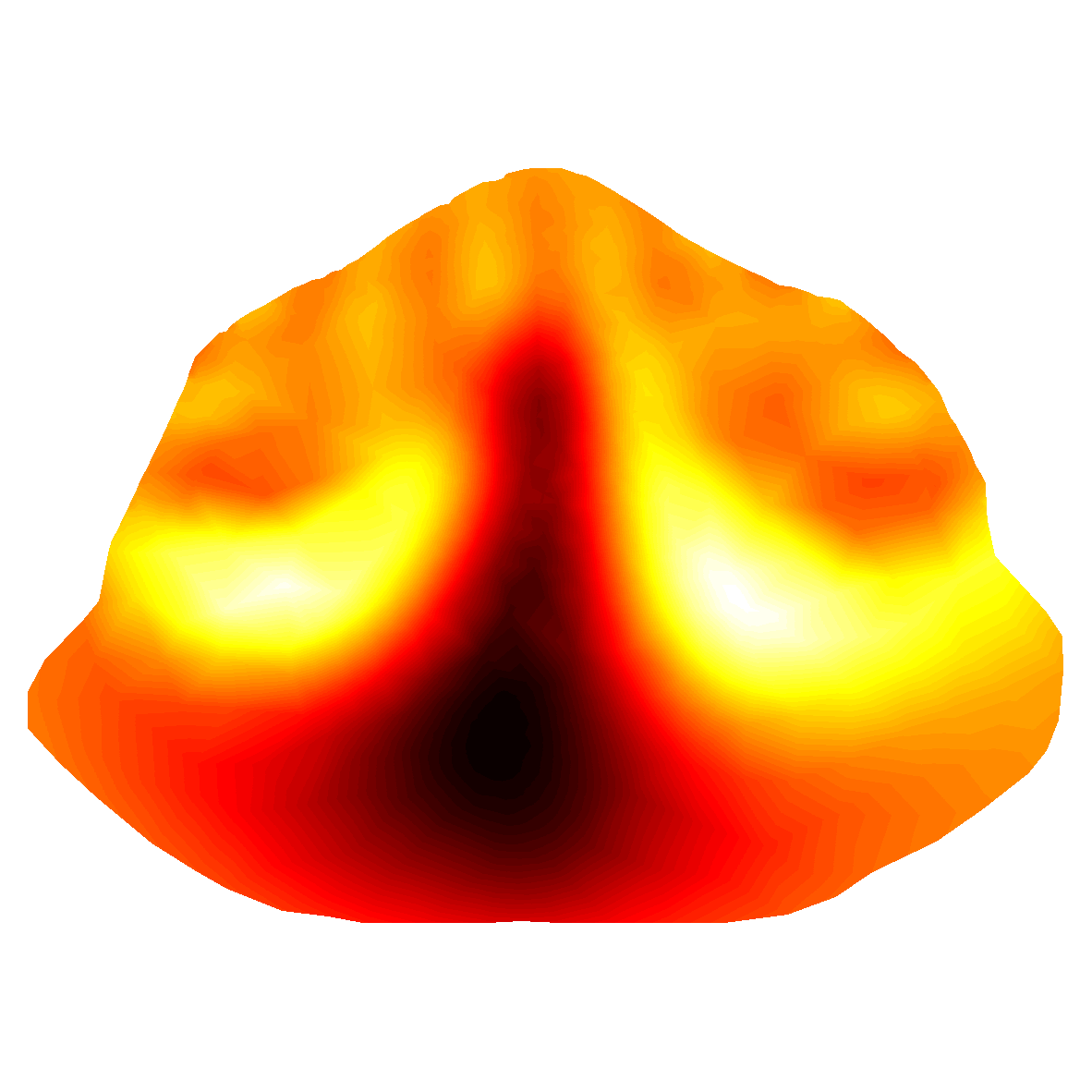}} \\[3 ex]
       %%%%%%%%%%%%%%%
            & &  &\parbox[c]{3cm}{\includegraphics[width=3cm]{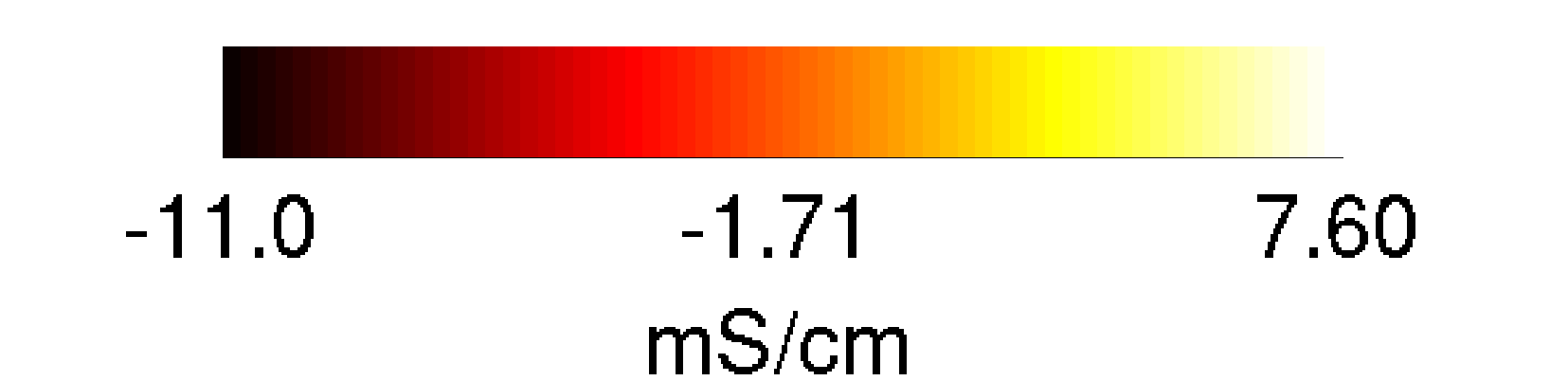}} \\[0 ex]
    \end{tabular}
    \caption{Case 4: Simulated test case of imaging of vocal folds. 
(E1)-(E4) refer to the estimates listed in
section \ref{estsec}. 
The color scale denotes the common minimum to maximum scale of the conductivity values.}
\label{case99}
    \end{figure}

%The results of Case 4 are shown in Fig. \ref{case99}. 
In Fig. \ref{case99},
the top row shows
the true initial conductivity $\sigma_1$ (left), 
the conductivity after the change $\sigma_2$ (middle) and the true difference
$\sigma_2 - \sigma_1$ (right). 
The simulation geometry was taken from a computerized tomography (CT) image of the neck area. 
The CT image was segmented into subdomains corresponding to
bone, bone marrow, soft tissues, cartilage and trachea. 
The width of the neck was 11.78 cm.
The conductivities of the tissues
were set as 0.65 mS for bone, $4.55$ mS for bone marrow, 
$6.5$ mS for soft tissues and $7.8$ mS for cartilage, 
corresponding to values found in the literature \cite{gabriel2009electrical,binette2004tetrapolar}.
Note, however, that in reality,
i) there are more fine structures across the neck than in our model case, 
ii) there is some variation in the tissue conductivity also within each organ, and
iii) the target is three-dimensional.
The initial state $\sigma_1$ corresponds to a case where the vocal folds are closed 
by holding breath.
The conductivity $\sigma_2$ after the change corresponds to a case
where the vocal folds are partially open, 
forming a thin non-conducting opening in the trachea.

%Our hypothesis is that setting the electrodes near the
%region where the change of the target is known to take place
%is beneficial for the proposed ROI based reconstruction,
%and it may yield to reconstructions with a high accuracy.
%For this reason,
%in this simulation
In this simulation, a partial boundary measurement geometry was employed such
that
$L= 12$ electrodes were placed on the frontal part of the neck boundary near the glottal area, 
the electrode array covering less than half of the boundary $\partial \Omega$.
The electrodes are illustrated with thick black lines
in Fig. \ref{case99} (top left).
The rightmost electrode was identified with electrode index $\ell=1$,
and the electrode indices increased in counter clockwise direction.
In the simulation of the measurement data,
pairwise current injections were employed
such that one of the electrodes was fixed as the sink and
currents were applied sequentally between the sink and each one of the
remaining 11 electrodes. This process was repeated using electrodes 
$\{1,3,5,7,9\}$ as the sink.
% $e_{i,i=1,3,5,7,9} \leftrightarrow e_{j, j=1,...,L, j\neq i}$.
% $e_3\rightarrow e_{i, i=2,...,L}$,
% $e_5\rightarrow e_{i, i=2,...,L}$,
% $e_7\rightarrow e_{i, i=2,...,L}$ and
% $e_9\rightarrow e_{i, i=2,...,L}$. 
For each current injection,
voltage measurements between electrode $1$ and
the remaining 11 electrodes were taken.
% $e_1 \leftrightarrow e_{i, i=2,...,L}$ 
The data was computed using a FE mesh with $11815$ nodes and $5824$ triangular elements. 
Random additive Gaussian noise with standard deviation of $0.25 \%$ of the computed noiseless 
voltages was added to the data.

In the computation of the reconstructions, 
the conductivity was approximated in a first order piecewise linear basis
with $932$ nodes, i.e. the $\sigma \in \mathbb{R}^{932}$. 
The parameters $a$, $b$ and $c$  in the construction of 
$\Gamma_\sigma$ (\ref{eq.sm.pr.covmat}) 
used in reconstructions (E1), (E2) and (E4)  were set as
$a= 15.15 \mathrm{m}\mathrm{S}^2$,
$b= 7.25\, {\rm cm}$
and
$c = 1.52\times10^{-2} \mathrm{m}\mathrm{S}^2$. 
As the expectation $\sigma^\ast$ in (E2) and (E4), 
the best homogeneous conductivity estimate 
based on data set $V_1$ was used.
The parameter $\alpha$ for the TV functional
was selected as $\alpha = 0.25$. 
For the parameter $\beta$ in the TV functional, we used 
value $\beta = 0.05$.

The standard difference reconstruction (E1) is shown in the bottom row in Fig \ref{case99}. 
The reconstruction of the change $\delta \sigma$ is highly erroneous 
and has elongated artefacts towards the back of neck domain. 
These artefacts are due to the inhomogeneous background conductivity 
and the partial boundary problem where measurement electrodes 
are placed only at a small portion of the boundary. 
The reconstructions (E2) and (E3), which are based on computing separate 
absolute reconstructions of $\sigma_1$ and $\sigma_2$ are 
shown in the second and third row in Fig \ref{case99}.
Although the resolutions of reconstructions  (E2) and (E3)
are not very high, they are clearly better than (E1).
Especially the reconstruction (E3) which uses TV regularization is quite feasible. 

The reconstruction (E4) with the proposed approach is shown on the fourth row
in Fig \ref{case99}.
This reconstruction clearly outperforms (E1)-(E3);
especially, the size of the glottal opening has been reconstructed with
a significantly better accuracy in (E4) than in the other estimates. 
The area and width estimates are given in Table \ref{tab.widthsandareas}.
The width estimates were computed as the half widths of the
reconstructed inclusions along a line cross-secting the glottis from the middle horizontally.
The estimated width corresponding to reconstruction (E4) is 0.47 cm,
which was relatively close to the true width 0.39 cm,
while the other width estimates vary between 1.11 cm and 1.17 cm.

In this simulation,
the electrodes were set only on the frontal part of the domain boundary
that is close to the region-of-interest where the change of the target was
known to take place.
The results demonstrate that the ROI based reconstruction can tolerate well such a partial boundary setting:
Indeed, although the background conductivity is rather complex, with small structures,
reconstruction (E4) estimates width of the glottis opening with less than 1 mm accuracy,
which is a significant improvement to the resolution of
the conventional reconstructions (E1)-(E3).
This finding indicates that one could employ existing multichannel EGG measurement geometry 
(see e.q. \cite{kob2009}) for EIT imaging of vocal folds 
and therefore the ROI based reconstruction could allow 
flexibility in design of a practical measurement system.
%The result suggests that restricting the conductivity 
%change to a region of interest can be 
%particularly beneficial in partial boundary problems. 

%

\section{Conclusions}
\label{conclusion}

In this paper, we proposed a novel approach
to EIT image reconstruction in cases where
EIT measurement of a time-varying target is available
before and after a change of the target. In the proposed
approach, the conductivity after the change is represented 
as a linear combination of the initial conductivity and the change, 
and the EIT inverse problem is formulated as simultaneous reconstruction
of the initial conductivity and the change  
based on the two EIT data sets using the regularized least squares formalism.
The approach enables the use of different spatial models for the initial conductivity and
the change by different regularization functionals and it also allows for the restriction of the conductivity
change to a region of interest in cases where the changes are {\em a priori} known to occur in
a certain subvolume of the body.  

The proposed approach was tested with three test cases using experimental data from  
laboratory set-up and one simulated test case related to EIT imaging of vocal folds. 
The proposed approach outperformed the conventional difference imaging approach and the
frame-by-frame absolute imaging approach in all test cases. 
The findings suggest that the proposed approach
can be useful in EIT applications where one is interested in 
detecting a change in the conductivity between two time instants,
especially when the conductivity change can be restricted to a
relatively small subdomain.
The results of the simulation study also suggest that 
the approach can be particularly beneficial in partial boundary data problems. 

In this study we computed results using 2D computational models. 
However, extension to 3D is straightforward.
Furthermore,
the proposed approach is based on solving the non-linear inverse problem, 
which is known to be prone to modelling errors,
such as poorly known electrode locations, boundary shape and domain truncation.
A few alternative methods for recovering from modeling errors exist,
% (see Section \ref{sect.Introduction} and the references within)
and combining those methods with the ROI reconstruction
proposed in this paper is a relatively straightforward task.

\section*{Acknowledgments} The authors thank Mika Mikkola and Rami Korhonen from
Department of Applied Physics, 
University of Eastern Finland,
for their support with the MIMICS 
software in the glottis simulation.

% You may incorporate your references as follows in your main tex file.
% Using BibTex is not recommended but can be handled.

\bibliographystyle{AIMS}
\bibliography{Ref_Sep2012B}

\medskip
% The data information below will be filled by AIMS editorial staff
Received xxxx 20xx; revised xxxx 20xx.
\medskip

\end{document}